\documentclass[12pt]{article}
\usepackage[a4paper, total={7in, 10in}]{geometry}
\usepackage{soul}
\usepackage[toc,page]{appendix}
\usepackage[utf8]{inputenc}
\usepackage[english]{babel}
\usepackage[dvipsnames,table,xcdraw]{xcolor} 
\usepackage{graphicx}
\usepackage{longtable}
\usepackage{subcaption}
\usepackage{url}
\usepackage[colorlinks=true, allcolors=blue]{hyperref}

\usepackage{enumitem}
\usepackage{multirow}
\usepackage{amsmath}
\usepackage{comment}
\usepackage{float}
\usepackage{svg}
\usepackage{MnSymbol}

\usepackage[backend=biber,style=ieee,refsegment=section,defernumbers=true, sorting=none]{biblatex}
\usepackage{csquotes}
\usepackage{chngcntr}
\usepackage{authblk}
\makeatletter
\let\blx@rerun@biber\relax
\makeatother

\providecommand{\keywords}[1]
{
  \small	
  \textbf{\textit{Keywords---}} #1
}

\title{Deriving and Evaluating a Detailed Taxonomy of Game Bugs}
\author[1]{Nigar Azhar Butt\thanks{Email: \texttt{i212842@nu.edu.pk}}\thanks{Corresponding author}}
\author[2]{Salman Sherin \thanks{Email: \texttt{salman.sherin@questlab.pk}}}
\author[1]{Muhammad Uzair Khan\thanks{Email: \texttt{uzair.khan@nu.edu.pk }}}
\author[1]{Atif Aftab Jilani \thanks{Email: \texttt{atif.jilani@nu.edu.pk }}}
\author[2]{Muhammad Zohaib Iqbal\thanks{Email: \texttt{zohaib.iqbal@questlab.pk }}}

\affil[1]{Department of Software Engineering, National University of Computer and Emerging Sciences, Islamabad, Pakistan}
\affil[2]{QUEST, Islamabad, Pakistan}

\date{}
\begin{document}

\maketitle

\begin{abstract}
 Game development has become an extremely competitive multi-billion-dollar industry. Many games fail even after years of development efforts because of game-breaking bugs that disrupt the game-play and ruin the player experience.
 The goal of this work is to provide a bug taxonomy for games that will help game developers in developing bug-resistant games, game testers in designing and executing fault-finding test cases, and researchers in evaluating game testing approaches.
 For this purpose, we performed a Multivocal Literature Review (MLR) by analyzing 436 sources, out of which 189 (78 academic and 111 grey) sources reporting bugs encountered in the game development industry were selected for analysis. We validate the proposed taxonomy by conducting a survey involving different game industry practitioners. 
 The MLR allowed us to finalize a detailed taxonomy of 63 game bug categories  in end-user perspective  including eight first-tier categories: Gaming Balance, Implementation Response, Network, Sound, Temporal, Unexpected Crash, Navigational, and Non-Temporal faults. 
 We observed that manual approaches towards game testing are still widely used. Only one of the approaches targets sound bugs whereas game balancing and how to incorporate machine learning in game testing is trending in the recent literature. Most of the game testing techniques are specialized and dependent on specific platforms.
\end{abstract}

\section{Introduction}
Game development is a multi-billion-dollar industry, with approximate revenue of \textdollar162.32 billion annually, and is expected to reach \textdollar256.97 billion by 2025  \cite{gameindustrystatistics}. Due to improvements in the computational power available on mobile devices, desktop systems, and gaming consoles, there has been a significant increase in the number of people who play games, leading to an increase in the number of games being developed. Consequently, the game development market is now highly competitive. 

Game-breaking bugs that disrupt the game-play have caused many games to fail commercially. One such example is the retraction of \textit{Cyberpunk 2077}\footnote{ \url{https://www.cyberpunk.net/}} from the PlayStation Store due to its poor quality despite seven years of development and millions of dollars of investment   \cite{Politowski2021survey}. According to Polish Business Insider, \textit{CD Projekt Red}'s stock value fell by more than 75\% after the game was launched\footnote{ \url{https://www.gizchina.com/2022/07/19/cyberpunk-2077-makes-cd-projekt-lose-75-of-its-market-value/}}. 

\par Automated testing techniques have significantly improved in the last 30 years, with various innovative techniques being published.
However, the most pervasive way of testing games in the industry is through manual playtesting of simple to complex scenarios on the game to test its functionality  \cite{Politowski2021gray}. The industry relies upon intrinsic knowledge accumulated by the playtesters to make the testing process fruitful  \cite{Politowski2021survey}. Issues in the testing process are considered to be one of the major causes of game failures  \cite{schultz2016game}. These include disorganized testing in an already constrained testing budget   \cite{Politowski2021gray}, lack of predefined goals for testing and misunderstanding the types of bugs that may appear.

\par Games are highly interactive software having a very dynamic development lifecycle and cross-cutting dependencies. There are rapid changes and updates involved which make them susceptible to gameplay errors. These gameplay errors are not necessarily software bugs in terms of the incorrect function value. Nonetheless, they have the potential to not only disrupt the gameplay experience but also result in  revenue losses for game developers. Such issues must be identified and resolved.

\par
Taxonomies in general have been used to provide a systematic method for increasing understanding of a diverse data. The fault taxonomies are well-known means of collecting and classifying the potential bugs in application domains. Bug taxonomies categorize software bugs based on characteristics like cause, severity, etc., and help in organizing and prioritizing bugs for efficient testing. The structured categorization also identifies patterns and common causes, which helps prevent similar bugs in the future. Thus, bug taxonomies and testing are interconnected, with the former providing a framework for the latter to improve the testing process  \cite{felderer2014usingtaxonomy}. Bug taxonomies are a way to guide testers by providing high-level goals to generate new testcases  \cite{marchetto2009usingtaxonomy} and achieve greater coverage for the system under test  \cite{Lewistaxonomy}. Game designers can benefit from these organized classifications by ensuring that the identified faults do not occur in their games. Test engineers use these to improve their test cases by ensuring that the test cases specifically test for the identified faults  \cite{felderer2014usingtaxonomy}. Researchers gain guidance from the taxonomies in designing their test strategies  \cite{marchetto2009usingtaxonomy}.
\par
Similarly, they can act as a warning for the developers to be aware of such faults in the first place. For researchers, bug taxonomies not only provide a method of validating proposed testing techniques by clearly stating types of prevalent bugs to test against  \cite{Lewistaxonomy}; but also open new avenues of research in designing mutation operators corresponding to a particular taxonomy category  \cite{RN237}, improvements in existing testing approaches and proposing new techniques  \cite{Lewistaxonomy}.
\par
In the game development and testing domain, taxonomies have been proposed for bug identification as well as the specification of vocabulary needed to explain the differences in failures that occur in games due to ambiguity in design  \cite{aytemiz2020taxonomy}. However, the bug taxonomy proposed by Lewis et al.  \cite{Lewistaxonomy} is limited to temporal and non-temporal implementation faults. It fails to cover the plethora of bugs that fall under navigation, sound, crash, gaming balance, and network faults.
\par
In this paper, we present an updated and detailed taxonomy of game bugs which extends the taxonomy of game failures by Lewis et al.  \cite{Lewistaxonomy} based on a  Multi-vocal Literature Review (MLR) to incorporate the perspective of  both researchers and industry practitioners.  Typically, MLR's are used when the  topic under focus has a distinct industrial relevance  \cite{wang2022improving}. We analyzed 436 sources in detail – including academic literature (143), postmortems (200), blogs and articles (43), and videos (50). Out of 436, we finally selected 189 sources i.e. academic literature (70), postmortems (62), blogs and articles (20), videos (24), and talk sessions (5) for reported bugs encountered in the games. The selected sources assist in the identification and classification of commonly appearing bugs that result in game failures even after launch. 
We have conducted a survey of game players and gaming industry professionals to validate that the proposed taxonomy can aid industry practitioners, such as developers and testers, and researchers by providing guidance for playtesting and for validation of new testing techniques  \cite{Lewistaxonomy}.
\par
The paper is structured as follows: Section \ref{sec:mlrmethod} describes the   multi-vocal
literature review method, that we used for the selection of resources. Section \ref{sec:mlrresults} shows the process of deriving, evolving, and verifying our detailed taxonomy. Section \ref{sec:taxonomy} presents our taxonomy. Section \ref{sec:validation} describes the method of validation of our taxonomy. Section \ref{sec:discussion}, provides a comprehensive discussion regarding our taxonomy. Section \ref{sec:threattovalidity} explain the threats to the validity of our study.  Section \ref{sec:relatedworks}, briefly describes related literature. \ref{sec:conclusion} concludes the paper with future work.

\section{Systematic Multivocal Literature Review Method} \label{sec:mlrmethod}

In order to ensure the thoroughness of our proposed taxonomy, we analyzed both white and grey literature. White literature includes peer-reviewed workshop, conference and journal studies   \cite{RN304}. Grey Literature consists of 
books, book chapters, government reports,
news articles, annual reports, presentations, wiki articles, videos,
blogs, thesis documents, emails and tweets, and letters  \cite{RN304}.
We performed the search for academic and grey literature till \textbf{2023/01/04} and considered the following:

\begin{enumerate}
\item \emph{Academic Literature}, including both white and grey literature studies such as:
\begin{enumerate}
\item peer-reviewed journal, conference, and workshop studies
\item non-peer-reviewed thesis, magazine articles, and book chapters.
\end{enumerate}
\item \emph{Grey Literature},  including:
\begin{enumerate}
\item	Game development postmortems, 
\item	Session talks and presentations from well-known game development conferences
\item Articles and blogs.
\item Videos
\end{enumerate}
\end{enumerate}

\begin{figure}
\centering
\begin{subfigure}{0.44\textwidth}
    \includegraphics[trim=2cm 17cm 10cm 3cm, clip, width=\textwidth]{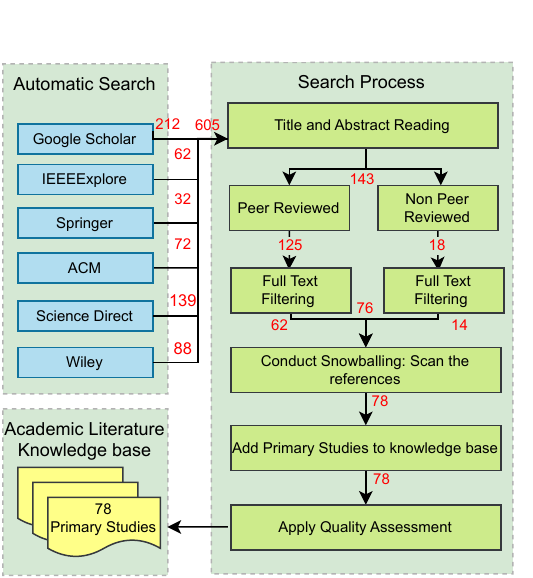}
   \caption{\label{fig:academicsearch} The methodology of systematic review of academic literature.}
\end{subfigure}
\hfill
\begin{subfigure}{0.55\textwidth}
    \includegraphics[width=\textwidth]{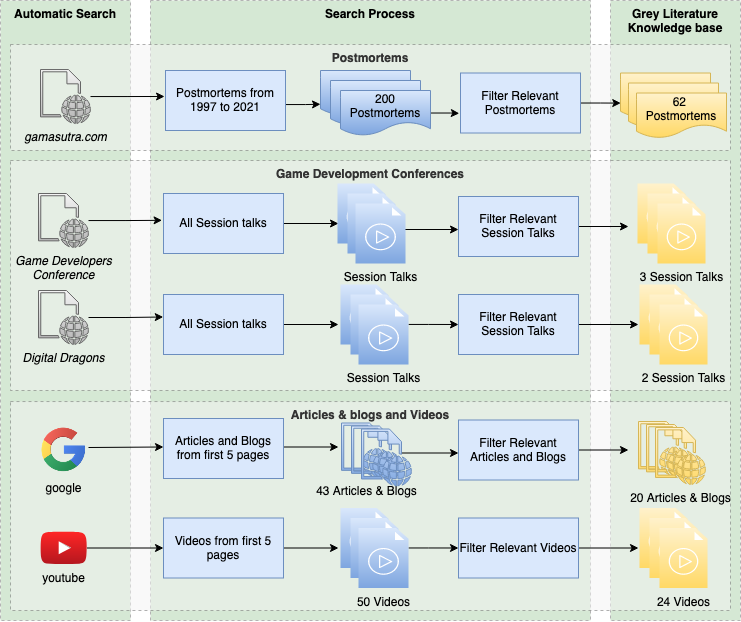}
    \caption{\label{fig:greysearch} The search process of the systematic survey of grey literature.}
\end{subfigure}
\hfill
\begin{subfigure}{0.7\textwidth}
    \includegraphics[width=\textwidth]{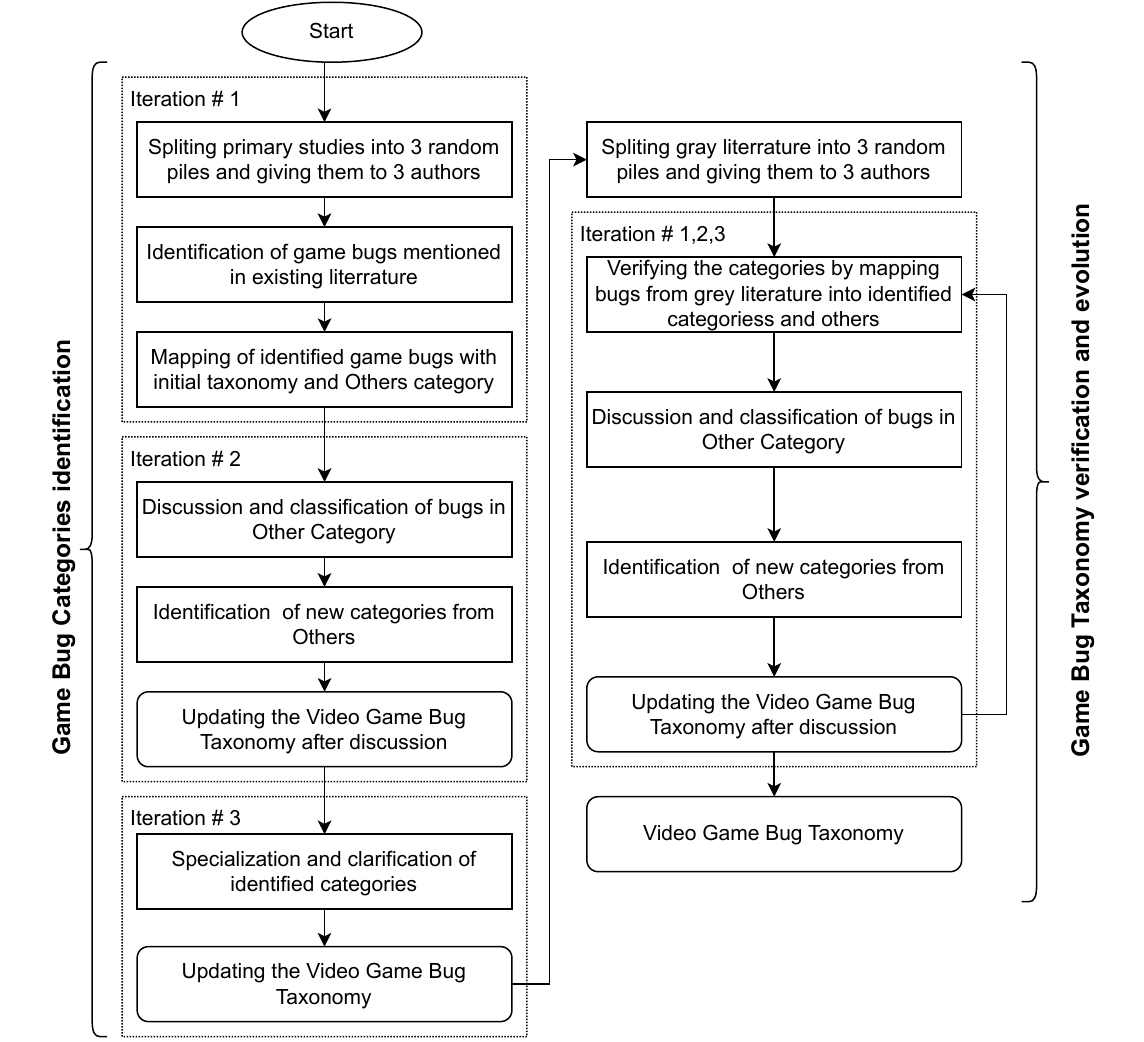}
    \caption{\label{fig:taxonomy_derivation} The process of derivation of the detailed bug taxonomy for games is depicted.}
\end{subfigure}
\caption{\label{fig:studies} Illustrates the phases of the search process and the number of primary studies in each phase along with the process of deriving the detailed bug taxonomy.}
\end{figure}

The goal for the SLR of Academic Literature was to find the game bugs and implementation faults that are currently being identified by the existing game testing techniques. We then performed a systematic survey of the grey literature to verify the identified game bug categories. We decided to opt for game testing approaches to identify game bug categories because testing approaches focus on implementation and functional faults including anomalies. Studies on game development highlight faults in the game development process rather than implementation faults. Similarly, game-play approaches do not focus on bug identification rather they highlight useability challenges as well as score maximization and goal completion aspects of the game-play  \cite{RN215}. 
The complete dataset is available at \url{https://doi.org/10.5281/zenodo.8425685}.

\subsection{Academic Literature}
We used six sources of academic literature:  Google Scholar\footnote{\url{https://scholar.google.com/}}, IEEE Explore\footnote{\url{https://ieeexplore.ieee.org/}}, Springer\footnote{\url{https://link.springer.com/}}, ACM\footnote{\url{https://www.acm.org/}}, Science Direct\footnote{\url{https://www.sciencedirect.com/}}, and Wiley\footnote{\url{https://onlinelibrary.wiley.com/}}. The following generic query was modified to be used in different digital libraries. 
Our objective was to make our query more inclusive to prevent the possibility of overlooking critical faults. 
\begin{center}
\textbf{((“approach” OR “technique” OR “method”) AND (“video game”) AND (“verify” OR “verification” OR  “playtest*” OR “test*” OR “debug*” OR “fault find*” OR “fault detect*” OR “anomaly detect*” OR “glitch detect*” OR “glitch find*”))
}
\end{center}

 Figure \ref{fig:academicsearch} illustrates the phases of the search process and the number of studies in each phase of the SLR of Academic Literature. We adapted our search process from  \cite{khan2019landscaping}. The preliminary search of digital libraries yielded a total of \textbf{605 studies} (Google Scholar(212), IEEE Explore(62), Springer(32), ACM(72), Science Direct(139), and Wiley(88)). The \textit{inclusion} and \textit{exclusion} criteria for the selection of primary studies is described as follows:
\begin{itemize}
    \item \textbf{Inclusion Criteria for Academic Literature:}
    \begin{enumerate} [label=IC\arabic*]  
    \setlength\itemsep{0em}
        \item Studies must be about game testing and fault finding.
        \item Studies must discuss a functional game testing approach.
        \item Full text must be available.
        \item Studies must be in English.
    \end{enumerate}
    
    \item \textbf{Exclusion Criteria for Academic Literature:}
    \begin{enumerate} [label=EC\arabic*]
        \item Studies only focused on automated gameplay.
        \item Studies that have a relevant extension.
        \item Studies simply re-applying an existing technique.
        \item Studies verifying game design at the design phase.
        \item Studies that provide an approach or platform to observe and analyze data for game testing rather than a game testing approach itself.
 \end{enumerate}
\end{itemize}

\subsubsection{Filtering: Title and Abstract Reading}
We read through the title and abstract of each of \textbf{605 studies}  extracted as a result of the initial query.
We filtered  \textbf{177} studies, including \textbf{122}  from Google Scholar, \textbf{34} from IEEE Explore,  \textbf{7} from Springer, \textbf{21} from ACM,  \textbf{4}  from Science Direct , and  \textbf{2} from Wiley digital library. Studies like \cite{albaghajati2022co}, for which fulltext was not available were filtered out at this step as well.
From the remaining relevant studies, we filtered duplications and finally ended with \textbf{143} relevant studies. 
\subsubsection {Separating Peer-reviewed and non-peer-reviewed Studies}
We divided the \textbf{143} relevant studies into peer-reviewed and non-peer-reviewed categories. We had \textbf{125}  peer-reviewed studies that were published in journals (28), conferences (80), workshops (10), and symposiums (7). And \textbf{18} non-peer-reviewed studies which included thesis (8), non-peer-reviewed articles (9), and magazine article (1).
\subsubsection {Filtering: Full-Text Reading}
We read and analyzed the full text of the studies we had previously filtered and comprehensively applied the inclusion and exclusion criteria. All studies that fulfilled the inclusion criteria were included.
Since our focus was on implementation faults, we reviewed and analyzed literature that focused on testing of functional aspects of games.
Studies like  \cite{RN306} were excluded based on EC1. AI-based automated game-playing is a complete research domain in itself. The focus of such studies is the successful completion of a pre-defined game scenario or maximizing the score. While such approaches can be used for triggering bugs, their inherent focus is not bug identification and were therefore excluded. Studies like  \cite{RN65} which have a comprehensive extension like [\textcolor{blue}{S7}], were excluded based on EC2, since including both would have resulted in duplication. Similarly, studies like  \cite{RN307}, were excluded despite being the most recent version because [\textcolor{blue}{S7}] contains the details of test sequence generation, modeling, and implementation fault identification that is congruent to our analysis. Studies that do not present a novel approach but rather report on using an existing method like  \cite{RN164} were excluded based on EC3. Game design verification can be considered game testing at the design phase; however, such studies do not focus on implementation faults or game bugs that appear in the end product during a later stage of the game development lifecycle. Hence, such studies  \cite{RN189} were excluded based on EC4. Studies that present an approach or platform to observe and analyze data for game testing rather than a game testing approach itself like  \cite{RN192} were excluded based on EC5. We only included those studies in our academic literature, the knowledge base of primary studies, that fulfilled the inclusion criteria. We obtained  \textbf{76 studies} including both peer-reviewed (62) and non-peer-reviewed (14) studies. Since the goal of our study is to devise a comprehensive taxonomy of 
game bugs, we have included non-peer-reviewed literature to ensure the completeness of the proposed taxonomy.

\subsubsection{Snowballing}
We performed the backward and forward snowballing on the 76 studies. We analyzed the title and abstract of each study again to verify it's relevance.  We found two more relevant studies during snowballing which we then added to our knowledge base. The final dataset has \textbf{78} studies including peer-reviewed (\textbf{64})( \cite{RN58},  \cite{RN56},  \cite{RN201},  \cite{RN208},  \cite{ RN171},  \cite{ RN87},  \cite{ RN245},  \cite{ RN165},  \cite{ RN115},  \cite{ RN214}, 
 \cite{ RN211},  \cite{ RN246},  \cite{ RN219},  \cite{ RN187},  \cite{ RN247},  \cite{ RN204},  \cite{ RN207},  \cite{ RN166},  \cite{ RN145},  \cite{ humansynthetic2019},  \cite{ RN202},  \cite{ RN209},  \cite{ RN220},  \cite{ RN91},  \cite{ RN100},  \cite{ RN249},  \cite{RN111},  \cite{ RN248},  \cite{ RN41}, 
 \cite{ RN37},  \cite{ RN212},  \cite{ RN60},  \cite{ RN184},  \cite{ RN188},  \cite{ RN200},  \cite{ RN110},  \cite{ RN80},  \cite{ RN102},  \cite{ RN250},  \cite{ RN218},  \cite{ RN251},  \cite{ RN252},  \cite{ RN253},  \cite{ RN254},  \cite{ RN255},  \cite{ RN256},  \cite{ RN52},  \cite{ RN257}, 
 \cite{ RN258},  \cite{ RN261},  \cite{ RN262},  \cite{ RN263},  \cite{ RN264},  \cite{ RN265},  \cite{ RN266},  \cite{ RN267},  \cite{ RN268}, \cite{rivergame}, \cite{goexplore}, \cite{usingrlforload}, \cite{lightweighthuman},  \cite{neuroevolution2022}, \cite{inspector2022liu}, \cite{gamebalance2021})
and non-peer-reviewed (\textbf{14}) studies ( \cite{RN270},  \cite{ RN271},  \cite{ RN272}, 
 \cite{ RN273},  \cite{ RN175},  \cite{ RN274},  \cite{ RN275},  \cite{ RN176},  \cite{ RN276},  \cite{ RN277},  \cite{ RN278},  \cite{ RN112},  \cite{ RN269}, \cite{efficient2022rrt}). 

\subsubsection{Quality Assessment}

In SLR's, quality assessment of selected studies is usually done for study selection, weighing each study, highlighting differences in quality of the studies, or to build confidence in results of the review paper \cite{zhou2015quality}. We have used quality assessment to weigh each individual study and to provide researchers and practitioners, confidence in the results and conclusions. We took inspiration for quality assessment criteria from  \cite{RN314}. The following  questions were used to assess the quality of our selected primary studies:

\begin{enumerate}[label=QC\arabic*]
  \setlength\itemsep{0em}
    \item Is/are the research objective/s clearly identified?
    \item Was the test generation method clearly explained?
    \item Was the test execution method clearly explained?
    \item Was the game bug detection mechanism clearly explained?
    \item Was the proposed approach demonstrated using a case study?
    \item Were game bugs discussed?
    \item Were the identified bugs described with an example?
\end{enumerate}

\par
\begin{figure*}[h]
\begin{subfigure}{.5\textwidth}
\centering
\includegraphics[width=\textwidth]{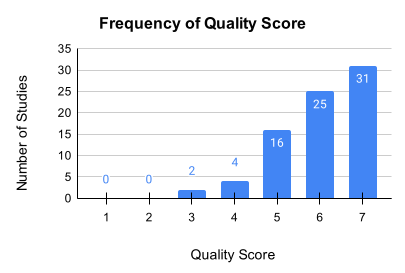}
\caption{\label{fig:qualittyscore}Illustrates frequency of Quality Score for the selected studies}
\end{subfigure}
\begin{subfigure}{.5\textwidth}
\centering
\includegraphics[width=\textwidth]{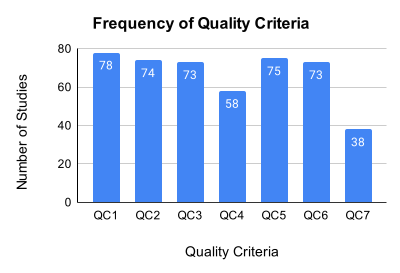}
\caption{\label{fig:qualitysummary}Illustrates how many selected studies fulfilled each Quality Criteria.}
\end{subfigure}
\caption{Illustrates Quality Assessment results of selected studies.}
\end{figure*}

 These questions were answered either ‘‘yes’’ or ‘‘no’’, rated as one or zero, respectively. The sum of the scores for all these questions was used to assess the quality of a primary study. We, however, did not exclude any study based on its quality score; rather, this score just depicts the quality rank of the primary studies included in this study. We observed that most of our selected studies scored well on the overall quality assessment criteria. Figure \ref{fig:qualittyscore} illustrates that 72 studies had quality scores greater than or equal to 5. We also observed that over half of the selected studies (40 studies) failed on QC7 and 20 studies 
for QC4 (figure \ref{fig:qualitysummary}). The detailed analysis of quality assessment is shown in table \ref{tab:qualityassesment} of appendix \ref{app:qualityassessment}.

\par We analyzed our primary studies in detail and compared the identified game bugs with the existing taxonomy \cite{Lewistaxonomy}. We found that it failed to cover all the bugs and hence we developed our updated game bug taxonomy and verified it further with the help of grey literature.

\subsection{Grey Literature}
The SLR of academic literature allowed us to identify the game bugs and implementation faults that are currently being identified by the existing game testing techniques. We then performed a systematic survey of the grey literature to not only verify whether the identified game bug categories are present in popular games and influence the gameplay experience of gamers, but also to add the perspective of industry professionals to our taxonomy. Hence, we modified the inclusion and exclusion criteria for the categories of reviewed grey literature. Figure \ref{fig:greysearch} illustrates the phases of the search process and the number of primary studies in each phase of the systematic survey of grey literature. The \textit{inclusion} and \textit{exclusion} criteria for the selection of primary web sources is described as follows:

\begin{itemize}
    \item \textbf{Inclusion Criteria for Grey Literature:}
    \begin{enumerate} [label=IC\arabic*]\addtocounter{enumi}{4}
        \item The source must be about bug identification in a game.
        \item Source is about an experience regarding an encounter with a bug in a game.
        \item Source must be available.
        \item Source must be in English.
    \end{enumerate}
    
    \item \textbf{Exclusion Criteria for Grey Literature:}
    \begin{enumerate} [label=EC\arabic*]\addtocounter{enumi}{8}
        \item Sources in a language other than English.
        \item Sources describing a game without mentioning an example of game bugs.
        \item Sources only describing a duplicate bug.
        \item Sources describing design flaws at the design phase.
 \end{enumerate}
\end{itemize}

\subsubsection{	Postmortems}
We read the postmortems available at the Gamasutra website\footnote{\url{http://www.gamasutra.com}}. It is a successor of the  Game Developers Magazine and is currently considered to be the most complete resource for digital-game postmortems \cite{RN231}. After applying the inclusion and exclusion criteria we selected 62 relevant postmortems ( \cite{PM1},  \cite{PM2}, 
 \cite{PM3},  \cite{PM4},  \cite{PM5},  \cite{PM6},  \cite{PM7},  \cite{PM8},  \cite{PM9},  \cite{PM10},  \cite{PM11},  \cite{PM12},  \cite{PM13},  \cite{PM14},  \cite{PM15},  \cite{PM16},  \cite{PM17}, 
 \cite{PM18},  \cite{PM19},  \cite{PM20},  \cite{PM21},  \cite{PM22},  \cite{PM23},  \cite{PM24},  \cite{PM25},  \cite{PM26},  \cite{PM27},  \cite{PM28},  \cite{PM29},  \cite{PM30},  \cite{PM31}, 
 \cite{PM32},  \cite{PM33},  \cite{PM34},  \cite{PM35},  \cite{PM36},  \cite{PM37},  \cite{PM38},  \cite{PM39},  \cite{PM40},  \cite{PM41},  \cite{PM42},  \cite{PM43},  \cite{PM44},  \cite{PM45}, 
 \cite{PM46},  \cite{PM47},  \cite{PM48},  \cite{PM49},  \cite{PM50},  \cite{PM51},  \cite{PM52},  \cite{PM53},  \cite{PM54},  \cite{PM55},  \cite{PM56},  \cite{PM57},  \cite{PM58},  \cite{PM59}, 
 \cite{PM60},  \cite{PM61},  \cite{PM62}) that discuss implementation faults in games whether they were pre-release or post-release. 

\subsubsection{	Game Development Conferences}
We searched the talk sessions and presentations regarding fault-finding in specialized conferences on game development like Digital Dragons\footnote{\url{http://digitaldragons.pl/ }}  and GDC\footnote{\url{https://www.gdconf.com/} } \cite{Politowski2021survey}. The five relevant talks are from GDC (  \cite{T1},  \cite{T2},  \cite{T3} ), and Digital Dragons (  \cite{T4},  \cite{T5} ) that we included in our grey literature knowledge base. We read transcripts and watched the presentations; and summarized the points related to implementation faults and used them to verify and evolve our taxonomy.
\subsubsection{	Web Articles and Blogs}
We queried Google Search to find articles about bugs in games on websites and blogs using keywords.
\begin{center}
    \textbf{“video game” AND (“glitch*” OR “fault*” OR “bug*” OR “flaw*”) AND “game-breaking”.
    }
\end{center}
 Our focus was to identify the game critical bugs that make it very difficult for gamers to play the games. To make the search as systematic as possible we followed the steps by Bajwa et al. \cite{RN795} and used the Google search in Google Chrome internet browser. We performed the following steps before running the query:
\begin{enumerate} [label=\roman*]
  \setlength\itemsep{0em}
\item Signed out from google\footnote{\url{https://www.google.com/ }}.
\item 	Cleared the browser search history.
\item 	Cleared out the browser cache.
\item 	Disabled the instant search predictions option from google.
\item 	Enabled the ‘100 results/links per page” in browser settings.
\item 	Added the plugin SEOquake \footnote{\url{https://www.seoquake.com/index.html }}  to the browser.
\end{enumerate}
We selected 20 most relevant articles after applying exclusion and inclusion criteria including seven  news articles (A), nine  blogs (B), two wiki articles (W), and two forum articles (F) ( \cite{W1},  \cite{W2},  \cite{W3},  \cite{W4},  \cite{W5} 
 \cite{W6},  \cite{W7},  \cite{W8},  \cite{W9},  \cite{W10},  \cite{W11},  \cite{W12},  \cite{W13},  \cite{W14},  \cite{W15},  \cite{W16},  \cite{W17},  \cite{W18},  \cite{W19},  \cite{W20}). 
 We read the complete articles and summarized the points related to game bugs and used them to verify and evolve our taxonomy. 

\subsubsection{	Videos}
We searched to find YouTube  videos about game bugs using the following keywords: 
\begin{center}
    \textbf{“video game” AND (“glitch*” OR “fault*” OR “bug*” OR “flaw*”) AND “game-breaking”.
    }
\end{center}
 YouTube provides the game community with context-rich bug information and is a great resource for bug identification  \cite{RN229}. Our focus was to identify the game-critical bugs that make it difficult for gamers to play games. To make the search as systematic as possible, we took inspiration from Bajwa et al.  \cite{RN795} and used the Google Search in the Google Chrome internet browser. We performed the steps described in the previous subsection to run the query.

The 24 most relevant videos were selected 
( \cite{V1},  \cite{V2},  \cite{V3},  \cite{V4},  \cite{V5},  \cite{V6},  \cite{V7},  \cite{V8},  \cite{V9},  \cite{V10},  \cite{V11},  \cite{V12}, 
 \cite{V13},  \cite{V14},  \cite{V15},  \cite{V16},  \cite{V17},  \cite{V18},  \cite{V19},  \cite{V20},  \cite{V21},  \cite{V22},  \cite{V23},  \cite{V24}) after applying exclusion and inclusion criteria over 50 retrieved videos. 
 We watched the complete videos having a total of 20 hours of watch time. We then summarized the points related to game bugs and used them to verify and evolve our taxonomy. 

\begin{figure} [h]
\begin{subfigure}{.45\textwidth}
\centering
\includegraphics[width=0.9\textwidth]{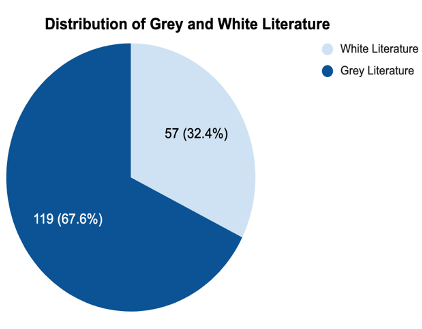}
\caption{\label{fig:greywhite}Illustrates the distribution of grey and white literature.}
\end{subfigure}
\begin{subfigure}{.45\textwidth}
\centering
\includegraphics[width=\textwidth]{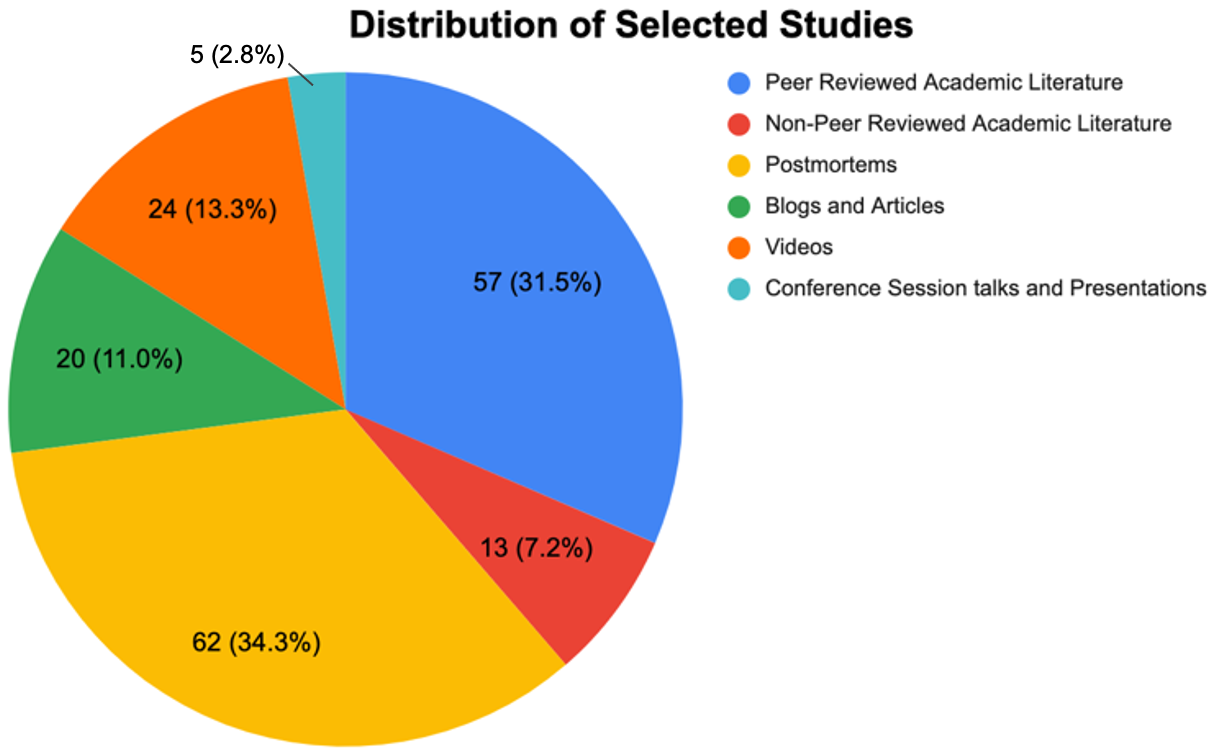}
\caption{\label{fig:studytypes}Illustrates the different types of selected sources.}
\end{subfigure}
\caption{\label{fig:selectedstudies} Overview of the distribution of selected sources.}
\end{figure}

\subsection{Deriving the Detailed Bug Taxonomy} \label{sec:mlrresults}
In this subsection, the process of derivation of the detailed bug taxonomy for games from the selected sources is described in detail as shown in figure \ref{fig:taxonomy_derivation}. Figure \ref{fig:selectedstudies} illustrates the distribution of selected studies in terms of white and grey literature (figure \ref{fig:greywhite}) and different types such as academic, postmortems, blogs, videos, and conference sessions (figure \ref{fig:studytypes}). 

\subsubsection{Game Bug Categories identification}
Once we had our primary studies from academic literature, we analyzed them in detail and compared the game bugs with the existing game bug taxonomy \cite{Lewistaxonomy}. We found that it failed to cover all the bugs and hence we developed our updated taxonomy and verified it further with the help of grey literature. This process was performed with the consensus of all authors. The selected academic literature was randomly divided into three sets of equal size for data extraction and classification. We performed multiple iterations to identify game bug categories.
\par In the first iteration, we started by extracting game bugs identified in each study and classifying them into the main categories of the initial taxonomy. The initial taxonomy consisted of the existing game bug taxonomy \cite{Lewistaxonomy}   (depicted by the underlined categories in figure \ref{fig:taxonomy}) as well as two new main categories Network Faults (subsection \ref{ssec:network})
and Gaming Balance Faults (subsection \ref{ssec:gamingbalance})
, which we identified during the preliminary analysis of the domain. Any bugs that did not fall into these categories were classified into the Others category with additional comments. 
\par In the second iteration, we discussed the bugs classified into the Others category and identified new categories. Examples of such categories are Unexpected Crash (subsection \ref{ssec:crash}) and Navigational Bugs (subsection \ref{ssec:navi}).
\par In the third iteration, we had discussions regarding existing categories and subcategories to finalize whether the subcategories were under the correct main categories. The Implementation Response Issues category was modified. In  \cite{Lewistaxonomy}, it was presented as a sub-category of Temporal faults and included scenarios in which game response speed did not match the expected speed. However, we converted it into a main-tier category. We observed that faulty system response was not only restricted to the speed of required response rather non-temporal elements were also involved such as faulty collision detection or reward estimation. Hence, it was decided that Implementation Response Faults will be a main-tier category.
\par
\begin{figure}
    \centering
    \includegraphics[width=0.7
    \textwidth]{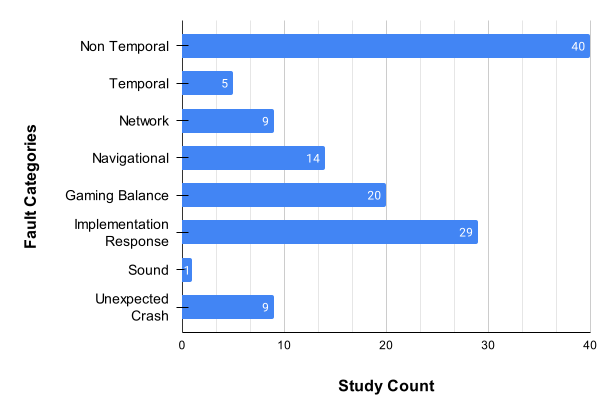}
    \caption{Distribution of Game bugs in academic literature.}
    \label{fig:bugbyacademiclit}
\end{figure}

 Lastly, we had discussions regarding the specialization of the categories based on examples identified from the selected studies. The seventh non-Temporal fault sub-category is Invalid Graphical Representation (subsection \ref{sssec:invalidgraphicalrepresentation}). It is one of the categories that we modified, which was originally proposed by Lewis et al. \cite{Lewistaxonomy}. We extended it into further sub-categories, which include Abnormal Text, 
Corrupted Frame, 
Extra Game Asset, 
Incorrect Visual State Transition, 
Missing Game Asset, 
and Slow loading. 
Figure \ref{fig:bugbyacademiclit}, shows the distribution of game bugs in academic literature.
Once we had the draft of our taxonomy, we further verified and evolved it with the help of grey literature. The detailed classification of selected studies with respect to game bug categories is available in \textit{Video Game Bug Taxonomy MLR-step6-Academic Literature-Taxonomy} file of our dataset  \cite{butt_nigar_azhar_2023_7631876} as well as in table \ref{tab:bug-src} of appendix \ref{app:summary}.

\subsubsection{Game Bug Taxonomy verification and evolution}
Once we had compiled the selection of grey literature into our knowledge base, we analyzed them in detail and identified and classified the game bugs into different categories. For this purpose, we followed the same process as done for academic literature. The selected grey literature was randomly divided into three sets of equal size for data extraction and classification. We performed multiple iterations to identify game bug categories. 
In the first iteration, we simply classified the bugs into an Others category if they did not fall into an existing category. Then we had discussions regarding bugs placed in the Others category. It was debated among authors whether to place it into an existing category or create a new one. We observed a
lack of work done in the identification of Sound Faults (subsection \ref{ssec:sound}) in games in the academic literature. We had only encountered one such study \cite{rivergame}, and hence had placed such faults in the Others category. However, our extensive survey of game postmortems and grey literature highlighted the presence of faults related to game sounds.
We repeated this process of classification and discussion for three iterations until all authors were satisfied with the taxonomy categories. The dataset  \cite{butt_nigar_azhar_2023_7631876}
provides the final iteration of the classification of the game bugs identified in grey literature into categories identified in the game bug taxonomy. The table \ref{tab:bug-src}
in appendix \ref{app:summary} 
shows the list of sources from where each of the main-tier bug categories were extracted and verified.
\par
\begin{figure*}
\noindent
  \makebox[\textwidth]{
    \includegraphics[trim={0.7cm 2cm 0.5cm 0},clip,width=1\textwidth]{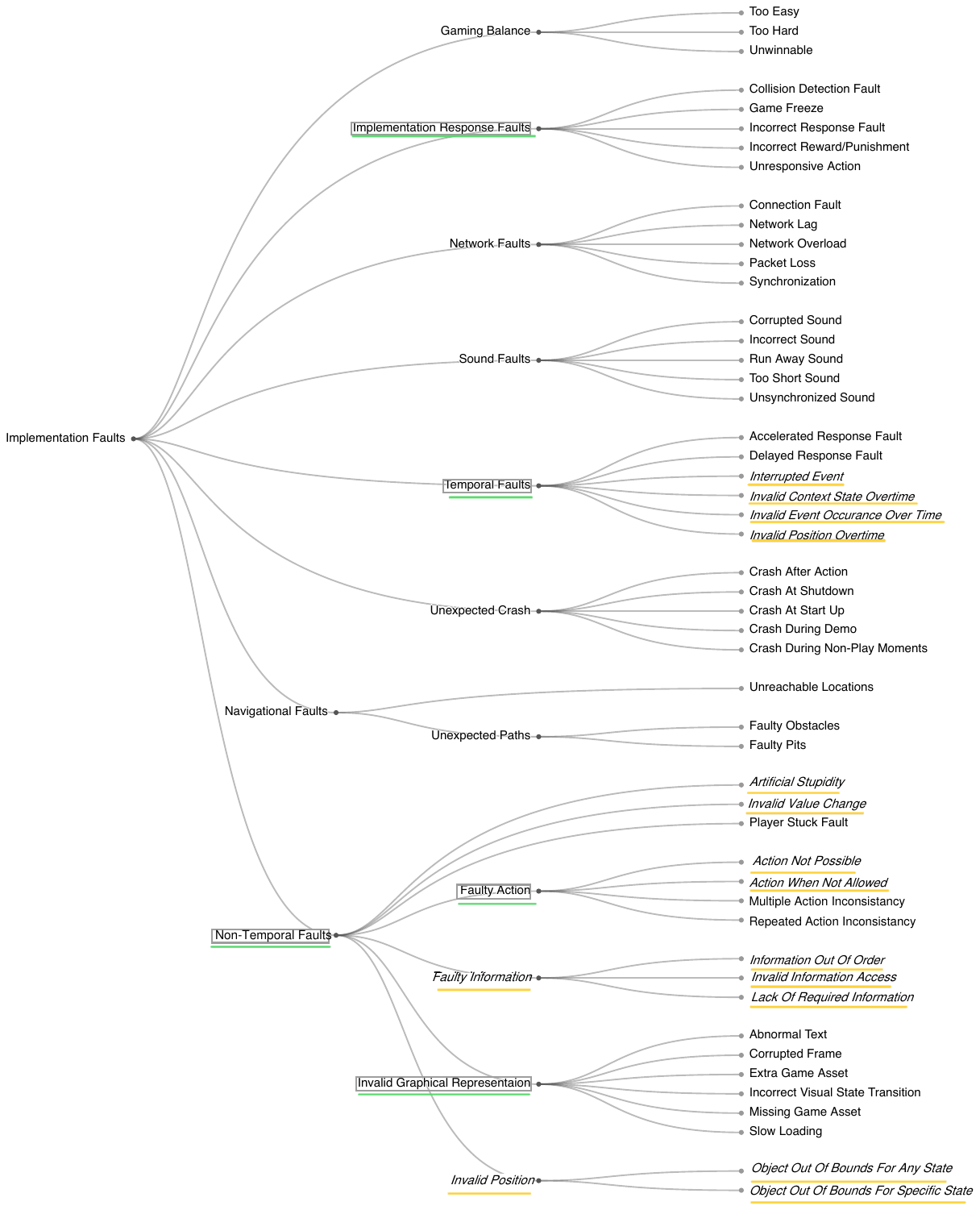}}
    \caption{\label{fig:taxonomy} The proposed Detailed Taxonomy of \textbf{63} Game Bug categories including the \textbf{five} categories enclosed in a box, that have been modified from existing literature while the italicized \textbf{15} categories are used without modification from existing literature.}
\end{figure*}

 Our proposed taxonomy is shown in figure \ref{fig:taxonomy}. It considers a total of 63 categories. These include the 20 categories proposed by Lewis et al. \cite{Lewistaxonomy}, of which we have modified five categories via the creation of new subcategories and one category (subsection \ref{sssec:invalideventoccurrenceovertime}) by extending its definition. 

\section{Taxonomy} \label{sec:taxonomy}

In this section, we will go over the taxonomy in depth. The focus of our taxonomy is \textbf{Implementation Faults} that occur during game execution. This starts from the time the game application is launched, or the game player logs in until the moment the game application is shut down or the game player logs out of the game application.  In accordance with ISO 24765 (2017), a \textbf{fault} occurs when the system behaves in a manner that is anomalous or inconsistent with the requirements or specifications  \cite{iso2017}.  It is also called a bug or a defect. 
The taxonomy has been designed from the end-user perspective and is based on user-observable behavior that is anomalous or inconsistent with user expectations. 
\par Implementation Faults are further divided at first-tier into eight categories: Gaming Balance, Implementation Response, Network, Sound, Temporal, Unexpected Crash, Navigational, and Non-Temporal faults. Table \ref{tab:taxonomy}
in appendix \ref{app:summary}
provides brief and concise descriptions with examples of the specialized new game bug categories from the taxonomy.We have also provided a video resource\footnote{\url{https://bit.ly/vgbtvideo}} to better help in understanding tricky game bug categories with visual examples. The categories present in our taxonomy are described in detail in the following sub-sections.

\subsection{Gaming Balance} \label{ssec:gamingbalance}
A balanced game is one in which the players that have equivalent or similar skillsets have an equal chance of winning, i.e., whether it is player-vs-player (PvP) or player-vs-environment (PvE). Additionally, this includes a gamer's ability to complete the game while matching its expected difficulty level, which means that the player should neither find it too easy to play nor too difficult. While such terms are incredibly subjective, this category refers to an excessive advantage or disadvantage to the player playing the game. Fairness is a universally sought-after attribute in games by players \cite{schell2008art}. We consider it to be an implementation fault because the system is behaving in a manner inconsistent with design expectations. In total 20 of 78 selected academic studies on game testing approaches focus on such scenarios.
\par This category specifically refers to unbalanced gameplay and is further divided into three sub-categories. 
It is a fault such that a player finds the game either too easy that it becomes boring and repetitive; or too hard that it triggers the desire to give up; or unwinnable that regardless of the player's effort the game is impossible to beat.


\subsubsection{	Too Easy}
This category includes gameplay faults due to which players find it too easy to play a game. It is critical if a particular set of parameters leads to unfair advantages in multiplayer games \cite{V7}. If a game level is easier than player expectations,  it is placed in this category \cite{PM25}. 
\par Additionally, this category includes bugs that make a game scenario easy to complete. For example, season 11 of \textit{League of Legends}\footnote{\url{https://www.leagueoflegends.com/}} contains a bug that allowed players to kill enemies including high-level bosses in a single attack with a hundred percent probability \cite{V16}.
\subsubsection{	Too Hard}
This category includes gameplay Faults such that players find it too difficult to play a game. This includes difficulty in finishing a task due to obscure design, ambiguous task requirements, or unreasonable constraints. 
For example, the counter-intuitive structure of the first level in \textit{Frozen Synapse}\footnote{\url{https://store.steampowered.com/app/98200/Frozen_Synapse/}} resulted in many players requiring an undesirably long amount of time to beat it  \cite{PM26}. 
\subsubsection{	Unwinnable}
This category includes gameplay faults such as players being unable to finish the game especially if it is due to an overly competent AI, convoluted game logic, game constraints such as small countdown timers, or impossible to obtain in-game items. Being unwinnable renders the game unplayable. In the game \textit{Political Machine}\footnote{\url{https://www.politicalmachine.com/}}, the AI opponent is too competent. When the players got up to George Washington, the opponent AI won every state making it impossible for the players to advance any further in the game  \cite{PM51}. 
\subsection{Implementation Response Faults} \label{ssec:implresp}
These types of faults were originally presented in the context of temporal faults and included scenarios where game response speed did not match the expected speed  \cite{Lewistaxonomy}. 
We observed that faulty system response was not restricted to the speed of required response, but also faulty responses to player actions.
\par We extended this category by adding five sub-categories into it including Collision Detection, Game Freeze, Incorrect Response, Incorrect Reward or Punishment, and Unresponsive Action. They are described in the following sub-categories. 
\subsubsection{	Collision Detection Fault}
Collision is a very common element in games. Its effects vary from interaction with benign components of the environment 
which at most block the path to helpful components 
to the dangerous components that can ‘kill’ the player.
\par Generally, in a video game, the collision of a player avatar with in-game elements or in-game elements with each other  \cite{PM12} has associated functionality that depends on correct and timely detection of that interaction. For example, if the game \textit{Mario Bros.}\footnote{\url{https://supermariobros.io/}} is  working correctly then, Mario 
stops when his collision is detected with the pipe and ‘dies’ when his collision is detected with the Goomba. 
In  \cite{RN56}, an open-source Mario game is seeded with a faulty collision detection so, Mario could pass through the Goomba without any damage. 

\subsubsection{	Game Freeze}
This fault occurs when the game becomes unresponsive. It can be complete unresponsiveness such that the game screen remains unchanged or partial unresponsiveness such that the player can change camera angles or view different aspects of the game but there is no response to any in-game actions.Such a fault may occur when the player causes the game to enter an abnormal state (section \ref{ssec:crash}).
\par 
For example in \textit{Fortnite}\footnote{\url{https://www.epicgames.com/fortnite/en-US/home}}, in the ‘build-a-brella’ steps after the player has completed customization and clicked purchase battle pass, the game stopped responding and the only way to make it work was to restart the game \cite{V6}. 
There are rare cases in multi-player games, in which a player can cause another player’s game to freeze or crash. In \textit{Tony Hawks Underground 2}\footnote{\url{https://tonyhawkgames.fandom.com/wiki/Tony\_Hawk's\_Underground\_2}}, players could cause other players' games to freeze by sending them messages longer than the restricted limit  \cite{W1}.
\subsubsection{	Incorrect Response Fault}
In-game actions have corresponding expected responses. When the response to an action is contrary to expectation then it is considered faulty such as a button that is expected to open a door instead closes it. \textit{ Final Fantasy V}\footnote{\url{https://store.steampowered.com/app/1173810/FINAL\_FANTASY\_V/}} had a weapon (the Chicken Knife), that would sometimes make the player run away from battle instead of doing an attack \cite{W2}. 
\subsubsection{	Incorrect Reward/Punishment Fault}
Certain actions within a game result in a reward that either increases score, game-specific attributes, or gifts items; or a punishment that deducts game-specific attributes or loss of items \cite{schell2008art}. This fault occurs when an in-game action results in less or more reward compared to expected, or conversely less or more punishment than expected. 
\par Such bugs can take forms like in \textit{Faxanadu}\footnote{\url{https://nintendo.fandom.com/wiki/Faxanadu}}, the reward for beating dungeons are items that can be used to clear certain blockades or obstacles in certain screens. The rewarded items disappear after being used. If the player leaves the screen, the blockade reappears, but the items needed to clear the blockades do not respawn, even after beating the dungeons again.Hence there is a lack of reward for an action that has an expected reward \cite{W2}.
\subsubsection{	Unresponsive Action Fault}
In-game actions have expected responses.When the action does not result in a response then it is considered faulty such as a button that is expected to open a door does not open the door. It differs from \textit{Incorrect Response} because this category covers a lack of response rather than a response that is different from expected. It also differs from \textit{Game freeze} which involves the complete game and all its actions becoming unresponsive. In this category, only one action becomes unresponsive. 
\par In \textit{Cyberpunk 2077}\footnote{\url{https://www.cyberpunk.net/}}, when a non-playing character (NPC) Takemura calls, the player takes the call but the NPC does not go away on end call action regardless of how many time the action is performed \cite{V21}. 
\subsection{		Network Faults}\label{ssec:network}
This category includes network faults in online games
The network problems were reported in  \cite{gamenetworkbugs} in the context of online, especially online multiplayer games like \textit{DOTA-2}\footnote{\url{https://www.dota2.com/home}}. This category encompasses problems that arise due to the host game application’s connection and communication with the game server as well as the stable performance of  game server during increased or unexpected network traffic. The academic literature focuses mostly on establishment of connection, 
and load testing; 
with some work focusing on the detection of lost data packets \cite{gamenetworkbugs}. 
However, the survey of the grey literature highlighted network problems that are caused by a lack of bandwidth, latency, and synchronization, especially in Massively Multiplayer Online Role-playing Games (MMORPG). Network problems are difficult to reproduce and isolate even when they are identified because they do not necessarily occur in the game code itself.
The Network Faults category is further divided into five sub-categories including Connection Fault \cite{gamenetworkbugs}, Network Lag \cite{gamenetworkbugs}, Network Overload \cite{gamenetworkbugs}, Packet Loss \cite{gamenetworkbugs}, and Synchronization. They are described in proceeding sub-categories.
\subsubsection{	Connection Fault}
The first requirement for playing an online game is creation of a connection to ensure stable communication. It is not limited to the initial connection and login to the game server. 
In \textit{Diablo 3}\footnote{\url{https://kr.diablo3.blizzard.com/en-us/}}, initial connection was established without any issues, only if a player switched shields with a certain Templar follower, server gave the player an error message and kicked them out of game. Furthermore, the server prevented the player from logging back in \cite{V2}.
\subsubsection{	Network lag}
This fault occurs when gameplay experience in online games is impacted by lagging response due to network speed constraints and data packet size constraints. Network Lag can cause a player’s in-game actions to be delayed. Consequently, by the time the action is received and executed by the server, the game state has changed. This lag between the game server and the player results in the player’s view of the game world being out of date, impeding the player’s ability to make proper decisions in the game. In First Person Shooter (FPS) games like \textit{DOTA 2}\footnote{\url{https://www.dota2.com/home}}, this lag would result in the players' shots missing. 
\subsubsection{	Network Overload}
A critical requirement for multiplayer online games is the ability to handle increased traffic and load with stable performance. It is especially necessary for games in the MMORPG genre. Load testing and stress testing are great tools for discovering bottlenecks before deployment and prevent situations like those encountered by \textit{Wireless Pets}\footnote{\url{https://www.gamedeveloper.com/programming/postmortem-games-kitchen-s-i-wireless-pets-i-}}.
It could not even handle 10,000 users playing it simultaneously before crashing \cite{PM17}. \textit{Diablo II}\footnote{\url{https://diablo2.blizzard.com/en-us/}} faced bugs that only appeared at much higher usage rates, like more than 100,000 players, to trigger certain buggy situations \cite{PM6}. 
\subsubsection{	Packet Loss}
Packet loss describes packets of data not reaching their destination after being transmitted across a network. Packet loss is commonly caused by network congestion, hardware issues, software bugs, and several other factors. These types of problems are extremely difficult to reproduce and isolate. Even when they are identified, they are often impossible to fix because they do not occur due to fault in the game code itself. A certain level of packet loss is expected during any type of communication via the Internet. 
However, if it causes obvious game-play issues like lack of execution of critical player commands; then it should be resolved. During the development of \textit{Toontown}\footnote{\url{https://www.toontownrewritten.com/}}, a variety of packet loss problems were encountered. Eventually, the development team decided to use Secure Sockets Layer(SSL) to send game data back and forth from the clients, to ensure security as well as to prevent the misinterpretation of game data by various pieces of networking hardware between \textit{Toonntown} servers and players on internet \cite{PM10}.
\subsubsection{	Synchronization }
Network-based synchronization faults are generally specific to multiplayer online games in which players with or against each other. In such games, game states must be synchronized across multiple players to allow them to play the game. Issues with synchronization can result in the player’s view of the game world being out of pace with other players. 
In \textit{Sins of a Solar Empire: Rebellion}\footnote{\url{https://www.sinsofasolarempire.com/}}, there were situations in which players became desynchronized during multiplayer matches. Partway through a match, players would find inconsistencies between game states for different players. A planet might seem to be owned by one player on one machine but owned by a second player on another machine. A battle could be raging in one corner of the galaxy between two players, while other players would see nothing \cite{PM49}.
\subsection{	Sound Fault}\label{ssec:sound}
Our extensive survey of game postmortems, and grey literature highlighted the presence of sound-related faults in games.
Game audio can be split into two separate categories, diegetic and non-diegetic sound. The non-diegetic sounds refer to a game’s soundtrack, background ambiance which is not directly generated by or linked to game environment, and narrative commentary like in \textit{Grand Theft Auto IV: The Lost and Damned}\footnote{\url{https://gta.fandom.com/wiki/The\_Lost\_and\_Damned}}, the player is thrown from a car and dies, initiating the ‘wasted’ sound effect. Alternatively, Diegetic sounds usually refer to a game’s sound effects, which are directly generated by game elements, and the dialogue that takes place between characters, like pressing the acceleration button in a racing game leads to accelerating car sound. 
Sound effects provide important gameplay cues for players. They clarify or reinforce player actions, giving feedback on players' decisions \cite{schell2008art},  \cite{zubek2020elements}.
It is especially critical for rhythm and dance games which require on-spot sound effects and synchronized music \cite{W2}. 
This category encapsulates faulty in-game sounds. This category is further divided into five sub-categories including Corrupt, Incorrect, Run Away, Too Short, and Unsynchronized sounds. They are described in proceeding sub-categories.
\subsubsection{	Corrupt sound }
These faults involve the in-game sounds including both diegetic and non-diegetic, becoming distorted, warped, pitched, garbled, or crackled. This can occur due to issues with sound settings in the game application itself, due to faults in the deployment platform as well as audio and video drivers. While the issues with settings in the game application like audio synchronization with different video and framerate per second are comparatively easy to identify, the issues that arise due to the numerous combinations of deployment platforms are not as easy to predict and debug. In \textit{Final Zone II}\footnote{\url{https://en.wikipedia.org/wiki/Final_Zone_II}}, a horrible buzzing sound would sometimes start after the intro cutscene and continue throughout the game \cite{W2}.
\subsubsection{	Incorrect sound}
These faults involve the correctness of diegetic sounds. This category encapsulates faults in which the sound behind an animation is incorrect or does not match the circumstances or is completely missing. It can be something obvious like dialog not matching sound or gunfire sounding like water dripping. In \textit{Whacked}\footnote{\url{https://en.wikipedia.org/wiki/Whacked!}}, missing sound resources led to either sounds missing completely or wrong sounds being played behind animations \cite{PM41}. 
\subsubsection{	Run Away sound}
When the sound continues even after the animation completes or the sound effect continues to play even after animation or scenario corresponding to the sound ends. It differs from the Unsynchronized sounds category because the sound starts at the right time but it ends too late as opposed to Unsynchronized sounds that may also start with a delay. In \textit{The Italian Job}\footnote{\url{https://en.wikipedia.org/wiki/The_Italian_Job_(2003_video_game)}}, due to a lack of sound resources available, the engine sounds looped very badly \cite{PM35}.
\subsubsection{	Too Short sound}
The sound ends before the animation or corresponding scenario completes. Non-diegetic sounds usually work fine, especially if they are just looping WAV files. However, sound effects are extremely difficult to get right if the audio person is not in-house experiencing the creation of the title firsthand. A sound having the wrong length is a serious problem. If the requirement is for a sound to go along with the swooshing pendulum blades, then it should be specified for how long the pendulum is going to swing \cite{PM41}.\textit{Whacked} and \textit{Descent: Freespace}\footnote{\url{https://en.wikipedia.org/wiki/Descent:_FreeSpace\_-\_The_Great_War}}, both encountered the problem of short sounds \cite{PM41}. It differs from the Unsynchronized sounds category because the sound starts at the right time but it ends too early as opposed to Unsynchronized sounds that may also start before the animation begins.
\subsubsection{	Unsynchronized sound}
This involves sounds that do not synchronize with their corresponding scenarios and animations. These include sounds starting or ending too early; or starting and ending too late; or skipping in between. Such faults are extremely critical in Rhythm and dance games in which players rely on audio cues to play the game. Similarly, in RTS games like \textit{DOTA 2} if the sounds are delayed then it becomes difficult for players to respond to threats in a timely manner. In MMORPG games, the sound synchronization issues can also be due to network problems. \textit{DJMAX Portable Black Squares}\footnote{\url{https://en.wikipedia.org/wiki/DJMax_Portable_Black_Square}} and \textit{Clazziquai Editions}\footnote{\url{https://en.wikipedia.org/wiki/DJMax_Portable_Clazziquai_Edition}}, background music had a bad habit of skipping and desynchronizing every now and then. In a Rhythm Game, this is a big problem, as it can make the song more difficult to play \cite{W2}.
\subsection{Temporal} \label{ssec:temporal}
This category includes those faults which require some knowledge of the previous game state to accurately categorize. The game state refers to the complete status of all game elements and their attributes at a specific moment in time \cite{zubek2020elements}.
For example, in Super Mario, Mario jumps up to the right height and continues to hover in mid-air (Invalid position over time). At any single point in time, Mario is at a valid height however, by analyzing the coordinates of Mario over multiple states, the fault can be identified.
This category is further divided into six sub-categories in which two sub-categories are new i.e. Delayed Response and Accelerated Response; one sub-category is redefined but originally taken from \cite{Lewistaxonomy}, Invalid Event Occurrence Over time. The remaining three of the subcategories from Lewis et al. \cite{Lewistaxonomy} are adopted as it is. They are Interrupted Event, Context State Over time, and Invalid Position Over time. They are described in the proceeding sub-sections.
\subsubsection{	Accelerated Response}
The game or player response to or consequence of an action is accelerated giving an unfair advantage or disadvantage, especially in real-time games. In \textit{King's Quest IV}, when Rosella (the character the player controls) is in the ogre's house and must reach the door before he catches her, the ogre travels across the screen too fast for a player to react \cite{W2}.
In essence, the response itself is not faulty at any single point in time but due to the accelerated nature of the response as compared to previous game states a fault can be identified.

\subsubsection{	Delayed Response}
The game or player response to or consequence of an action is delayed or halted giving an unfair advantage or disadvantage, especially in real-time games. The response itself is not faulty at any single point in time but due to the delayed nature of the response as compared to previous game states a fault can be identified. This fault is not necessarily caused by the configuration of the game application itself but could be a consequence of network problems resulting in lag in online games; or even limitation of deployment platform in terms of limited RAM etc. 
In \textit{King's Quest IV}\footnote{\url{https://playclassic.games/games/adventure-dos-games-online/play-kings-quest-iv-perils-rosella-online/}}, at some random intervals the response to Rosella (the character the player controls) and other characters would lag especially in less computationally powerful computers which can be very damaging for players in real-time games \cite{W2}. 
\subsubsection{	Interrupted Event}
This category includes any event that stopped unexpectedly. These include a sound effect cutting off or an enemy stopping mid-attack for no visible reason. If the enemy stopped because someone or something applied a freeze effect on it then it is not an Interrupted Event because it meets the expectations of the player in the game. The system's in-game response is only faulty because it results in an event termination before its expected time. 
\subsubsection{	Invalid Context State Over time}
State in this case refers to the user-observable display of the game rather than game code flags and values. This category includes faults in which a valid in-game element state occurs before or after the expected time; or continues longer than or terminates earlier than the expected time. In \textit{Super Mario}, when Mario catches a star, its image starts flickering and Mario enters an invulnerable state for 20 seconds. If the invincible state lasts less than or more than 20 seconds then it will fall into this category.
\subsubsection{	Invalid Event Occurrence Over time}\label{sssec:invalideventoccurrenceovertime}
 This category classifies discrete events that occur too often or too seldom. They include incidents such as firing a gun as discrete events. We have expanded it to include faults in which a valid in-game event occurs before or after expected time; or continues longer than or terminates earlier than expected time. Events in our case include game events such as \textit{Hell Event} in \textit{Lords Mobile}\footnote{\url{http://lordsmobile.igg.com/}}. They are daily Turf Events which run for 55 minutes and refresh every hour. Players complete corresponding tasks to win points and get rewards. If the \textit{Hell Event} starts at the wrong time or terminates before the expected 55 minutes, then it will fall into this category.
\subsubsection{	Invalid Position Over time}
This category includes faults in which an in-game element retains a position that is valid in any single point in time but invalid or faulty in context with previous states. In other words, the element is in a valid position for less than or more than a valid amount of time. In \textit{Super Mario}, if Mario jumps up to the right height and continues to hover in mid-air. At any single point in time Mario is at a valid height however, by analyzing the coordinates of Mario over multiple state, the fault can be identified.
\subsection{	Unexpected Crash} \label{ssec:crash}
It is a new category added in the updated taxonomy. 
This type of fault causes the entire game application to crash and exit \cite{RN87}. The common causes for Unexpected Crash to occur are memory leaks, division by zero, and recursive function calls. They are considered critical bugs that need to be resolved on a priority basis.
Unfortunately, the occurrence of such faults can sometimes be non-deterministic and unrepeatable. For example, if the game crashes unexpectedly due to memory leaks, then it will not always occur at a single point in game \cite{PM33}. The \textit{Knightly Adventure}\footnote{\url{https://www.metacritic.com/game/pc/knightly-adventure}} game, had problems with devices like iPod touch 4th generation and iPhone 3 which led to memory crashes \cite{PM33}.
This category is further divided into five sub-categories based on time of occurrence. They include Crash After Action, Crash at Shutdown, Crash at Start Up, Crash During Demo, and Crash During Non-Play Moments. They are described in the proceeding sub-sections.
\subsubsection{	Crash After Action}
One of the most obvious types of bugs in the game are the ones that result in the game unexpectedly crashing after an in-game action has been performed \cite{ RN165}.
In \textit{Cuphead}\footnote{\url{http://www.cupheadgame.com/}}, game crashed on the parry jump action \cite{PM9}.When the player successfully performed the parry action, took a shot after a pause, and then died, the game would crash. The testers found that the function responsible for the execution of the parry action had a dependency on the player module which was unresolved if the player died. The team had a lot of trouble reproducing the crash scenario. They recommended using breakpoints in Integrated Development Environments (IDEs) to stop the game in run-time and analyze game state via code.
In \textit{Mighty No.9}\footnote{\url{http://www.mightyno9.com/}}, the game unexpectedly crashed when the player fired a weapon or returned to the main menu during gameplay \cite{V8}. 
\subsubsection{	Crash At Shutdown}
It is critical to ensure that the procedures involved in the shutdown of game in console and PC games, and player exit from game in online games are executed correctly. These shutdown procedures generally include saving the player’s game state as well as certain game-specific actions like putting player resources to hibernate. If a game application crashes during shutdown down this can result in loss of player data and issues in initializing the game especially if the faulty shutdown results in game data corruption. In \textit{Super Meat Boy}\footnote{\url{http://www.supermeatboy.com/}}, one of the biggest faults was the game crash during shutdown \cite{PM55}. 
\subsubsection{	Crash At Start Up}
It is critical to ensure that the procedures involved in initialization of game applications in console and PC games, and player login to games in online games are executed correctly. These start-up procedures generally include correctly initializing game applications, retrieving saved game states, restoring players' game data, and in multiplayer game synchronization with other players. If a game application crashes during start-up, then it means that the game has failed to initialize. In \textit{Grand Theft Auto: Vice City}\footnote{\url{https://en.wikipedia.org/wiki/Grand_Theft_Auto:_Vice_City}}, the game crashed at startup if the player had saved game state at the ice-cream factory save point while reliving the crime kingpin fantasy, which corrupted the save file. Hence, when player reloaded the game from corrupted save file the game crashed at startup \cite{V3}. A similar save file corruption resulted in the same start-up fault in Prey \cite{V2}.
\subsubsection{	Crash During Demo}
Game demos generally have two variations: playable and non-playable. For this category, we focus on playable demos while non-playable demos are covered in the next sub-section (Crash During Non-Play Moments). Playable demos generally have the same gameplay as the upcoming full game with the purpose of acting as a tutorial to teach new functionalities to the players \cite{zubek2020elements}. While the demo appears to be just a short version of the full-fledged game, that is usually not the case from the development and coding perspective. Hence, even if certain functionalities have been tested during the actual gameplay, they must be tested for the demo separately. Since the demos are generally coded separately from the game base code. In \textit{PONCHO}\footnote{\url{https://en.wikipedia.org/wiki/Poncho_(video_game)}}, there was a crash bug in the game demo, which required game developers to reboot the game \cite{PM36}.
\subsubsection{	Crash During Non-Play Moments}
Games generally have in-game cinematic events called cut-scenes or event scenes which are non-play non-interactive game sequences that interrupt gameplay. They are used to reward players, display conversations between characters, consequences of player actions or foreshadow future events. A non-playable demo is essentially the gaming equivalent of a teaser trailer and does not require any player response other than to skip if possible. These can be pre-made videos or rendered during gameplay especially if the cut-scenes require visualization of a customizable character. A game can crash during non-play moments such as during the introduction sequence of a loop boss, Throne II in \textit{Nuclear Throne}\footnote{\url{https://en.wikipedia.org/wiki/Nuclear_Throne}} on windows \cite{W1}. Testing for such bugs separately is important because testers generally have animations and cutscenes completely disabled or they skip them.
\subsection{	Navigational Faults}\label{ssec:navi}
It is a navigational fault when players are unable to navigate to a place where they should be able to reach or conversely, they are able to navigate to a location they should not be able to access. It can be something simple like the height of a platform being so high that a player cannot reach it. In \textit{Vampire: The Masquerade}\footnote{\url{https://www.worldofdarkness.com/vampire-the-masquerade}}, one of the major problems identified in the development process was the problem of path-finding and navigation of variably-sized characters across a completely free-form 3D environment \cite{PM61}. 
The navigational faults category is further divided into two sub-categories as 
described in the proceeding sub-sections.
\subsubsection{	Unreachable Locations}
it is an  Unreachable Location fault when players are unable to navigate to a place where they should be able to navigate to. It can be something simple like the height of a platform being so high that a player cannot reach it. 
If a visible platform was unreachable for Mario but Mario needed to climb it to continue with the game progression then it would be a navigational fault. In \textit{Rastan}\footnote{\url{https://en.wikipedia.org/wiki/Rastan_(video_game)}} on the Commodore 64 computer, at the second level, it is impossible to make a jump over a flaming pit over two ropes to continue with the game progression \cite{W2}. Similarly, an enemy in a certain room in \textit{Beyond Good and Evil}\footnote{\url{https://en.wikipedia.org/wiki/Beyond_Good_\%26_Evil_(video_game)}} drops a key when it is defeated. However, depending on how the player defeats the enemy, the key can spawn in unreachable locations like inside corners, or in the ceiling, or even slightly beneath the floor \cite{W2}.
\subsubsection{	Unexpected Paths}
When players can navigate to a location, they should not be able to access it falls into this category. If in Super Mario, Mario can walk through the green pipe 
when that should not be possible, it would be an unexpected path’s fault. In only faulty collision detection, Mario would not be able to completely pass through the pipe. In \textit{Fallout}, players were sometimes able to reach places and levels that were still being developed and were not meant to be accessed \cite{V2}. These faults can occur due to design and implementation conflicts and are a result of two situations, Faulty Obstacles and Faulty Pits.
\par \textbf{\textit{	Faulty Obstacles:}}
This fault is a combination of faulty collision detection with obstacles and an exploit that allows players to move through the obstacle. So, players can navigate across obstacles against which they should collide. In \textit{Mass Effect}, Matriarch Benezia's use of Biotic powers would toss the main character through the wall and out into the empty void surrounding the rendered game area \cite{W2}. 
\par \textbf{\textit{	Faulty Pits:}}
This fault allows players to navigate over a pit or empty space through which they should fall. They are a combination of faulty collision detection with the environment and an exploit that allows players to move on air apparently when they should not be able to. If the player can fly or has the ability to hover then falling is not expected, then it would not be a fault. If in \textit{Super Mario Galaxy}\footnote{\url{https://en.wikipedia.org/wiki/Super_Mario_Galaxy}}, Mario is in flying Mario form then, its ability to hover over pits is not a fault. But if Mario is in any other form and still does not fall through a pit then it is faulty. 
\subsection{	Non-Temporal}\label{ssec:nontemporal}
This category includes those faults which can be found by inspecting the game state at any point in time. For example, in \textit{Super Mario}, Mario jumps up to a height higher than should be permitted.
This category is further divided into seven sub-categories. One sub-category is new, Player Stuck Fault. We have accepted four of the subcategories by Lewis et al. \cite{Lewistaxonomy} as is, such as Artificial Stupidity, Invalid Value Change, Invalid Position, and Information. We have extended two of the subcategories via the creation of new sub-categories. They are Faulty Action and Invalid Graphical Representation. They are described in the proceeding sub-sections.
\subsubsection{	Player Stuck Fault}
When a player reaches a game state such that there is no means of progression and the game does not terminate, it falls into this category. It can be something obvious like Mario in \textit{Super Mario}, falling into a pit that does not kill him, but its boundaries are high enough that Mario cannot jump out of the pit. Similarly, in Guacamelee, the player could get locked in a fight arena unable to exit it or make any further progression in the game \cite{PM13}.
It can also be subtle like in \textit{Paper Mario: The Origami King}\footnote{\url{https://en.wikipedia.org/wiki/Paper_Mario:_The_Origami_King}}, the \textit{Spring of Rainbows - VIP} card in the game’s Shangri-Spa area is needed to get to unlock a secret stage. When the player enters that stage the card is automatically used. If the player exits the stage without completing the quest then there is no way to get the card again and the player is essentially stuck \cite{V23}.
\subsubsection{	Artificial Stupidity}
This category includes faults related to undesirable NPC behavior which disillusioned the player to NPC’s lack of intelligence. It includes cases in which NPCs blocked the players path or became unresponsive to gameplay interactions. Basically, behaving, as the name implies, in a stupid manner. In \textit{Tresspassers}\footnote{\url{https://en.wikipedia.org/wiki/Trespasser_(video_game)}}, Dinosaurs were governed by a set of emotions that theoretically should have prompted them to pick appropriate responses at any time. However, in practice they oscillated rapidly between many activities, sometimes even standing still, and twitching as they tried to decide what to do  \cite{PM12}.
This category specifically refers to game AI and elements controlled by it.
\subsubsection{	Invalid Value Change}
This category includes faults that impact the values of counters in an unexpected manner. It includes timers counting down too fast, score increasing unexpectedly, an attack causing an unexpected loss in health or conversely a `heal’ spell increasing health unexpectedly. The \textit{Mafia II}\footnote{\url{https://en.wikipedia.org/wiki/Mafia_II}} game, the health bug is a notorious example of this fault. It caused the player’s health to continuously decrease without any reprieve or cause  \cite{V15}.
\subsubsection{	Faulty Information}
This category includes all the faults related to in-game information that the user has or should have access to including information about the player as well as other players if the game requires such communication.
\par \textbf{\textit{	Invalid Information access:}}
This category includes the faults related to players having access to information; they should not have. In single-player games, this can ability to see through walls and obstacles, or information regarding enemy locations or game maps that should only be accessible in later stages of games. In multiplayer games, this includes information regarding statistics and movements of other players.
\par \textbf{\textit{	Lack of required information:}}
This category includes the faults related to players lacking access to information that they should have. In single-player games, this can include the inability to see open pathways (a malfunction of in-game camera), or information regarding enemy locations or game maps that should have become accessible in specific stages of games. In multiplayer games, this can include information regarding statistics and movements of allied players needed to make strategic decisions.
\par \textbf{\textit{	Information out of order:}}
This category includes the faults in which players receive information in an unexpected or faulty order. It is most relevant in role-playing games where players have many alternatives of receiving certain information but it is likely that other sources of such information are not updated to reflect that information access. In \textit{F.E.A.R.}\footnote{\url{https://en.wikipedia.org/wiki/F.E.A.R._(video_game)}} game, at one point, player is asked to download data from a laptop.  There are multiple laptops with downloadable data to get the backstory. However, the player is told to download this data from a particular laptop, a long time after encountering this laptop. However once the data has been downloaded from any machine, it cannot be done again and the mission objectives are not updated to show task completion. The player basically must restart from the last save and wait for the objective to be given, then download the data  \cite{W2}.
\subsubsection{	Invalid Position}
It includes faults in which an in-game item is in an invalid position. It differs from Invalid Position Over time in which an in-game element retains a position that is valid at any single point in time but invalid or faulty in context with previous states. If Mario in \textit{Super Mario bros.} jumps to an invalid height it is an invalid position. Instead, if Mario hovers at the right height for too long then it is Invalid Position Over time.
\par \textbf{\textit{	Object out of bounds for any State:}}
This category includes faults in which an in-game element is in an invalid position regardless of game state. In \textit{Project Gotham Racing 3}\footnote{\url{https://en.wikipedia.org/wiki/Project_Gotham_Racing_3}}, there was a bug where cars started 30 feet in the air, facing the wrong way around \cite{PM37}.
\par \textbf{\textit{	Object out of bounds for a specific state:}}
This category includes faults in which an in-game element is in an invalid position only because of the corresponding game state. If the flying Mario in \textit{Super Mario Bros.} cannot be underwater, and it falls below the water surface, then it is out of bounds for that specific state. If it was not in flying Mario form then it would not be considered a fault.
\subsubsection{	 Faulty Action}
This category includes the faulty execution of an in-game action by a player either due to the inability to act when permissible or ability to take impermissible action.
Actions are the “verbs” of game mechanics. They are base in-game operations a player can perform \cite{schell2008art}.
This category is further divided into four sub-categories including two sub-categories that are newly proposed after discussion amongst the authors - Multiple Action Inconsistency and Repeated Action Inconsistency. We have accepted two of the subcategories by Lewis et al. \cite{Lewistaxonomy} as is - Action Not Possible and Action When Not Allowed. They are described in the proceeding sub-sections.
\par \textbf{\textit{	Action Not Possible:}}
 In simple terms, a player is unable to take an in-game action in a specific game state that should be permissible in that game state. In \textit{Sanitarium}\footnote{\url{https://en.wikipedia.org/wiki/Sanitarium_(video_game)}} game, Max (the character the player controls) would get stuck around corners and claim “Can’t go that way”, even when path was available \cite{PM11}. Similarly, \textit{Crysis}\footnote{\url{https://www.crysis.com/}} had a bug that made the final boss randomly become ‘untargetable’ (and thus invulnerable) which meant that no form of action was possible \cite{W2}.\textit{Space Station Silicon Valley}\footnote{\url{https://en.wikipedia.org/wiki/Space_Station_Silicon_Valley}} (for Nintendo 64) is impossible to win with Hundred-Percent Completion because an action required to pick up the critical item is not possible \cite{W2}. In \textit{Sphynx and the cursed mummy}\footnote{\url{https://en.wikipedia.org/wiki/Sphinx_and_the_Cursed_Mummy}}, if a player saves game at midpoint save point a gate needed to progress is closed permanently and open-door action is not possible \cite{W1}.
\par \textbf{\textit{	Action When Not Allowed:}}
In this category, a player is able to take an in-game action in a specific game state that should not be permissible in that game state. \textit{Global Agenda's} 1.3.2 patch\footnote{\url{https://en.wikipedia.org/wiki/Global_Agenda}} contained a major bug in the auction house that allowed players to effectively create money from nothing \cite{W2}. \textit{League of Legends: Season 11} as a bug that allows players to one-shot enemies using all types of weapons even those that do not permit this action \cite{V16}.
\par \textbf{\textit{	Multiple Action Inconsistency:}}
It encompasses faults that occur only when multiple different actions are taken simultaneously or in a certain order. Such faults are deterministic and repeatable. When Mario is made to fire an attack while jumping then the player is taking two actions simultaneously. Mario’s fire attack could be working as intended in isolation but it is possible that it does not work while Mario is jumping. In \textit{Runescape 3}\footnote{\url{https://play.runescape.com/}}, a game-breaking bug allows players to instant kill any boss in the game without using up any ‘death touched’ darts. This bug only occurs when certain weapons are equipped and used in a particular order \cite{V1}.
\par \textbf{\textit{	Repeated Action Inconsistency:}}
It encompasses faults that occur only when an action is repeated. Such faults are deterministic and repeatable. When Mario is made to jump continuously it is meant to make small hops. Mario’s jump could be working as intended in isolation but it is possible that it does not work while Mario is made to jump multiple times repeatedly.\textit{Psychonauts}\footnote{\url{https://en.wikipedia.org/wiki/Psychonauts}} had a quirk where a player became unable to use double jump in a level where it was needed to jump between flaming grates while the water level is rapidly rising \cite{W2}.
\subsubsection{	Invalid Graphical Representation}\label{sssec:invalidgraphicalrepresentation}
These faults occur when game graphics are faulty. In other words, the graphical representation of game elements is faulty either due to rendering or location of appearance. When Mario in \textit{Super Mario Bros.} catches a fire flower, he transforms into Fire Mario. If the graphical representation of Mario remains unchanged or becomes something else then it will fall into this category of faults. 
The reasons for such faults can vary from the defects of hardware (e.g., GPU-related issues) to game application settings (e.g., the wrong setting of rendering special effects) to source-code bugs (e.g., incorrect transitions) \cite{ RN211}. Such issues are most found in graphically demanding games such as \textit{Assasin’s Creed, Red Dead redemption}\footnote{\url{https://en.wikipedia.org/wiki/Red_Dead_Redemption}}, and \textit{Cyber punk}.
This category is further divided into six new sub-categories that we have added to our updated taxonomy, including Abnormal Text \cite{ RN211},  \cite{ RN209},  \cite{ RN251},  \cite{ RN256}, Corrupted Frame \cite{ RN211},  \cite{ RN209},  \cite{ RN37},  \cite{ RN251},  \cite{ RN256},  \cite{ RN175}, Extra Game Asset \cite{ RN209},  \cite{ RN251},  \cite{ RN256},  \cite{ RN274}, Incorrect Visual State Transition \cite{ RN37},  \cite{ RN253},  \cite{ RN256},  \cite{ RN274},  \cite{ RN276}, Missing Game Asset \cite{ RN211},  \cite{ RN209},  \cite{ RN251},  \cite{ RN253},  \cite{ RN256},  \cite{ RN274}, and Slow loading \cite{ RN251},  \cite{ RN256},  \cite{ RN175}. They are described as follows.
\par \textbf{\textit{	Abnormal Text:}}
This fault encompasses issues with the appearance of text in the game application. Text is an important aspect of many modern games whether it is in form of dialog between in-game characters, instruction for a player to help with the game progression or communication between different players in multi-player games. This fault can take many forms. Text can appear in the wrong location; it may even cover a character or any other game object. Text can be incomprehensible either due to being blurry or due to low contrast with background or simply displaying wrong information \cite{ RN211}. In \textit{Pokémon Red, Blue, Green and Yellow}\footnote{\url{https://awesomegames.miraheze.org/wiki/Pok\%C3\%A9mon_Red,_Blue,_Green_and_Yellow}}, the text describing \textit{MISSINGNO} pokemon was often glitchy \cite{V3}.
\par \textbf{\textit{	Corrupted Frame:}}
The most common example of a corrupted frame is the game screen becoming pixelated. 
Pixelation is caused by displaying a bitmap or a section of a bitmap at such a large size that individual pixels, small single-colored square display elements that comprise the bitmap, are visible. The image frame could become scrambled if incorrect post-processing effects have been added, or resolution requirements of the game applications are not compatible with the deployment platform. This could occur for the complete frame or just part of the frame. A more subtle case of this fault can occur if the frame as a whole has an unexpected colored tint. In \textit{Amnesia: A Machine for Pigs}\footnote{\url{https://www.aamfp.com/}}, earlier players noticed a blue ‘fog’ during gameplay which was actually due to compromised color-grading \cite{PM56}.

\par \textbf{\textit{	Extra Game Asset:}}
As the name implies an in-game element that should not be in a particular game state is visible. In \textit{Jak X: Combat Racing}\footnote{\url{https://en.wikipedia.org/wiki/Jak_X:_Combat_Racing}}, has a fault known as \textit{saving glitch} where the save icon stays visible on the game at all times. Forcing the player to hard reset, if they want to get rid of it \cite{W1}.
\par \textbf{\textit{	Incorrect Visual State Transition:}}
An in-game element transitions to an incorrect state. The animation for a jumping Mario in Super Mario Bros., is different to that of a swimming Mario. If swimming Mario animation is made visible when Mario is jumping then it would fall into this category. In \textit{Slow Down, Bull}\footnote{\url{https://store.steampowered.com/app/333580/Slow_Down_Bull/}}, Annette the bull catcher was notorious for not transitioning properly through her various tell states, even when all the variables were set, due to the un-intuitive transition settings in the animators overriding what animation was being told to play in the state machine \cite{PM2}
\par \textbf{\textit{	Missing Game Asset:}}
An in-game element that should be visible in a particular game state is not visible. In Sins of a Solar Empire: Rebellion, A battle could be raging in one corner of the galaxy between two players, while the third player would see nothing \cite{PM49}. In Donkey Kong Country 2, if a player picks up the barrel as it breaks, then the avatar is seen carrying an invisible barrel \cite{V3}. In \textit{Pokémon Red, Blue, Green and Yellow}, a player could try to fight or capture a glitchy pokemon \textit{MISSINGNO} which only appeared as a collection of corrupted pixels \cite{V3}.
\par \textbf{\textit{	Slow loading:}}
The in-game graphics experience a delay in rendering on the screen.Most common example of Slow Loading is low frame rate. It also manifests in delays in game asset renderings especially in online games. These issues are not necessarily encoded in the game application, rather can also arise due to limitations of the host machine and network speed. They are commonly found in graphically demanding games.\textit{Elder Scrolls V: Skyrim} was notorious for its rendering issues \cite{V4}. Similarly, \textit{ Mighty No.9}, also fell victim to slow frame rate \cite{V8}. 

\section{Validation} \label{sec:validation}
In this section, we evaluate the developed game bug taxonomy which we have described in the previous section. We performed a survey to gain insights from the participants, using google forms , regarding the frequency and severity of the game bugs as well as the priority with which they should be fixed. Such surveys have been widely used in the literature for validation purposes as they are an effective means to validate proposed fault models directly from the experts \cite{unterkalmsteiner2011evaluation}. In the following sub-sections, we have described the procedure by which we designed our survey questionnaire, the process of response collection and finally the result synthesis.

\subsection{Survey Design}
In this sub-section, we have described the procedure by which we designed our survey questionnaire. It consists of 60 questions.
The details on the survey questionnaire are given in appendix \ref{app:survey} via table \ref{tab:survey}. 
The actual questionnaire can be found at \url{https://bit.ly/vgbtsurveypdf}. We have followed the state-of-the-art survey guidelines for developing the online survey questionnaire presented in  \cite{punter2003conducting}.   These guidelines specified that anonymity of the respondent must be ensured. They suggested use of likert-scales as well as provision of necessary instructions support readability and understanding.
\par The Question 1-9 cover the background information regarding the participants as well as their area of expertise. We used Likert scales to gauge the frequency, severity, and priority of our identified game bug categories, as follows:
\begin{enumerate}
\item	5-point Likert scale for frequency of game bugs following the guidelines in  \cite{brown2010likert} 

\item	6-point Likert scale for severity is used in the survey questionnaire for game bugs following the guidelines in  \cite{zhao2014empirical}.

\item	6-point Likert scale for priority that should be allocated to fixing a certain type of bug in games following the guidelines in  \cite{gomes2019bug}.

\end{enumerate}

\par The questions related to severity and priority of bugs are correlated. Their presence helps us recognize quality responses. For example, Q11 asks the participants to rate severity of Gaming Balance Faults while playing games while Q12 asks the participants to prioritize the need to fix such faults. If the participant considers the bug to be critical then congruently its priority should be high.

\subsection{Pilot Survey}
To help ensure the understand-ability of the survey, all the authors went through the survey. We then conducted a pilot survey with five participants. A pilot survey is a common tool to test the research tools including the questions, and survey structure \cite{chakraborty2018understanding}.
We  asked two less frequent game players and three avid gamers to review the survey and give their response to ensure the questions were clear and complete. For this purpose we used the think aloud technique to get the feedback of participants regarding survey design. They only suggested minor edits. The changes we made include: adding descriptive information regarding answers where selection of multiple options was allowed, adding clarifying description for sub-categories of game bugs, adding an image of taxonomy  before final remarks, adding three commonly played game genres to the relevant question.

\subsection{Survey Execution}
In this sub-section, we have described the process of response collection. We have followed the guidelines presented in  \cite{punter2003conducting} for conducting our online survey. The participants of the survey population were identified from LinkedIn\footnote{\url{https://www.linkedin.com}} following the recommendations in  \cite{imtiaz2020framework}. We provided the descriptions of all the taxonomy terms as well as the explanations of the likert scale terms used to rate severity, frequency, and priority of game bugs. We included the descriptions in the survey which we sent to the survey population.

\subsection{Evaluation of the game bug taxonomy}
In this sub-section, we have described the process of synthesis of the result.
We received 168 responses to our survey. The details of expertise of the respondents are shown in figures \ref{fig:gameplay}, \ref{fig:gametesting}, and \ref{fig:gamedev}
of appendix \ref{app:studyfig}. The complete set of anonymized survey responses are available at \url{https://bit.ly/vgbtsurvey}.

\par  The survey showed that on average 96\% of survey respondents encountered our main-tier game bug categories (figure \ref{fig:ecounterfreq} of appendix \ref{app:studyfig}). Network faults are considered most frequent (with 10.2\%, 43.5\%, 36.4\% and 9\% respondents encountering them always, often, sometimes, and rarely respectively) while Sound faults are considered least frequent  (with 6\%, 16.1\%, 32.2\%, 35.2\%, and 10.8\% respondents encountering them always, often, sometimes, rarely, and never respectively). This is depicted in figure \ref{fig:freq}  of appendix \ref{app:studyfig} 
which graphically  presents the survey data. Similarly, \textbf{60\%}(average percentage of respondents that consider main-tier faults with major or higher severity) of the respondents agree  that all the identified main-tier game bug categories have a Major or higher severity with exception of sound and navigational faults (with only 48.9\% and 47.2\% considering them major and higher respectively) which are considered by most to be of minor or low severity. Alternately, network faults are considered most severe  (with 10.8\%, 37\%, 30.3\%, 17.3\%, and 4\% respondents considering them blocker, critical, major, minor and low respectively). This is depicted in figure \ref{fig:sev} of appendix \ref{app:studyfig}
Alternatively,  on average \textbf{77.5\%} of the respondents agree  that all the identified main-tier game bug categories should have atleast medium priority fix with exception of sound and navigational faults (depicted in figure \ref{fig:pri} of appendix \ref{app:studyfig})

In terms of frequency, less than 4\% respondents stated that they had never encountered the identified bugs. Comparatively, over 96\% of the respondents agreed that they had encountered such bugs. It should also be noted that even the subcategories least encountered were encountered by over 11\% of the respondents. This is reflected in figure \ref{fig:ecounterfreq} of the appendix \ref{app:studyfig}.
The four percent can be explained by the genre of games that were played by the particular respondents for example, the respondent who claimed to not have encountered sound faults mostly played casual games like candy crush or racing, and platform games for such games sounds are not a critical part of gameplay.


\par Moreover, to identify relationships between our defined metrics (Frequency, Severity and Priority), we performed correlation analysis to quantify the degree to which our defined metrics are related and presented our findings in table \ref{tab:corr}. We used Spearman's rank correlation to find the correlation between Frequency, Severity and Priority of game bug categories as we gathered ordinal data corresponding to each metric. The table \ref{tab:corr} presents Spearman correlation coefficient values corresponding to each category. The Spearman correlation coefficient, $r_s$, can take values from +1 to -1. The $r_s$ of +1 indicates a perfect association of ranks, while $r_s$ of zero indicates no association between ranks and $r_s$ of -1 indicates a perfect negative association of ranks. The closer $r_s$ is to zero, the weaker the association between the ranks. If 0.00  $\leq r_s \leq $ 0.20, then the correlation is considered negligible(highlighted in red). If 0.21 $\leq r_s \leq $ 0.40, then the correlation is considered weak(highlighted in yellow). This means that there is a weak tendency for the two variables to increase or decrease together, but there are also many exceptions. If 0.41 $\leq r_s \leq $ 0.60, then the correlation is considered moderate(highlighted in green).

\par The tests show that the correlation between the three variables (frequency, severity and priority) is mostly weak to moderately \textbf{positive} as shown by values highlighted in yellow and green in the table \ref{tab:corr}. For example, correlation between severity of navigation faults with the priority to fix it is 0.4701. The reason for this could be that game developers tend to prioritize fixing issues that are severe, and navigation faults in games can have a significant impact on the player's experience. Therefore, if a navigation fault is severe, it is likely to be given higher priority for fixing.
The values highlighted in yellow depict weak positive correlation. For example, correlation between frequency of occurrence of network faults with the priority to fix it is 0.3458. This can be due to the fact, that network faults are not necessarily revealed at developers' end rather they arise at players' side due to issues in internet connection, network traffic etc \cite{PM6}. The priority to fix them is not as high as the frequency of occurrence of network faults. The values highlighted in red depict a negligible correlation. The correlation between Frequency and severity is mostly \textbf{weakly-positive} as shown by values highlighted in yellow in table \ref{tab:corr} with exception of Unexpected Crash which is negligible. The reason for this could be that not all crash faults lead to catastrophic consequences, such as loss of progress or data corruption. Some crash faults may simply result in the game closing or freezing temporarily, which may be frustrating but not necessarily severe. Additionally, some crash faults may only occur under specific circumstances or with specific hardware configurations, leading to a high frequency of occurrence among a subset of players but not across the entire player base.

The correlation between Frequency and Priority is \textbf{weakly-positive} or \textbf{negligible} as shown by values highlighted in yellow and red in table \ref{tab:corr}.
This is because the bugs that occur most frequently are not necessarily the most severe or impactful ones. Therefore, the bugs that are less frequent but have a higher impact on the gameplay experience would be prioritized. Additionally, other factors such as the complexity of fixing the bug, available resources, and time constraints may also influence the priority assigned to fixing a particular bug.

The correlation between Severity and Priority is \textbf{moderately-positive} as shown by values highlighted in green table \ref{tab:corr}. The higher correlation between the severity and priority of game bugs suggests that the severity of a bug is an important factor in determining its priority for fixing. Additionally, bugs with a higher severity rating may also be more urgent to fix because they could potentially cause players to stop playing the game.

\par None of the values of correlation are greater than 0.8 which implies a very strong positive correlation. The results of the statistical tests confirm the implications of the MLR itself. The frequency of occurrence of a game bug does not necessarily imply severity and priority. Rather severity depends on the impact the bug has on a player's overall ability to complete a gameplay task. Any specific bug category, in general, is not the most critical rather it varies from game to game.

\begin{table} [H]
\centering
\captionsetup{font=normalsize}
\caption{Illustrates the spearmen's rank correlation between frequency of occurrence of bug with its perceived severity, frequency of occurrence of bug with its perceived priority and a bugs perceived severity with its perceived priority respectively.}
\label{tab:corr}
\scriptsize
\begin{tabular}{|l|l|l|l|} 
\hline
\textbf{Bug Category}                     & \begin{tabular}[c]{@{}l@{}}\textbf{Frequency \&}\\\textbf{Severity}\end{tabular} & \begin{tabular}[c]{@{}l@{}}\textbf{Frequency \&}\\\textbf{Priority}\end{tabular} & \begin{tabular}[c]{@{}l@{}}\textbf{Severity \&}\\\textbf{Priority}\end{tabular}  \\ 
\hline
\textbf{Gaming Balance}                   & {\cellcolor[rgb]{1,0.922,0.612}}0.3394                                           & {\cellcolor[rgb]{1,0.78,0.808}}0.1870                                            & {\cellcolor[rgb]{0.776,0.937,0.808}}0.4812                                       \\ 
\hline
\textbf{Implementation Response}          & {\cellcolor[rgb]{1,0.922,0.612}}0.3505                                           & {\cellcolor[rgb]{1,0.78,0.808}}0.0275                                            & {\cellcolor[rgb]{1,0.922,0.612}}0.2860                                           \\ 
\hline
\textbf{Network}                          & {\cellcolor[rgb]{1,0.922,0.612}}0.3550                                           & {\cellcolor[rgb]{1,0.922,0.612}}0.3458                                           & {\cellcolor[rgb]{0.776,0.937,0.808}}0.5443                                       \\ 
\hline
\textbf{Sound}                            & {\cellcolor[rgb]{0.776,0.937,0.808}}0.4162                                       & {\cellcolor[rgb]{1,0.922,0.612}}0.2449                                           & {\cellcolor[rgb]{0.776,0.937,0.808}}0.5715                                       \\ 
\hline
\textbf{Temporal}                         & {\cellcolor[rgb]{0.776,0.937,0.808}}0.4369                                       & {\cellcolor[rgb]{1,0.922,0.612}}~0.3691                                          & {\cellcolor[rgb]{0.776,0.937,0.808}}0.4931                                       \\ 
\hline
\textbf{Unexpected Crash}                 & {\cellcolor[rgb]{1,0.78,0.808}}0.0940                                            & {\cellcolor[rgb]{1,0.78,0.808}}0.1536                                            & {\cellcolor[rgb]{0.776,0.937,0.808}}0.4890                                       \\ 
\hline
\textbf{Navigation}                       & {\cellcolor[rgb]{0.776,0.937,0.808}}0.4883                                       & {\cellcolor[rgb]{1,0.922,0.612}}~0.3120                                          & {\cellcolor[rgb]{0.776,0.937,0.808}}0.4701                                       \\ 
\hline
\textbf{Non Temporal}                     & {\cellcolor[rgb]{1,0.922,0.612}}0.3417                                           & {\cellcolor[rgb]{1,0.78,0.808}}0.1952                                            & {\cellcolor[rgb]{0.776,0.937,0.808}}0.5276                                       \\ 
\hline
\end{tabular}
\end{table}

\section{Discussion} \label{sec:discussion}
This section summarizes the  lessons learned. We discuss the implications for each game bug category and provide an overview of research trends.

\subsection{Gaming Bugs}

\subsubsection{Gaming Balance}

There appears to be an upward trend in focus on gaming balance issues in academic literature, especially with the rise in interest in MMO  games where variation in parameters of one archetype or model may influence game-play experience of all archetypes or  models. This is also consistent with our survey, as 32.1$\%$ of the respondents considered gaming balance issues to have major severity, 22.6$\%$ considered them critical and 4.2$\%$ considered them to possess blocker severity.
Most frequently encountered Gaming Balance faults were too easy and too hard, encountered by  46.7\% of respondents.


\subsubsection{Sound Faults}

\par On the contrary, there is a lack of research related to the identification of sound bugs in games.  We only found one approach  (i.e., \cite{rivergame}) in the academic literature  that focused on identification of sound faults mentioned in Section \ref{ssec:sound}.
Sound effects (SFX) provide important gameplay cues for players where they clarify or reinforce player actions, giving feedback on the player’s decisions. Game sound is important as it gives the game context, individuality, and depth. Iconic games are synonymous with their soundtracks and sound effects, and it is beyond doubt that a strong palette of sound is integral to the player’s overall experience of a game. Inversely, issues in-game sounds can lead to bad user experience. It is especially critical for rhythm and dance games which require on-spot sound effects and synchronized music \cite{W2}. 
\par 
The results of our survey show that while most of the respondents considered the sound faults to be a minor inconvenience, the players that made use of sounds for tactics (e.g. calculate distance of enemy from sound of footsteps in DOTA) considered it a severe issue.

\subsubsection{Implementation Response Faults}

In total  29 out of 78 selected academic studies on automated game testing approaches focus on implementation response faults. Most frequently encountered Implementation response issue is Game Freeze and it is encountered by over  65\% (or 109) of respondents as compared to second most frequent which was collision encountered by 50 respondents (29.7\%). However, only 4 and 6 academic studies focus on them respectively. Alternatively, incorrect response is covered by 16 studies but only 27\% (46) of respondents claim to have encountered them

\subsubsection{Network Faults}

Network faults are considered most severe  (with 10.8\%, 37\%, 30.3\%, 17.3\%, and 4\% respondents considering them blocker, critical, major, minor and low respectively).
However, only  9 out of 78 selected academic studies focused on network faults. The respondents who stated that they always encountered navigational faults mostly played MOBA and action  genre games.

Most frequently encountered Network faults is Network lag and it is encountered by  63.1\% (106) of respondents followed by connection fault which was encountered by 42.2\% (71) of the respondents. However, none of the academic studies cover network lag.

\subsubsection{Temporal Faults}

Most frequently encountered Temporal faults are Delayed Response and it is encountered by  49.4\% (83) of respondents followed by Interupted Event  which was encountered by 30.9\% (52) of the respondents. Unfortunately, neither of these two categories is covered in academic literature.  

\subsubsection{Non-Temporal Faults}

Most frequently encountered Non-Temporal faults are Player Stuck issue and it is encountered by  41.6\% (70) of respondents followed by Faulty Action which was encountered by 32.9\% (55) of the respondents. Contrarily only 4 studies in the academic literature cover the Player Stuck issue  while 12 studies cover faulty actions.

\subsubsection{Navigational Faults}

 Certain bugs like being able to move through walls, are sometimes considered tactic features. Michał Madej \cite{T4}  in his talk in \textit{Digital Dragons} 2014 discussed how certain bugs in their game were not reported by the testers because those bugs were being used as strategy tactics \cite{T4}. The respondents who stated that they \textit{always} encountered navigational faults mostly played MOBA and action genre games. Contrarily, many of the respondents who stated \textit{never} were players of casual games like candy crush and angry birds that do not have an in-game navigation element.
 
\subsubsection{Unexpected Crash Faults}

The survey shows that correlation between frequency and priority, and frequency and severity of such faults is nearly negligble. The reason for this could be that  the frequency of game crashing faults may not necessarily indicate their impact on gameplay or the user experience. Some crashes may occur more frequently than others but may not have a significant impact on the overall gameplay or user experience. Alternatively, there is a moderately positive correlation between severity and priority. Most frequently encountered Crashes are Crash as Start Up and Crash after Action encountered by  49.4\% (83) and 44.6\% (75) of respondents respectively.

\subsection{Research Trends}

\par In recent years, there has been an increased focus on using machine learning for automated game testing and bug identification as concluded by \cite{assessmentframework} and shown in 
 \cite{RN201},  \cite{RN208},  \cite{ RN171},  \cite{ RN87},  \cite{ RN211},  \cite{ RN247},  \cite{ RN202},  \cite{ RN251}.  A major reason for this rise, is the capacity of machine learning techniques to handle unexpected scenarios during dynamic game-play. The dynamic game-play is a consequence of presence of intelligent agents in PvE and PvP games. 

\par Consequently, the popularity of dynamic game-play makes the automated generation of test oracle very difficult. In most of the existing game testing techniques, test oracle generation is manual. It is done either via human observation; by simply checking for game crashes; by testing for visual glitches; or by adding game-specific test assertion into the game source code  \cite{RN87} like checking for navigability on game-map; or by modeling game environment in a deterministic form like state machines  \cite{RN56},  \cite{RN214}.

\par However, a serious limitation of automated oracle generation techniques is that they are specific to games. Model-based techniques can permit a certain level of generalization, provided a game-specific model is defined. The model can be in form of a domain-specific language like VGDL  \cite{ humansynthetic2019}, a state machine \cite{RN56},  \cite{RN214}, or any other modeling language that can be used to model game scenarios like the graphical modeling language used in  \cite{RN245} that allows testers to  model action sequences needed to complete a game quest scenario in a role-playing game. Unfortunately, exhaustive modeling of complex MMO games like \textit{Elder Scrolls V: Skyrim} is a challenging avenue.

\subsection{Game Testing Techniques}

\par
\begin{figure*}
\noindent
  \makebox[\textwidth]{
  
    \includegraphics[width=1.3\textwidth, angle=90]{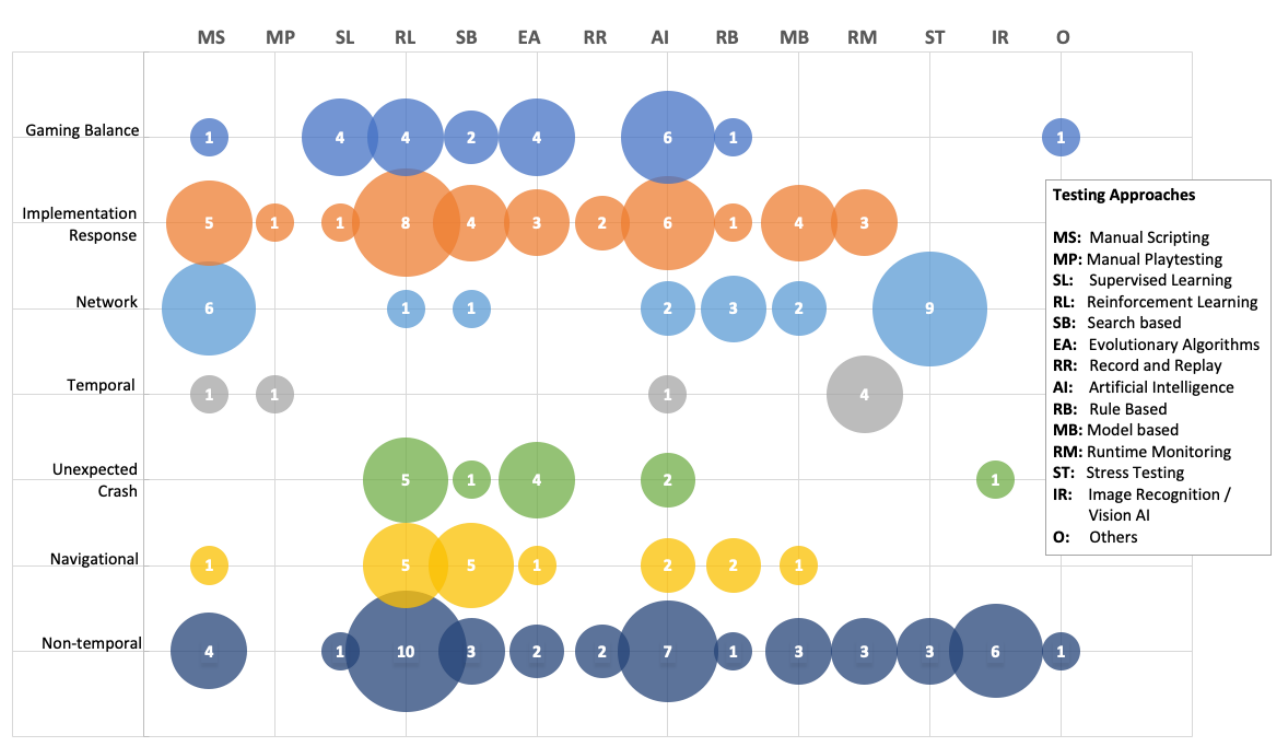}}
    \caption{\label{fig:bubble} Illustrates the types of game testing techniques used for identification of game bug.}
\end{figure*}

  The trend of game-testing techniques proposed in academic literature  shows that most of the work is focused on AI-based techniques.Gaming balance faults are captured using RL \cite{ RN247},  \cite{ RN249},  \cite{ RN110},  \cite{ RN218}, and AI based approaches \cite{ RN188},  \cite{ RN258},  \cite{ RN263},  \cite{ RN265},  \cite{ RN266},  \cite{ RN267}. 
Implementation faults are captured using manual scripting \cite{ RN219},  \cite{ RN91},  \cite{ RN41},  \cite{RN270},  \cite{RN273} and AI based approaches \cite{ RN165},  \cite{ RN91},  \cite{ RN258}, \cite{ RN176}. Network faults are identified using Stress testing. Temporal and Navigation faults are least focused. Temporal faults are mostly handled by using runtime monitoring systems and navigational ones are handled using RL and Search based approaches. Unexpected crashes are  handled using RL and EA-based approaches. Non-temporal faults are captured using RL, AI and ML-based approaches like image recognition(figure \ref{fig:bubble}).  The detailed categorization of selected studies based on testing techniques versus bug category is present in \textit{Video Game Bug Taxonomy MLR - bug capturing techniques} in the dataset as well as table \ref{tab:bugcap} in appendix \ref{app:summary}.

\par Difficulty in generalizing an automated testing technique across genres and  platforms is also a major cause of lack of automation \cite{Politowski2021survey}. One  difficulty  is  that  different  test   techniques   usually   require   different   pieces   of   information   about   the   system,   e.g.,   specifications   in   different  forms,  or  may  target  different  types of  systems as well as different goals \cite{briand2004empirical}. An approach designed for 3-match games (like Candy Crush) will not map to a platform game (like Super Mario) which are popular genres according to our survey. On the other hand, an approach designed to test reachability of a player's avatar to a particular location for platform games (like Super Mario) could map to arcade or role-playing games (like Legend of Zelda).  

\par Majority of the game testing approaches target system-level testing, which is also highlighted in postmortems. Most of the testing is done in final stages of development which leads to time crunch and bugs being passed to post-production phase. Our conclusion is  early integration of testing is needed which aligns with findings drawn by \cite{Politowski2021gray},  \cite{schultz2016game},  \cite{PM15}.

\subsection{Lessons Learned from Grey Literature}
\par During analysis of grey literature, we learned of  an interesting aspect of presence of game bugs; their exploitation by game players to the extent that some are treated as features. Certain bugs like being able to move through walls are sometimes considered tactic features. Michał Madej \cite{T4}  in his talk in \textit{Digital Dragons} 2014 discussed how certain bugs in their game were not reported by the testers because those bugs were being used as strategy tactics \cite{T4}. This introduces the concepts of \textit{bugs as features}, which is a prevalent theme in MMO games. These \textit{bugs} can belong to any of the proposed bug categories. Major MMORPG game servers like \textit{Runescape-3} and \textit{Elite Dungeons} place temporary bans on players caught abusing bugs as well as penalize them via removal of accumulated wealth and bonuses \cite{V1}.

\par The impact of different game bugs is not consistent during gameplay experience for games of different genres. Not all bugs equally impact gameplay experience. Most of the respondents considered the  severity  of network issues to be high. 
The results of our survey show that while most of the respondents considered the sound faults to be a minor inconvenience, the players that made use of sounds for tactics (e.g. calculate distance of enemy from sound of footsteps in DOTA) considered it a severe issue.


\par A major challenge highlighted in the grey literature, especially in postmortems related to game testing is the difficulty in reproducibility of game bugs. Video games are complex software with cross-cutting dependencies which makes reproducing a bug difficult \cite{PM9},  \cite{PM10},  \cite{PM49} in such situations log files and bug reports were found to be the most important resources. Postmortem analysis showed that industry mostly used external playtesters to test their games at last stages of development. Very rarely automated testing techniques were used and even those focused more on automated execution of scripted or recorded scenarios. 

\subsection{Discussion Regarding Proposed Taxonomy}

\par 
An advantage of bug taxonomies is the identification of bugs that can be seeded into software to evaluate testing approaches. This opens another avenue of research, the proposition of mutation operators corresponding to categories in bugs. An example of this can be removing collision detection from an obstacle.

\par According to Lough et al.  \cite{RN309}, a taxonomy should be comprehensible, complete, deterministic, mutually exclusive, repeatable, well defined, unambiguous and useful. We have designed our taxonomy keeping to the characteristics defined by  \cite{RN309}. However, we have observed that in the domain of games, mutual exclusivity is not possible due complexity of system interactions and sequence combinations. The resulting inevitability of cross-cutting and overlapping means that bugs may be placed in various categories. For  example, a player's avatar is able to pass through a blocking obstacle. This creates a bug that falls into category of both “Collision Detection Fault” and “Faulty Obstacle”. Hence, in such a scenario, it will depend on the observer perspective where the bug is placed. If the bug is categorized from perspective of the avatar then it would be collision detection fault; alternatively if the focus is the new obstacle then it could be  faulty obstacle.
\par The focus of the taxonomy is the Implementation Faults that occur during game execution from the time game application is launched or game player logs in to the game, until it is shut down or game player logs out. Hence at this stage, we do not consider installation or deletion faults; or development or deployment platform compatibility faults within the scope of this taxonomy. Such faults can be the focus of future work.

\section{Threats to Validity} \label{sec:threattovalidity}
In this section, the threats to validity are explained in detail along with steps followed to minimize their influence.
\subsection{Internal Threats Validity} 
The selection of literature is an internal validity threat in this study. To mitigate, we used a systematic search method to find the white and grey literature. We searched the major digital libraries with different query phrases and utilized a uniform inclusion and exclusion criteria for final academic study selection. Despite using a systematic approach, there is still a potential of missing a relevant study, and there may also be studies published in languages other than English. However, we limited our search to studies that were published in English. 
Another threat to internal validity is researcher bias. We attempted to mitigate it by having each manuscript evaluated by at least two authors and having all conflicts in study selection discussed and resolved by several reviews and group meetings among the authors.
\subsection{Construct Threats Validity}
 We used the findings from the MLR when we refined the questionnaire following many discussions among the authors, in order to offset difficulties with construct validity. 
Another threat to construct validity was whether the survey participants correctly understood the identified game bug categories. To achieve this, we supplied detailed descriptions of the identified categories, for each of the relevant questions. To Further ensure objectivity of the participants, we assured all participants that their identities would be kept anonymous.

\subsection{Conclusions Threats Validity}
 To validate the treatment's dependability, all studies were thoroughly examined by at least two authors to eliminate bias in data extraction, which can lead to inaccurate conclusions. We also evaluated the quality of included studies which confirmed that most of them are of good quality based on the quality assessment criteria. Any disagreements over the retrieved data were settled by consensus among the authors. To ensure data traceability, the provided graphs and tables are created from the extracted data in a spreadsheet.

\section{Related Work} \label{sec:relatedworks}
In this section, we discuss related literature, highlighting the differences from each study. Politowski et al. \cite{Politowski2021gray} and Washburn et al. \cite{RN230} analyzed postmortems from the video-game development industry, with a focus on management and production aspects, and best practices and difficulties in the development process, respectively. Our analysis of postmortems from \textit{Gamesutra.com} focuses on implementation faults resulting in gameplay bugs, with extended and specialized categories, including Graphics/sound issues and Game mechanics/system issues. 
Politowski et al. \cite{Politowski2021survey} surveyed the game testing practices in the industry and found that manual playtesting and intrinsic knowledge are the most commonly used methods. Albaghajati et al. \cite{assessmentframework}
conducted a survey of academic literature to provide a framework for game testing approaches. Redavid et al. \cite{gametestingoverview}
presented an overview of game testing practices from a software engineering perspective. We also surveyed academic and grey literature, with a focus on identifying limitations of existing automated game testing approaches and classifying bugs encountered by game players. Our proposed taxonomy provides guidance for playtesters and researchers in creating robust automated game testing approaches. Other related studies include diagnostic taxonomies of game failures from usability perspective \cite{aytemiz2020taxonomy}, challenges specific to network computer games \cite{gamenetworkbugs}, and a bug database for comparative study of testing techniques \cite{li2022gbgallery}. 

\par We have extended the taxonomy presented in   \cite{Lewistaxonomy} after an MLR to identify and classify commonly appearing bugs that result in game failures even after launch. We have added major classifications of Navigational bugs, Network problems, and Gaming Balance and further expanded the non-temporal implementation failures class to include sound issues, unexpected crashes and System response issues. We have also added to other classes described in detail in the following sections. We have validated our taxonomy from professionals.  

\section{Conclusion} \label{sec:conclusion}
This paper proposed detailed game bug taxonomy and discussed essential characteristics of the existing game testing approaches. The taxonomy is based on comprehensive analysis of game bugs and existing game testing approaches by conducting systematic review of white literature and grey literature. The classification of bugs into different categories is performed by thoroughly analyzing the existing classification schemes for bugs and game-testing approaches. The taxonomy provides classification for \textbf{63} different types of \textbf{implementation faults} found 189 different sources of relevant literature. These 63 faults are classified into eight high-level categories i.e. Gaming Balance, Implementation response, Network, Sound, Temporal, Unexpected Crash, Navigational and Non-Temporal Faults. Five out of 63 faults have been observed and modified from existing literature and \textbf{15} categories are used from existing literature as is. While \textbf{43} categories are newly added as a result of  meticulous MLR process.
We validated the proposed taxonomy by conducting survey involving different game industry practitioners such as game developers, game testers, and game players. The results of the survey ratify the validity of bug taxonomy. We also performed correlation analysis between the defined metrics, which show moderately positive correlation between severity and priority and weakly positive correlation between frequency and priority of the bugs. 

\par We observed that manual approaches toward game testing are still dominating in the industry. There is very little work done for automated detection of sound bugs in games. Conversely, game balancing is the most studied topic in recent literature. Furthermore, most game testing techniques are specialized and depend on specific platforms. Like other sub-areas of testing, the issue of automated oracle also exists in game testing. Most recent approaches in game testing are incorporating machine learning techniques.  In future, we plan to focus on installation and deletion faults as well as  development and deployment platform compatibility faults.

\medskip

\scriptsize
\medskip

\printbibliography[keyword={referred},title={References}]
\newrefcontext[labelprefix=S]
\printbibliography[keyword={selected},title={Selected Studies},]
\newrefcontext[labelprefix=PM]
\printbibliography[keyword={postmortems},title={Selected Postmortems},]
\newrefcontext[labelprefix=T]
\printbibliography[keyword={talks},title={Selected Talk Sessions},]
\newrefcontext[labelprefix=W]
\printbibliography[keyword={web},title={Selected Web Sources},]
\newrefcontext[labelprefix=V]
\printbibliography[keyword={video},title={Selected Video Sources},]

\break
 \appendix
\begin{Huge}

\textbf{Appendix}

\end{Huge}

\counterwithin{figure}{section}
\renewcommand\thefigure{\thesection.\arabic{figure}}  

\counterwithin{table}{section}
 \renewcommand\thetable{\thesection.\arabic{table}} 
 \section{Quality Assessment} \label{app:qualityassessment}
 This appendix contains the table with detailed quality assessment of selected academic studies.
\begin{longtable}[c]{|r|l|l|l|l|l|l|l|r|}
\caption{Quality Assessment of Selected Academic Literature}
\label{tab:qualityassesment}\\
\hline
\multicolumn{1}{|l|}{\textbf{ID}} &
  \textbf{QC1} &
  \textbf{QC2} &
  \textbf{QC3} &
  \textbf{QC4} &
  \textbf{QC5} &
  \textbf{QC6} &
  \textbf{QC7} &
  \multicolumn{1}{l|}{\textbf{Quality Score}} \\ \hline
\endfirsthead
\multicolumn{9}{c}%
{{\bfseries Table \thetable\ continued from previous page}} \\
\hline
\multicolumn{1}{|l|}{\textbf{ID}} &
  \textbf{QC1} &
  \textbf{QC2} &
  \textbf{QC3} &
  \textbf{QC4} &
  \textbf{QC5} &
  \textbf{QC6} &
  \textbf{QC7} &
  \multicolumn{1}{l|}{\textbf{Quality Score}} \\ \hline
\endhead
\textbf{S1}  & $\checkmark$ & $\checkmark$ & $\checkmark$ & $\checkmark$ & $\checkmark$ & $\checkmark$ & X            & 6 \\ \hline
\textbf{S2}  & $\checkmark$ & $\checkmark$ & $\checkmark$ & $\checkmark$ & $\checkmark$ & $\checkmark$ & $\checkmark$ & 7 \\ \hline
\textbf{S3}  & $\checkmark$ & $\checkmark$ & $\checkmark$ & $\checkmark$ & $\checkmark$ & $\checkmark$ & $\checkmark$ & 7 \\ \hline
\textbf{S4}  & $\checkmark$ & $\checkmark$ & $\checkmark$ & $\checkmark$ & $\checkmark$ & $\checkmark$ & $\checkmark$ & 7 \\ \hline
\textbf{S5}  & $\checkmark$ & $\checkmark$ & $\checkmark$ & $\checkmark$ & $\checkmark$ & $\checkmark$ & $\checkmark$ & 7 \\ \hline
\textbf{S6}  & $\checkmark$ & $\checkmark$ & $\checkmark$ & $\checkmark$ & $\checkmark$ & $\checkmark$ & $\checkmark$ & 7 \\ \hline
\textbf{S7}  & $\checkmark$ & $\checkmark$ & $\checkmark$ & X            & $\checkmark$ & $\checkmark$ & X            & 5 \\ \hline
\textbf{S8}  & $\checkmark$ & $\checkmark$ & $\checkmark$ & $\checkmark$ & $\checkmark$ & $\checkmark$ & $\checkmark$ & 7 \\ \hline
\textbf{S9}  & $\checkmark$ & $\checkmark$ & $\checkmark$ & X            & $\checkmark$ & $\checkmark$ & X            & 5 \\ \hline
\textbf{S10} & $\checkmark$ & $\checkmark$ & $\checkmark$ & X            & $\checkmark$ & $\checkmark$ & X            & 5 \\ \hline
\textbf{S11} & $\checkmark$ & X            & X            & $\checkmark$ & $\checkmark$ & $\checkmark$ & $\checkmark$ & 5 \\ \hline
\textbf{S12} & $\checkmark$ & $\checkmark$ & $\checkmark$ & $\checkmark$ & $\checkmark$ & $\checkmark$ & $\checkmark$ & 7 \\ \hline
\textbf{S13} & $\checkmark$ & $\checkmark$ & $\checkmark$ & $\checkmark$ & $\checkmark$ & $\checkmark$ & X            & 6 \\ \hline
\textbf{S14} & $\checkmark$ & $\checkmark$ & $\checkmark$ & X            & $\checkmark$ & $\checkmark$ & X            & 5 \\ \hline
\textbf{S15} & $\checkmark$ & $\checkmark$ & $\checkmark$ & X            & $\checkmark$ & $\checkmark$ & X            & 5 \\ \hline
\textbf{S16} & $\checkmark$ & $\checkmark$ & $\checkmark$ & X            & X            & X            & X            & 3 \\ \hline
\textbf{S17} & $\checkmark$ & $\checkmark$ & $\checkmark$ & X            & X            & X            & X            & 3 \\ \hline
\textbf{S18} & $\checkmark$ & $\checkmark$ & $\checkmark$ & $\checkmark$ & $\checkmark$ & $\checkmark$ & $\checkmark$ & 7 \\ \hline
\textbf{S19} & $\checkmark$ & $\checkmark$ & $\checkmark$ & $\checkmark$ & $\checkmark$ & $\checkmark$ & X            & 6 \\ \hline
\textbf{S20} & $\checkmark$ & $\checkmark$ & $\checkmark$ & $\checkmark$ & $\checkmark$ & $\checkmark$ & $\checkmark$ & 7 \\ \hline
\textbf{S21} & $\checkmark$ & $\checkmark$ & $\checkmark$ & $\checkmark$ & $\checkmark$ & $\checkmark$ & $\checkmark$ & 7 \\ \hline
\textbf{S22} & $\checkmark$ & $\checkmark$ & $\checkmark$ & $\checkmark$ & $\checkmark$ & $\checkmark$ & $\checkmark$ & 7 \\ \hline
\textbf{S23} & $\checkmark$ & $\checkmark$ & $\checkmark$ & X            & $\checkmark$ & $\checkmark$ & X            & 5 \\ \hline
\textbf{S24} & $\checkmark$ & $\checkmark$ & $\checkmark$ & $\checkmark$ & $\checkmark$ & $\checkmark$ & X            & 6 \\ \hline
\textbf{S25} & $\checkmark$ & $\checkmark$ & $\checkmark$ & $\checkmark$ & $\checkmark$ & $\checkmark$ & $\checkmark$ & 7 \\ \hline
\textbf{S26} & $\checkmark$ & $\checkmark$ & $\checkmark$ & $\checkmark$ & $\checkmark$ & $\checkmark$ & X            & 6 \\ \hline
\textbf{S27} & $\checkmark$ & $\checkmark$ & $\checkmark$ & X            & $\checkmark$ & $\checkmark$ & $\checkmark$ & 6 \\ \hline
\textbf{S28} & $\checkmark$ & $\checkmark$ & $\checkmark$ & $\checkmark$ & $\checkmark$ & $\checkmark$ & $\checkmark$ & 7 \\ \hline
\textbf{S29} & $\checkmark$ & $\checkmark$ & $\checkmark$ & $\checkmark$ & $\checkmark$ & $\checkmark$ & X            & 6 \\ \hline
\textbf{S30} & $\checkmark$ & $\checkmark$ & $\checkmark$ & $\checkmark$ & $\checkmark$ & $\checkmark$ & $\checkmark$ & 7 \\ \hline
\textbf{S31} & $\checkmark$ & $\checkmark$ & $\checkmark$ & $\checkmark$ & $\checkmark$ & $\checkmark$ & $\checkmark$ & 7 \\ \hline
\textbf{S32} & $\checkmark$ & $\checkmark$ & $\checkmark$ & $\checkmark$ & $\checkmark$ & $\checkmark$ & X            & 6 \\ \hline
\textbf{S33} & $\checkmark$ & $\checkmark$ & $\checkmark$ & X            & $\checkmark$ & $\checkmark$ & X            & 5 \\ \hline
\textbf{S34} & $\checkmark$ & $\checkmark$ & $\checkmark$ & $\checkmark$ & $\checkmark$ & $\checkmark$ & $\checkmark$ & 7 \\ \hline
\textbf{S35} & $\checkmark$ & $\checkmark$ & $\checkmark$ & $\checkmark$ & $\checkmark$ & $\checkmark$ & X            & 6 \\ \hline
\textbf{S36} & $\checkmark$ & $\checkmark$ & $\checkmark$ & $\checkmark$ & $\checkmark$ & $\checkmark$ & X            & 6 \\ \hline
\textbf{S37} & $\checkmark$ & $\checkmark$ & $\checkmark$ & $\checkmark$ & $\checkmark$ & $\checkmark$ & X            & 6 \\ \hline
\textbf{S38} & $\checkmark$ & $\checkmark$ & $\checkmark$ & $\checkmark$ & $\checkmark$ & $\checkmark$ & $\checkmark$ & 7 \\ \hline
\textbf{S39} & $\checkmark$ & $\checkmark$ & $\checkmark$ & $\checkmark$ & $\checkmark$ & $\checkmark$ & $\checkmark$ & 7 \\ \hline
\textbf{S40} & $\checkmark$ & $\checkmark$ & $\checkmark$ & $\checkmark$ & $\checkmark$ & $\checkmark$ & X            & 6 \\ \hline
\textbf{S41} & $\checkmark$ & X            & X            & $\checkmark$ & $\checkmark$ & $\checkmark$ & $\checkmark$ & 5 \\ \hline
\textbf{S42} & $\checkmark$ & $\checkmark$ & X            & X            & $\checkmark$ & $\checkmark$ & X            & 4 \\ \hline
\textbf{S43} & $\checkmark$ & $\checkmark$ & $\checkmark$ & X            & $\checkmark$ & $\checkmark$ & X            & 5 \\ \hline
\textbf{S44} & $\checkmark$ & $\checkmark$ & $\checkmark$ & $\checkmark$ & $\checkmark$ & $\checkmark$ & $\checkmark$ & 7 \\ \hline
\textbf{S45} & $\checkmark$ & $\checkmark$ & $\checkmark$ & $\checkmark$ & $\checkmark$ & $\checkmark$ & X            & 6 \\ \hline
\textbf{S46} & $\checkmark$ & X            & X            & $\checkmark$ & $\checkmark$ & $\checkmark$ & $\checkmark$ & 5 \\ \hline
\textbf{S47} & $\checkmark$ & $\checkmark$ & $\checkmark$ & $\checkmark$ & X            & X            & X            & 4 \\ \hline
\textbf{S48} & $\checkmark$ & $\checkmark$ & $\checkmark$ & X            & $\checkmark$ & $\checkmark$ & X            & 5 \\ \hline
\textbf{S49} & $\checkmark$ & $\checkmark$ & $\checkmark$ & X            & $\checkmark$ & $\checkmark$ & X            & 5 \\ \hline
\textbf{S50} & $\checkmark$ & $\checkmark$ & $\checkmark$ & X            & $\checkmark$ & $\checkmark$ & X            & 5 \\ \hline
\textbf{S51} & $\checkmark$ & $\checkmark$ & $\checkmark$ & $\checkmark$ & $\checkmark$ & $\checkmark$ & X            & 6 \\ \hline
\textbf{S52} & $\checkmark$ & $\checkmark$ & $\checkmark$ & $\checkmark$ & $\checkmark$ & $\checkmark$ & X            & 6 \\ \hline
\textbf{S53} & $\checkmark$ & $\checkmark$ & $\checkmark$ & $\checkmark$ & $\checkmark$ & $\checkmark$ & X            & 6 \\ \hline
\textbf{S54} & $\checkmark$ & $\checkmark$ & $\checkmark$ & $\checkmark$ & $\checkmark$ & $\checkmark$ & X            & 6 \\ \hline
\textbf{S55} & $\checkmark$ & $\checkmark$ & $\checkmark$ & $\checkmark$ & $\checkmark$ & $\checkmark$ & X            & 6 \\ \hline
\textbf{S56} & $\checkmark$ & $\checkmark$ & $\checkmark$ & $\checkmark$ & $\checkmark$ & $\checkmark$ & X            & 6 \\ \hline
\textbf{S57} & $\checkmark$ & $\checkmark$ & $\checkmark$ & $\checkmark$ & $\checkmark$ & $\checkmark$ & X            & 6 \\ \hline
\textbf{S58} & $\checkmark$ & $\checkmark$ & $\checkmark$ & $\checkmark$ & $\checkmark$ & $\checkmark$ & $\checkmark$ & 7 \\ \hline
\textbf{S59} & $\checkmark$ & $\checkmark$ & $\checkmark$ & $\checkmark$ & $\checkmark$ & $\checkmark$ & $\checkmark$ & 7 \\ \hline
\textbf{S60} & $\checkmark$ & $\checkmark$ & $\checkmark$ & $\checkmark$ & $\checkmark$ & $\checkmark$ & $\checkmark$ & 7 \\ \hline
\textbf{S61} & $\checkmark$ & $\checkmark$ & $\checkmark$ & $\checkmark$ & $\checkmark$ & $\checkmark$ & X            & 6 \\ \hline
\textbf{S62} & $\checkmark$ & $\checkmark$ & $\checkmark$ & $\checkmark$ & $\checkmark$ & $\checkmark$ & $\checkmark$ & 7 \\ \hline
\textbf{S63} & $\checkmark$ & $\checkmark$ & $\checkmark$ & $\checkmark$ & $\checkmark$ & $\checkmark$ & $\checkmark$ & 7 \\ \hline
\textbf{S64} & $\checkmark$ & $\checkmark$ & $\checkmark$ & $\checkmark$ & $\checkmark$ & $\checkmark$ & $\checkmark$ & 7 \\ \hline
\textbf{S65} & $\checkmark$ & $\checkmark$ & $\checkmark$ & $\checkmark$ & $\checkmark$ & $\checkmark$ & $\checkmark$ & 7 \\ \hline
\textbf{S66} & $\checkmark$ & $\checkmark$ & $\checkmark$ & X            & $\checkmark$ & $\checkmark$ & X            & 5 \\ \hline
\textbf{S67} & $\checkmark$ & $\checkmark$ & $\checkmark$ & $\checkmark$ & $\checkmark$ & $\checkmark$ & X            & 6 \\ \hline
\textbf{S68} & $\checkmark$ & $\checkmark$ & $\checkmark$ & $\checkmark$ & $\checkmark$ & $\checkmark$ & X            & 6 \\ \hline
\textbf{S69} & $\checkmark$ &              &              & $\checkmark$ & $\checkmark$ & $\checkmark$ & $\checkmark$ & 5 \\ \hline
\textbf{S70} & $\checkmark$ & $\checkmark$ & $\checkmark$ & $\checkmark$ & $\checkmark$ & $\checkmark$ & $\checkmark$ & 7 \\ \hline
\textbf{S71} & $\checkmark$ & $\checkmark$ & $\checkmark$ & $\checkmark$ & $\checkmark$ & $\checkmark$ & $\checkmark$ & 7 \\ \hline
\textbf{S72} & $\checkmark$ & $\checkmark$ & $\checkmark$ & $\checkmark$ & $\checkmark$ & $\checkmark$ & $\checkmark$ & 7 \\ \hline
\textbf{S73} & $\checkmark$ & $\checkmark$ & $\checkmark$ & $\checkmark$ & $\checkmark$ & $\checkmark$ & $\checkmark$ & 7 \\ \hline
\textbf{S74} & $\checkmark$ & $\checkmark$ & $\checkmark$ & X            & $\checkmark$ & $\checkmark$ & $\checkmark$ & 6 \\ \hline
\textbf{S75} & $\checkmark$ & $\checkmark$ & $\checkmark$ & X            & $\checkmark$ & $\checkmark$ & $\checkmark$ & 6 \\ \hline
\textbf{S76} & $\checkmark$ & $\checkmark$ & $\checkmark$ & $\checkmark$ & $\checkmark$ & $\checkmark$ & $\checkmark$ & 7 \\ \hline
\textbf{S77} & $\checkmark$ & $\checkmark$ & $\checkmark$ & X            & $\checkmark$ & X            & X            & 4 \\ \hline
\textbf{S78} & $\checkmark$ & $\checkmark$ & $\checkmark$ & X            & $\checkmark$ & X            & X            & 4 \\ \hline
\multicolumn{1}{|l|}{\textbf{\%}} &
  \multicolumn{1}{r|}{\textbf{100}} &
  \multicolumn{1}{r|}{\textbf{94.87}} &
  \multicolumn{1}{r|}{\textbf{93.59}} &
  \multicolumn{1}{r|}{\textbf{74.36}} &
  \multicolumn{1}{r|}{\textbf{96.16}} &
  \multicolumn{1}{r|}{\textbf{93.59}} &
  \multicolumn{1}{r|}{\textbf{48.72}} &
  \textbf{85.9} \\ \hline
\end{longtable}

 \newpage
 \section{Taxonomy Details} \label{app:summary}

\begin{center}

\scriptsize
\begin{longtable}{|l|l|p{10cm}|}
\captionsetup{font=normalsize}
\caption{\textbf{Sources for the derived game bugs main-tier categories}}
\label{tab:bug-src}\\
\hline
\textbf{Bug Categories} &
  \textbf{Source Type} &
  \textbf{Sources} \\ \hline \hline
\endfirsthead
\multicolumn{3}{c}%
{{ Table \thetable\ continued from previous page}} \\
\hline
\textbf{Bug Categories} &
  \textbf{Source Type} &
  \textbf{Sources} \\ \hline
\endhead
\multirow{5}{*}{\textbf{\begin{tabular}[c]{@{}l@{}}Gaming \\ Balance\end{tabular}}} &
  Academic &
  S3, S15, S26, S33, S34, S35, S36, S40, S48, S49, S51, S52, S53, S54, S55, S56, S57, S64, S74, S77 \\ \cline{2-3} 
 &
  Postmortems &
  PM1, PM2, PM14, PM16, PM20, PM24, PM25, PM26, PM29,   PM30, PM32, PM34, PM51, PM53, PM56, PM57, PM62 \\ \cline{2-3} 
 &
Talks   &
  N/A \\ \cline{2-3} 
 &
  Web &
  W1, W2, W9, W17 \\ \cline{2-3} 
 &
  Video &
  V7,   V16 \\ \hline
\multirow{5}{*}{\textbf{\begin{tabular}[c]{@{}l@{}}Implementation \\ Response\end{tabular}}} &
  Academic &
  S1, S2, S6, S8, S12, S13, S18, S20, S21, S24, S29,   S31, S33, S38, S39, S43, S44, S49, S58, S61, S62, S63, S65, S66, S68, S70, S72, S75, S76 \\ \cline{2-3} 
 &
  Postmortems &
  PM4, PM7, PM12, PM21, PM22, PM23, PM39, PM42, PM45,   PM47, PM54 \\ \cline{2-3} 
 &
  Talks &
  T1,   T2, T4 \\ \cline{2-3} 
 &
  Web &
  W1, W2, W4, W5, W10, W13, W14, W18, W20 \\ \cline{2-3} 
 &
  Video &
  V2, V3, V5, V10, V13, V20 \\ \hline
\multirow{5}{*}{\textbf{Network}} &
  Academic &
  S12, S23, S25, S27, S37, S45, S50, S60, S71 \\ \cline{2-3} 
 &
  Postmortems &
  PM3, PM6, PM10, PM14, PM17, PM21, PM27, PM28, PM31,   PM43, PM46, PM49, PM52 \\ \cline{2-3} 
 &
  Talks &
  T4 \\ \cline{2-3} 
 &
  Web &
  W2 \\ \cline{2-3} 
 &
  Video &
  V2,   V14 \\ \hline
\multirow{5}{*}{\textbf{Sound}} &
  Academic &
  S58\\ \cline{2-3} 
 &
  Postmortems &
  PM35, PM41 \\ \cline{2-3} 
 &
  Talks &
  T3 \\ \cline{2-3} 
 &
  Web &
  W2 \\ \cline{2-3} 
 &
  Video &
  N/A \\ \hline
\multirow{5}{*}{\textbf{Temporal}} &
  Academic &
  S18, S24, S38, S44, S76 \\ \cline{2-3} 
 &
  Postmortems &
  PM18 \\ \cline{2-3} 
 &
  Talks &
  T1 \\ \cline{2-3} 
 &
  Web &
  W2,   W8 \\ \cline{2-3} 
 &
  Video &
  V3 \\ \hline
\multirow{5}{*}{\textbf{\begin{tabular}[c]{@{}l@{}}Unexpected \\ Crash\end{tabular}}} &
  Academic &
  S6, S8, S11, S20, S31, S48, S67, S73, S75 \\ \cline{2-3} 
 &
  Postmortems &
  PM4, PM8, PM9, PM30, PM33, PM36, PM40, PM44, PM47,   PM50, PM55 \\ \cline{2-3} 
 &
  Talks &
  T2,   T5 \\ \cline{2-3}
  
  &
   Web &
  W1, W2, W3, W4, W19 \\
\cline{2-3}
  
 &

  Video &
  V2, V3, V4, V6, V10, V22 \\ \hline
\multirow{5}{*}{\textbf{Navigational}} &
  Academic &
  S4, S5, S9, S20, S21, S34, S39, S59, S63,  S65, S66, S70, S72, S73, S78 \\ \cline{2-3} 
 &
  Postmortems &
  PM5, PM11, PM61 \\ \cline{2-3} 
 &
  Talks &
  T3 \\ \cline{2-3} 
 &
  Web &
  W1,   W2, W4 \\ \cline{2-3} 
 &
  Video &
  V2,   V10 \\ \hline
\multirow{5}{*}{\textbf{Non-Temporal}} &
  Academic &
  S1, S2, S6, S7, S8, S11, S13, S18, S19, S20, S21,   S22, S24, S25, S28, S30, S32, S38, S39, S41, S43, S44, S45, S46, S49, S58, S60, S61, S62, S63, S65,   S66, S67, S69, S70, S72, S73, S74, S75, S76 \\ \cline{2-3} 
 &
  Postmortems &
  PM2, PM8, PM11, PM12, PM13, PM15, PM19, PM21, PM23,   PM37, PM38, PM47, PM48, PM49, PM54, PM56, PM58, PM59, PM60 \\ \cline{2-3} 
 &
  Talks &
  T1,   T2, T3 \\ \cline{2-3} 
 &
  Web &
  W1, W2, W4, W5, W6, W8, W10, W11, W12, W15, W16 \\ \cline{2-3} 
 &
  Video &
  V1, V2, V3, V4, V7, V8, V10, V11, V12, V15, V16, V17,   V18, V19, V23 \\ \hline
\end{longtable}
    
\end{center}

\begin{longtable}[H]{|p{2cm}|p{2cm}|p{4.5cm}|p{5cm}|c|}
\captionsetup{font=normalsize}
\caption{\textbf{Descriptions and examples of specialized new game bug categories}}
\label{tab:taxonomy}\\ \hline
\textbf{Categories} &
  \textbf{Parent Category} &
  \textbf{Descriptions} &
  \textbf{Game Example} &
  \textbf{Ref}  \\ \hline
\endfirsthead
\multicolumn{5}{c}%
{{ Table \thetable\ continued from previous page}} \\ \hline
\textbf{Categories} &
  \textbf{Parent Category} &
  \textbf{Descriptions} &
  \textbf{Game Example} &
  \textbf{Ref} \\ \hline
\endhead
Player Stuck Fault &
  Non Temporal Fault &
  Player is unable to progress in   the game either due to inability to move from location or inability to   complete a critical task. &
  In Guacamelee,   the player could get locked in the fight arena. &
  PM13 \\ \hline
Accelerated Response Fault &
  Temporal Fault &
  Consequence of an action is   accelerated giving unfair advantage or disadvantage especially in real time   games &
  In King's   Quest IV, when Rosella is in the ogre's house and   must reach the door before he catches her, the ogre travels across the screen   in seconds. &
  W2 \\ \hline
Delayed Response Fault &
  Temporal Fault &
  There is an unexpected delay in   response to an action or a scenario. It is especially critical in real-time   games. &
  In King's   Quest IV, at some random intervals the response to   Rosella and other characters would lag &
  W2 \\ \hline
Network Lag &
  Network   Faults &
  Game play experience is impacted   by lagging response in network games due to network speed constraints and   Packet size constraints &
  In Fireteam, lag would result in the players shots missing, by the time the   shot has been fired the intended target has gone around a corner &
  PM27 \\ \hline
Synchronization &
  Network Faults &
  Different game states are visible   to different players in a multi-player game. A game event starts according to   an incorrect or different time zone. &
  In Sins of a   Solar Empire: Rebellion, partway through a match,   players would find inconsistencies between the game states for different   players &
  PM49 \\ \hline
Connection Fault &
  Network Faults &
  Player is unable to connect with   the game server. &
  In Fallout 76, if multiple mannequins were dressed in same outfit it   resulted in the players being logged out and unable to establish a connection   again. &
  V14 \\ \hline
Network Overload &
  Network Faults &
  Game play experience of player is   negatively impacted once number of connected players exceeds a specified   amount. &
  \textit{Wireless Pets   could not even handle 10,000 users playing it simultaneously} &
  PM17 \\ \hline
Packet Loss &
  Network Faults &
  A variety of packet loss or   corruption problems that happen between game servers and client applications   due to unreliable or unsustainable connection especially required for   MMORPGs. &
  In Toontown, a variety of packet loss problems were encountered &
  PM10 \\ \hline
Unreachable Locations &
  Navigational Faults &
  A player is unable to reach a   location in game or a position on screen that should be accessible. &
  In Rastan on   the Commodore 64, on the second level, it is   impossible to make a jump over a flaming pit overtwo ropes to continue with   the game progression &
  W2 \\ \hline
Unexpected Paths &
  Navigational Faults &
  A player can reach a location in   game or a position on screen that should be inaccessible. &
  In Fallout 77, players were able to reach places and levels that were still   being developed. &
  V2 \\ \hline
Too Easy &
  Gaming Balance &
  Players find it too easy to play a   game. It is especially critical if a particular set of parameters allows for   unfair advantages in multiplayer games. &
  In Hearthstone if a player plays   \textit{jerry rig carpenter}, it reduces the mana value needed to play   cards to zero which is equivalent to handing a free win &
  V7 \\ \hline
Too Hard &
  Gaming Balance &
  Players find it too difficult to   finish the game. This includes difficulty in finishing a task due to obscure   design. &
  In Fallout Tactics, there are some   missions where the whole map depends on finding one key in an obscure place. &
  PM24 \\ \hline
Unwinnable &
  Gaming Balance &
  Players are unable to finish the   game especially if it is due to an overly competent AI or convoluted game   logic or game constraints such as small timers or impossible to obtain   in-game items. &
  In the Political Machine, the AI   was too competent. When player got up to George Washington to play. The   opponent AI won every state. &
  PM51 \\ \hline
Corrupted Frame &
  Invalid Graphical Representaion &
  Corrupted frames are visible. &
  \textit{Elder Scrolls V: Skyrim}, had a number rendering problems that resulted in corrupted   frames &
  V4 \\ \hline
Slow Loading &
  Invalid Graphical Representaion &
  Frame loading rate is too slow &
  \textit{Mighty No.9},   also fell victim to slow frame rate &
  V8 \\ \hline
Extra Game Asset &
  Invalid Graphical Representaion &
  An in-game element that should not   be displayed is visible. &
  In Jak X:   Combat Racing, has a fault known as   \textit{saving glitch} where the save icon hangs on the game &
  W1 \\ \hline
Missing Game Asset &
  Invalid Graphical Representaion &
  An in-game element that should be   displayed is not present. &
  In Donkey Kong   Country II, if a player picks up the barrel as it   breaks, then the avatar is seen carrying an invisible barrel &
  V3 \\ \hline
Abnormal   Text &
  Invalid   Graphical Representaion &
  Text appears at wrong location or   is incomprehensible or may cover a characters or other objects. &
  In Pokémon   Red, Blue, Green and Yellow, the text describing   ‘MISSINGNO’ pokemon was often glitchy. &
  V3 \\ \hline
Incorrect Visual State Transition &
  Invalid Graphical Representaion &
  An in-game element transitions to   an incorrect state &
  In Slow Down,   Bull, Annette the bull catcher was notorious for   not transitioning properly through her various tell states &
  PM2 \\ \hline
Incorrect Sound &
  Sound Faults &
  The sound behind an animation is   incorrect or does not match the circumstances &
  In Whacked, missing sound resources led to either sounds missing or wrong   sounds being played behind animations &
  PM41 \\ \hline
Corrupted Sound &
  Sound Faults &
  The sound behind an animation is   corrupted or warped. &
  In Final Zone   II, a buzzing sound would start after the intro   cutscene and continue throughout the game. &
  W2 \\ \hline
Unsynchronized Sound &
  Sound Faults &
  The sound behind an animation is   not correctly synchronized especially critical for Rhythm games. &
  In Portable   Black Squares and Clazziquai, Edition’s music   became unsynchronized. &
  W2 \\ \hline
Too Short Sound &
  Sound Faults &
  The sound ends before the   animation completes &
  In Whacked, the sound of pendulum swing cut of before the animation ended &
  PM41 \\ \hline
Run Away Sound &
  Sound Faults &
  The sound continues even after the   animation completes &
  In The Italian   Job, due to lack of sound resources available, the   engines sounds looped very badly &
  PM35 \\ \hline
Multiple Action Inconsistancy &
  Action &
  An in-game action only appears   faulty after it is combined with another action. &
  In The Legend   of Zelda: The Wind Waker, there is a glitch where   if a player jumps and attacks on the back of the chest in the Ghost Ship, the   game will crash &
  W1 \\ \hline
Repeated Action Inconsistancy &
  Action &
  An in-game action only appears   faulty after it is taken repeatedly. &
  \textit{Psychonauts}   had a quirk where a player became unable to use double jump in a speccific   level &
  W2 \\ \hline
Game Freeze &
  Implementation Response Fault &
  Game stops responding to any user   action making progress impossible &
  \textit{In Majora’s Mask},   the game freezes when the player pushes the open icon button that appears on   an underwater chest in Termina Field &
  W1 \\ \hline
Unresponsive Action &
  Implementation Response Fault &
  Player performs an in-game action   successfully but the consequence or result of that action does not appear. &
  In Cyberpunk   2077, when Takemura calls, the player takes the   call but the NPC does not go away on end call. &
  V21 \\ \hline
Incorrect   Reward / Punishment &
  Implementation   Response Fault &
  An in-game action results in less   or more reward compared to expected, conversely less or more punishment than   expected is also possible. &
  In Super   Mario Bros. Deluxe for Game Boy Color, if a player earns the Yoshi Medal at   the same time as the Red Coins or High Score medals in Challenge mode, the   player only receives the Yoshi medal, and the others are lost forever. &
  W2 \\ \hline
Incorrect Response Fault &
  Implementation Response Fault &
  Player performs an in-game action   successfully but the consequence or result of that action is incorrect. &
  \textit{Final Fantasy V}   had a weapon (the Chicken Knife), that would sometimes make the player run   away from battle instead of doing an attack &
  W2 \\ \hline
Collision Detection Fault &
  Implementation Response Fault &
  all instances of faulty collision   detection &
  In PUBG, a faulty location on a   specific map allowed player avatars to disappear into the map texture &
  V2 \\ \hline
Crash During Demo &
  Unexpected Crash &
  Game crashes unexpectedly during   demo or practice sequence animation. &
  \textit{PONCHO} has a   crash bug in the game demo. &
  PM36 \\ \hline
Crash During Non-Play Moments &
  Unexpected Crash &
  Game crashes unexpectedly during   game Non-Play Moments. &
  In Nuclear   Throne on windows, game crashes during   introduction sequence of Loop Boss, Throne II. &
  W1 \\ \hline
Crash At Start Up &
  Unexpected Crash &
  Game crashes unexpectedly during   game start up sequence. &
  Save file corruption results in   the game crash at start up fault in Prey. &
  V2 \\ \hline
Crash After Action &
  Unexpected Crash &
  Game crashes unexpectedly after an   in-game action is executed. &
  In Mighty No.9, game unexpected crashed when player fired a weapon. &
  V8 \\ \hline
Crash At Shutdown &
  Unexpected Crash &
  Game crashes unexpectedly during   game shut down sequence. &
  In Super Meat   Boy, the game crash during shut down. &
  PM55 \\ \hline
Faulty Pits &
  Unexpected Paths &
  Players can navigate over a pit   through which they should fall &
  In Super Mario, Mario did not fall through a pit. &
  S2 \\ \hline
Faulty Obstacles &
  Unexpected Paths &
  Players can navigate across   obstacles against which they should collide. &
  In Mass Effect, Matriarch Benezia's use of Biotic powers would toss the main   character through the wall &
  W2 \\ \hline
\end{longtable}

\begin{table}[]
\centering
\scriptsize
\caption{Contains the detailed categorization of selected studies based on testing techniques versus bug category.}
\label{tab:bugcap}
\begin{tabular}{|p{3cm}|p{1.5cm}|p{1.5cm}|p{1.5cm}|p{1.5cm}|p{1.5cm}|p{1.5cm}|p{1.5cm}|}
\hline
 &
  \textbf{Gaming Balance} &
  \textbf{Impl. Response} &
  \textbf{Network} &
  \textbf{Temporal} &
  \textbf{Unexpect-ed Crash} &
  \textbf{Navigat-ional} &
  \textbf{Non-temporal} \\ \hline
\textbf{Manual scripting} &
  S36 &
  S13, S24, S29, S65, S68 &
  S23, S27, S37, S45, S50, S71 &
  S24 &
   &
  S65 &
  S24, S32, S45, S65 \\ \hline
\textbf{Manual Playtesting} &
   &
  S18 &
   &
  S18 &
   &
   &
   \\ \hline
\textbf{Supervised Learning} &
  S3, S36, S64, S77&
  S43 &
   &
   &
   &
   &
  S43 \\ \hline
\textbf{Reinforcement learning} &
  S15, S26, S36, S40 &
  S6, S20, S21, S31, S62, S63, S66&
   S60&
   &
  S6, S20, S31, S67, S73 &
  S5, S20,  S63, S73&
  S6, S20, S21, S60, S62, S63, S66, S67, S73\\ \hline
\textbf{Search-based} &
  S40, S48 &
  S2, S20, S21, S43 &
  S12 &
   &
  S48 &
  S9, S20, S21, S59, S78&
  S20, S21, S43 \\ \hline
\textbf{Evolutionary Algorithms}&
  S33, S48, S51, S57 &
  S6, S33, S39 &
   &
   &
  S6, S33, S39, S48 &
  S39 &
  S6, S39 \\ \hline
\textbf{Record \& Replay} &
   &
  S1, S29 &
   &
   &
   &
   &
  S1, S32 \\ \hline
\textbf{AI} &
  S34, S49, S52, S54, S55, S56 &
  S8, S24, S49, S58, S61,  S72, S75&
  S45, S54 &
  S24 &
  S8, S75 &
  S4, S34 &
  S24, S30, S45, S49, S58, S61, S72, S74\\ \hline
\textbf{Game grammer / game rules / VGDL} &
  S34 &
  S70 &
  S25, S27, S37 &
   &
   &
  S34, S70 &
  S25 \\ \hline
\textbf{Model based} &
   &
  S1, S2, S12, S38 &
  S12, S37 &
   &
   &
S4   &
  S1, S2, S38 \\ \hline
\textbf{Runtime monitoring} &
   &
  S38, S44, S76 &
   &
  S18, S38, S44, S76 &
   &
   &
  S38, S44, S76 \\ \hline
\textbf{Stress testing} &
   &
   &
  S12, S23, S25, S27, S37, S45, S50, S60, S71&
   &
   &
   &
  S25, S45, S60\\ \hline
\textbf{Computer Vision} &
   &
   &
   &
   &
  S11 &
   &
  S22, S32, S41, S46, S69, S70 \\ \hline
\textbf{Others} &
  S53 &
   &
   &
   &
   &
   &
  S28 \\ \hline
\end{tabular}
\end{table}

\newpage
\section{Survey Questionnaire Details} \label{app:survey}

\begin{longtable}{|c|p{7cm}|p{1cm}p{1cm}p{0.5cm}p{0.75cm}p{0.75cm}|}
\captionsetup{font=normalsize}
\caption{\textbf{Survey Questionnaire}}
\label{tab:survey}\\
\hline
\multirow{2}{*}{Q\#} &
  \multirow{2}{*}{Survey Question} &
  \multicolumn{5}{p{4cm}|}{Type of Answers} \\ \cline{3-7} 
 &
   &
  \multicolumn{1}{p{1cm}|}{Single Answer from a list} &
  \multicolumn{1}{p{1cm}|}{Multiple Answers from a list} &
  \multicolumn{1}{p{1cm}|}{Free text field} &
  \multicolumn{1}{p{0.75cm}|}{Numeric field} &
  \multicolumn{1}{p{0.75cm}|}{Likert Scale} \\ \hline
\endfirsthead
\multicolumn{7}{c}%
{{ Table \thetable\ continued from previous page}} \\
\hline
\multirow{2}{*}{Q\#} &
  \multirow{2}{*}{Survey Question} &
   \multicolumn{5}{p{4cm}|}{Type of Answers} \\ \cline{3-7} 
 &
   &
  \multicolumn{1}{p{1cm}|}{Single Answer from a list} &
  \multicolumn{1}{p{1cm}|}{Multiple Answers from a list} &
  \multicolumn{1}{p{0.5cm}|}{Free text field} &
  \multicolumn{1}{p{0.75cm}|}{Numeric field} &
  \multicolumn{1}{p{0.5cm}|}{Likert Scale} \\ \hline
\endhead
1 &
  Your Name &
  \multicolumn{1}{l|}{} &
  \multicolumn{1}{l|}{} &
  \multicolumn{1}{l|}{$\checkmark$} &
  \multicolumn{1}{l|}{} &
   \\ \hline
2 &
  Name of your organization &
  \multicolumn{1}{l|}{} &
  \multicolumn{1}{l|}{} &
  \multicolumn{1}{l|}{$\checkmark$} &
  \multicolumn{1}{l|}{} &
   \\ \hline
3 &
  Your Job Title &
  \multicolumn{1}{l|}{} &
  \multicolumn{1}{l|}{} &
  \multicolumn{1}{l|}{$\checkmark$} &
  \multicolumn{1}{l|}{} &
   \\ \hline
4 &
  Which game do you play most often? &
  \multicolumn{1}{l|}{} &
  \multicolumn{1}{l|}{} &
  \multicolumn{1}{l|}{$\checkmark$} &
  \multicolumn{1}{l|}{} &
   \\ \hline
5 &
  Which of the following is your area of   expertise? &
  \multicolumn{1}{l|}{} &
  \multicolumn{1}{l|}{$\checkmark$} &
  \multicolumn{1}{l|}{} &
  \multicolumn{1}{l|}{} &
   \\ \hline
6 &
  How long have you been playing video   games? (in years) &
  \multicolumn{1}{l|}{} &
  \multicolumn{1}{l|}{} &
  \multicolumn{1}{l|}{} &
  \multicolumn{1}{l|}{$\checkmark$} &
   \\ \hline
7 &
  How long have you been developing video   games? (in years) &
  \multicolumn{1}{l|}{} &
  \multicolumn{1}{l|}{} &
  \multicolumn{1}{l|}{} &
  \multicolumn{1}{l|}{$\checkmark$} &
   \\ \hline
8 &
  How long have you been testing video   games? (in years) &
  \multicolumn{1}{l|}{} &
  \multicolumn{1}{l|}{} &
  \multicolumn{1}{l|}{} &
  \multicolumn{1}{l|}{$\checkmark$} &
   \\ \hline
9 &
  What genre of video games do you play? &
  \multicolumn{1}{l|}{} &
  \multicolumn{1}{l|}{$\checkmark$} &
  \multicolumn{1}{l|}{} &
  \multicolumn{1}{l|}{} &
   \\ \hline
10 &
  How often do you encounter Gaming   Balance Faults while playing video games? &
  \multicolumn{1}{l|}{} &
  \multicolumn{1}{l|}{} &
  \multicolumn{1}{l|}{} &
  \multicolumn{1}{l|}{} &
  $\checkmark$ \\ \hline
11 &
  How would you rate severity of Gaming   Balance Faults while playing video games? &
  \multicolumn{1}{l|}{} &
  \multicolumn{1}{l|}{} &
  \multicolumn{1}{l|}{} &
  \multicolumn{1}{l|}{} &
  $\checkmark$ \\ \hline
12 &
  How would you prioritize the need to fix   Gaming Balance Faults encountered while playing video games? &
  \multicolumn{1}{l|}{} &
  \multicolumn{1}{l|}{} &
  \multicolumn{1}{l|}{} &
  \multicolumn{1}{l|}{} &
  $\checkmark$ \\ \hline
13 &
  What type of Gaming Balance Faults have   you encountered while playing video games? &
  \multicolumn{1}{l|}{} &
  \multicolumn{1}{l|}{$\checkmark$} &
  \multicolumn{1}{l|}{} &
  \multicolumn{1}{l|}{} &
   \\ \hline
14 &
  Which of the Gaming Balance   sub-categories do you find unnecessary and why? &
  \multicolumn{1}{l|}{} &
  \multicolumn{1}{l|}{$\checkmark$} &
  \multicolumn{1}{l|}{$\checkmark$} &
  \multicolumn{1}{l|}{} &
   \\ \hline
15 &
  How often do you encounter Implementation   Response Faults while playing video games? &
  \multicolumn{1}{l|}{} &
  \multicolumn{1}{l|}{} &
  \multicolumn{1}{l|}{} &
  \multicolumn{1}{l|}{} &
  $\checkmark$ \\\hline
16 &
  How would you rate severity of Implementation   Response Faults while playing video games? &
  \multicolumn{1}{l|}{} &
  \multicolumn{1}{l|}{} &
  \multicolumn{1}{l|}{} &
  \multicolumn{1}{l|}{} &
  $\checkmark$ \\ \hline
17 &
  How would you prioritize the need to fix   Implementation Response Faults encountered while playing video games? &
  \multicolumn{1}{l|}{} &
  \multicolumn{1}{l|}{} &
  \multicolumn{1}{l|}{} &
  \multicolumn{1}{l|}{} &
  $\checkmark$ \\ \hline
18 &
  What type of Implementation Response   Faults have you encountered while playing video games? &
  \multicolumn{1}{l|}{} &
  \multicolumn{1}{l|}{$\checkmark$} &
  \multicolumn{1}{l|}{} &
  \multicolumn{1}{l|}{} &
   \\ \hline
19 &
  Which of the Implementation Response   Faults sub-categories do you find unnecessary and why? &
  \multicolumn{1}{l|}{} &
  \multicolumn{1}{l|}{$\checkmark$} &
  \multicolumn{1}{l|}{$\checkmark$} &
  \multicolumn{1}{l|}{} &
   \\ \hline
20 &
  How often do you encounter Network   related Faults while playing video games? &
  \multicolumn{1}{l|}{} &
  \multicolumn{1}{l|}{} &
  \multicolumn{1}{l|}{} &
  \multicolumn{1}{l|}{} &
  $\checkmark$ \\ \hline
21 &
  How would you rate severity of Network    Faults while playing video games? &
  \multicolumn{1}{l|}{} &
  \multicolumn{1}{l|}{} &
  \multicolumn{1}{l|}{} &
  \multicolumn{1}{l|}{} &
  $\checkmark$ \\ \hline
22 &
  How would you prioritize the need to fix   Network  Faults encountered while playing video games? &
  \multicolumn{1}{l|}{} &
  \multicolumn{1}{l|}{} &
  \multicolumn{1}{l|}{} &
  \multicolumn{1}{l|}{} &
  $\checkmark$ \\ \hline
23 &
  What type of Network  Faults have   you encountered while playing video games? &
  \multicolumn{1}{l|}{} &
  \multicolumn{1}{l|}{$\checkmark$} &
  \multicolumn{1}{l|}{} &
  \multicolumn{1}{l|}{} &
   \\ \hline
24 &
  Which of the Network related sub-categories   do you find unnecessary and why? &
  \multicolumn{1}{l|}{} &
  \multicolumn{1}{l|}{$\checkmark$} &
  \multicolumn{1}{l|}{$\checkmark$} &
  \multicolumn{1}{l|}{} &
   \\ \hline
25 &
  How often do you encounter Sound Faults while   playing video games? &
  \multicolumn{1}{l|}{} &
  \multicolumn{1}{l|}{} &
  \multicolumn{1}{l|}{} &
  \multicolumn{1}{l|}{} &
  $\checkmark$ \\ \hline
26 &
  How would you rate severity of Sound   Faults while playing video games? &
  \multicolumn{1}{l|}{} &
  \multicolumn{1}{l|}{} &
  \multicolumn{1}{l|}{} &
  \multicolumn{1}{l|}{} &
  $\checkmark$ \\ \hline
27 &
  How would you prioritize the need to fix   Sound Faults encountered while playing video games? &
  \multicolumn{1}{l|}{} &
  \multicolumn{1}{l|}{} &
  \multicolumn{1}{l|}{} &
  \multicolumn{1}{l|}{} &
  $\checkmark$ \\ \hline
28 &
  What type of Sound Faults have you   encountered while playing video games? &
  \multicolumn{1}{l|}{} &
  \multicolumn{1}{l|}{$\checkmark$} &
  \multicolumn{1}{l|}{} &
  \multicolumn{1}{l|}{} &
   \\ \hline
29 &
  Which of the Sound Faults sub-categories   do you find unnecessary and why? &
  \multicolumn{1}{l|}{} &
  \multicolumn{1}{l|}{$\checkmark$} &
  \multicolumn{1}{l|}{$\checkmark$} &
  \multicolumn{1}{l|}{} &
   \\ \hline
30 &
  How often do you encounter Temporal   Faults while playing video games? &
  \multicolumn{1}{l|}{} &
  \multicolumn{1}{l|}{} &
  \multicolumn{1}{l|}{} &
  \multicolumn{1}{l|}{} &
  $\checkmark$ \\ \hline
31 &
  How would you rate severity of Temporal   Faults while playing video games? &
  \multicolumn{1}{l|}{} &
  \multicolumn{1}{l|}{} &
  \multicolumn{1}{l|}{} &
  \multicolumn{1}{l|}{} &
  $\checkmark$ \\ \hline
32 &
  How would you prioritize the need to fix   Temporal Faults encountered while playing video games? &
  \multicolumn{1}{l|}{} &
  \multicolumn{1}{l|}{} &
  \multicolumn{1}{l|}{} &
  \multicolumn{1}{l|}{} &
  $\checkmark$ \\ \hline
33 &
  What type of Temporal Faults have you   encountered while playing video games? &
  \multicolumn{1}{l|}{} &
  \multicolumn{1}{l|}{$\checkmark$} &
  \multicolumn{1}{l|}{} &
  \multicolumn{1}{l|}{} &
   \\ \hline
34 &
  Which of the Temporal Faults sub-categories   do you find unnecessary and why? &
  \multicolumn{1}{l|}{} &
  \multicolumn{1}{l|}{$\checkmark$} &
  \multicolumn{1}{l|}{$\checkmark$} &
  \multicolumn{1}{l|}{} &
   \\ \hline
35 &
  How often do you encounter Unexpected   Crash Faults while playing video games? &
  \multicolumn{1}{l|}{} &
  \multicolumn{1}{l|}{} &
  \multicolumn{1}{l|}{} &
  \multicolumn{1}{l|}{} &
  $\checkmark$ \\ \hline
36 &
  How would you rate severity of Unexpected   Crash Faults while playing video games? &
  \multicolumn{1}{l|}{} &
  \multicolumn{1}{l|}{} &
  \multicolumn{1}{l|}{} &
  \multicolumn{1}{l|}{} &
  $\checkmark$ \\\hline
37 &
  How would you prioritize the need to fix   Unexpected Crash Faults encountered while playing video games? &
  \multicolumn{1}{l|}{} &
  \multicolumn{1}{l|}{} &
  \multicolumn{1}{l|}{} &
  \multicolumn{1}{l|}{} &
  $\checkmark$ \\\hline
38 &
  What type of Unexpected Crash Faults have   you encountered while playing video games? &
  \multicolumn{1}{l|}{} &
  \multicolumn{1}{l|}{$\checkmark$} &
  \multicolumn{1}{l|}{} &
  \multicolumn{1}{l|}{} &
   \\ \hline
39 &
  Which of the Unexpected Crash sub-categories   do you find unnecessary and why? &
  \multicolumn{1}{l|}{} &
  \multicolumn{1}{l|}{$\checkmark$} &
  \multicolumn{1}{l|}{$\checkmark$} &
  \multicolumn{1}{l|}{} &
   \\ \hline
40 &
  How often do you encounter Navigational   Faults while playing video games? &
  \multicolumn{1}{l|}{} &
  \multicolumn{1}{l|}{} &
  \multicolumn{1}{l|}{} &
  \multicolumn{1}{l|}{} &
  $\checkmark$ \\\hline
41 &
  How would you rate severity of Navigational   Faults while playing video games? &
  \multicolumn{1}{l|}{} &
  \multicolumn{1}{l|}{} &
  \multicolumn{1}{l|}{} &
  \multicolumn{1}{l|}{} &
  $\checkmark$ \\\hline
42 &
  How would you prioritize the need to fix   Navigational Faults encountered while playing video games? &
  \multicolumn{1}{l|}{} &
  \multicolumn{1}{l|}{} &
  \multicolumn{1}{l|}{} &
  \multicolumn{1}{l|}{} &
  $\checkmark$ \\\hline
43 &
  What type of Navigational Faults have   you encountered while playing video games? &
  \multicolumn{1}{l|}{} &
  \multicolumn{1}{l|}{$\checkmark$} &
  \multicolumn{1}{l|}{} &
  \multicolumn{1}{l|}{} &
   \\ \hline
44 &
  Which of the Navigational sub-categories   do you find unnecessary and why? &
  \multicolumn{1}{l|}{} &
  \multicolumn{1}{l|}{$\checkmark$} &
  \multicolumn{1}{l|}{$\checkmark$} &
  \multicolumn{1}{l|}{} &
   \\ \hline
45 &
  How often do you encounter Non-Temporal   Faults while playing video games? &
  \multicolumn{1}{l|}{} &
  \multicolumn{1}{l|}{} &
  \multicolumn{1}{l|}{} &
  \multicolumn{1}{l|}{} &
  $\checkmark$ \\\hline
46 &
  How would you rate severity of Non-Temporal   Faults while playing video games? &
  \multicolumn{1}{l|}{} &
  \multicolumn{1}{l|}{} &
  \multicolumn{1}{l|}{} &
  \multicolumn{1}{l|}{} &
  $\checkmark$ \\\hline
47 &
  How would you prioritize the need to fix   Non-Temporal Faults encountered while playing video games? &
  \multicolumn{1}{l|}{} &
  \multicolumn{1}{l|}{} &
  \multicolumn{1}{l|}{} &
  \multicolumn{1}{l|}{} &
  $\checkmark$ \\\hline
48 &
  What type of Non-Temporal Faults have   you encountered while playing video games? &
  \multicolumn{1}{l|}{} &
  \multicolumn{1}{l|}{$\checkmark$} &
  \multicolumn{1}{l|}{} &
  \multicolumn{1}{l|}{} &
   \\ \hline
49 &
  Which of the Non-Temporal Faults sub-categories   do you find unnecessary and why? &
  \multicolumn{1}{l|}{} &
  \multicolumn{1}{l|}{$\checkmark$} &
  \multicolumn{1}{l|}{$\checkmark$} &
  \multicolumn{1}{l|}{} &
   \\ \hline
50 &
  How often do you encounter Action Faults   while playing video games? &
  \multicolumn{1}{l|}{} &
  \multicolumn{1}{l|}{} &
  \multicolumn{1}{l|}{} &
  \multicolumn{1}{l|}{} &
  $\checkmark$ \\\hline
51 &
  How would you rate severity of Action   Faults while playing video games? &
  \multicolumn{1}{l|}{} &
  \multicolumn{1}{l|}{} &
  \multicolumn{1}{l|}{} &
  \multicolumn{1}{l|}{} &
  $\checkmark$ \\\hline
52 &
  How would you prioritize the need to fix   Action Faults encountered while playing video games? &
  \multicolumn{1}{l|}{} &
  \multicolumn{1}{l|}{} &
  \multicolumn{1}{l|}{} &
  \multicolumn{1}{l|}{} &
  $\checkmark$ \\\hline
53 &
  What type of Action Faults have you   encountered while playing video games? &
  \multicolumn{1}{l|}{} &
  \multicolumn{1}{l|}{$\checkmark$} &
  \multicolumn{1}{l|}{} &
  \multicolumn{1}{l|}{} &
   \\ \hline
54 &
  Which of the Action Faults sub-categories   do you find unnecessary and why? &
  \multicolumn{1}{l|}{} &
  \multicolumn{1}{l|}{$\checkmark$} &
  \multicolumn{1}{l|}{$\checkmark$} &
  \multicolumn{1}{l|}{} &
   \\ \hline
55 &
  How often do you encounter Invalid   Graphical Representation Faults while playing video games? &
  \multicolumn{1}{l|}{} &
  \multicolumn{1}{l|}{} &
  \multicolumn{1}{l|}{} &
  \multicolumn{1}{l|}{} &
  $\checkmark$ \\\hline
56 &
  How would you rate severity of Invalid   Graphical Representation Faults while playing video games? &
  \multicolumn{1}{l|}{} &
  \multicolumn{1}{l|}{} &
  \multicolumn{1}{l|}{} &
  \multicolumn{1}{l|}{} &
  $\checkmark$ \\\hline
57 &
  How would you prioritize the need to fix   Invalid Graphical Representation Faults encountered while playing video   games? &
  \multicolumn{1}{l|}{} &
  \multicolumn{1}{l|}{} &
  \multicolumn{1}{l|}{} &
  \multicolumn{1}{l|}{} &
  $\checkmark$ \\\hline
58 &
  What type of Invalid Graphical   Representation Faults have you encountered while playing video games? &
  \multicolumn{1}{l|}{} &
  \multicolumn{1}{l|}{$\checkmark$} &
  \multicolumn{1}{l|}{} &
  \multicolumn{1}{l|}{} &
   \\ \hline
59 &
  Which of the Invalid Graphical   Representation sub-categories do you find unnecessary and why? &
  \multicolumn{1}{l|}{} &
  \multicolumn{1}{l|}{$\checkmark$} &
  \multicolumn{1}{l|}{$\checkmark$} &
  \multicolumn{1}{l|}{} &
   \\ \hline
60 &
  What major faults/errors/failures have   you encountered during gameplay in video games other than the ones already   classified? &
  \multicolumn{1}{l|}{} &
  \multicolumn{1}{l|}{} &
  \multicolumn{1}{l|}{$\checkmark$} &
  \multicolumn{1}{l|}{} &
   \\ \hline
\end{longtable}
\newpage
\section{Survey Response Visualization} \label{app:studyfig}

\scriptsize

This appendix present visual depiction of raw data from the survey responses used for derivation of conclusions in our paper.

\begin{figure} [H]
\begin{subfigure}{.18\textwidth}
\centering
\includegraphics[clip, trim=3cm 0.5cm 8cm 2cm,width=1.2\textwidth]{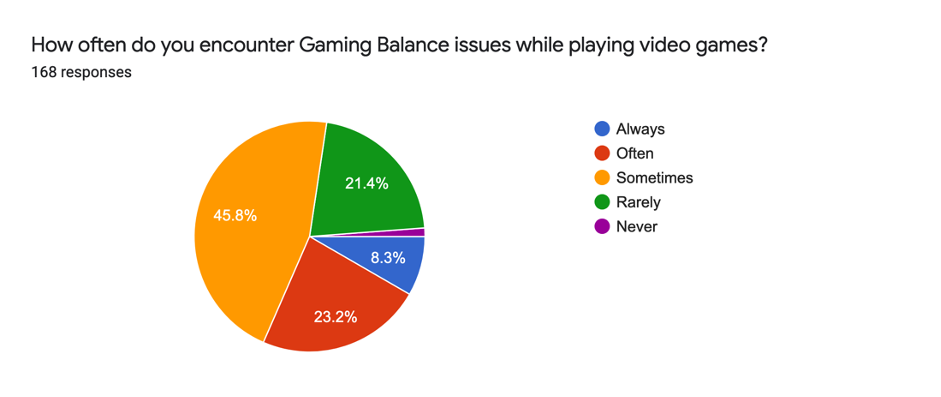}
\caption{\label{fig:fgb}Gaming Balance}
\end{subfigure}
\begin{subfigure}{.18\textwidth}
\centering
\includegraphics[clip, trim=3cm 0.5cm 8cm 2cm,width=1.2\textwidth]{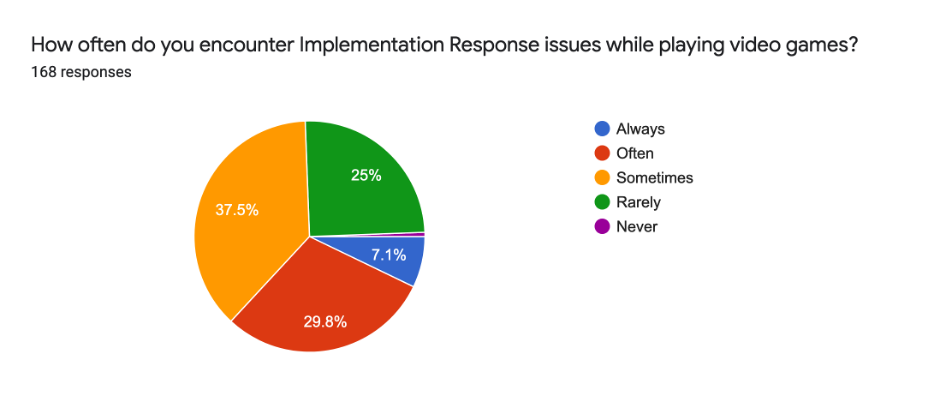}
\caption{\label{fig:fimplss}Impl. Response Faults}
\end{subfigure}
\begin{subfigure}{.18\textwidth}
\centering
\includegraphics[clip, trim=3cm 0cm 8cm 2cm,width=1.2\textwidth]{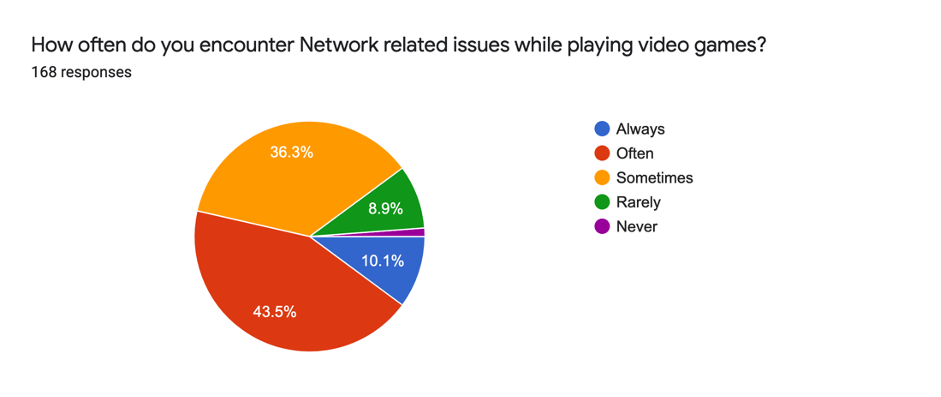}
\caption{\label{fig:fnets}Network   Faults}
\end{subfigure}
\begin{subfigure}{.18\textwidth}
\centering
\includegraphics[clip, trim=3cm 0cm 8cm 2cm,width=1.2\textwidth]{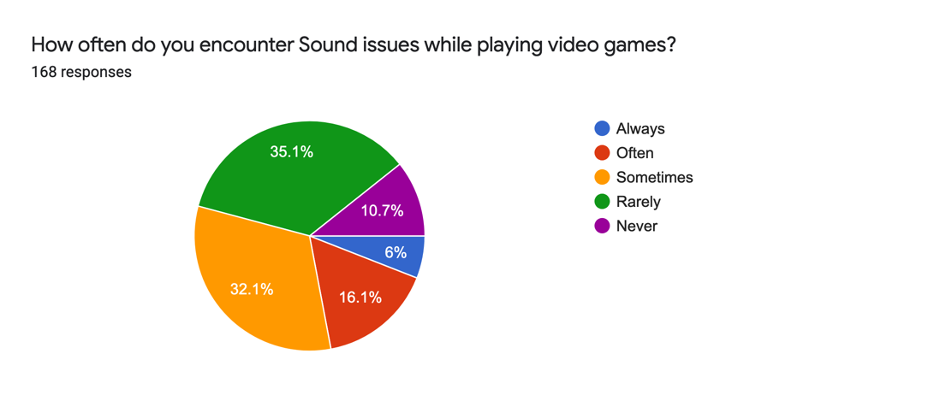}
\caption{\label{fig:fsound}Sound Faults.}
\end{subfigure}
\begin{subfigure}{.18\textwidth}
\centering
\includegraphics[clip, trim=3cm 0.5cm 8cm 2cm,width=1.2\textwidth]{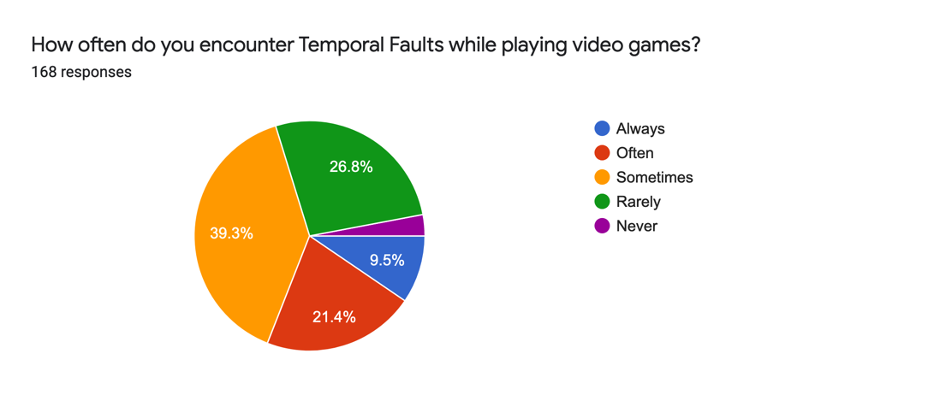}
\caption{\label{fig:fttemppp}Temporal Faults.}
\end{subfigure}

\begin{subfigure}{.18\textwidth}
\centering
\includegraphics[clip, trim=3cm 0.5cm 8cm 2cm,width=1.2\textwidth]{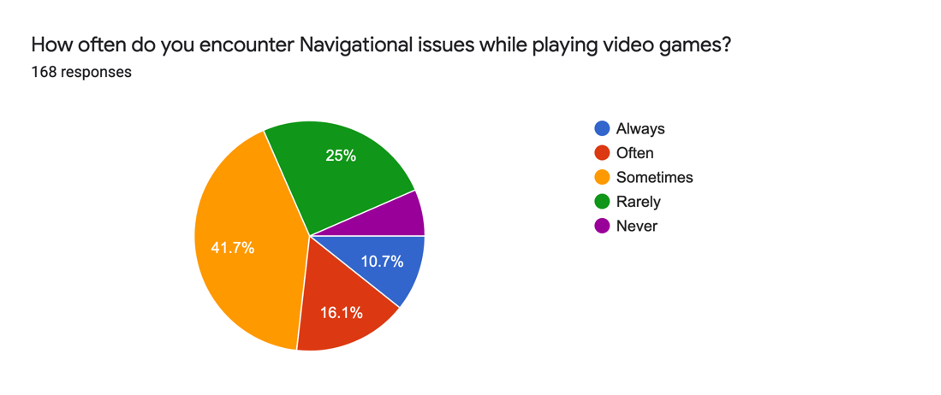}
\caption{\label{fig:fnavii}Navigational Faults.}
\end{subfigure}
\begin{subfigure}{.18\textwidth}
\centering
\includegraphics[clip, trim=3cm 0.5cm 8cm 2cm,width=1.2\textwidth]{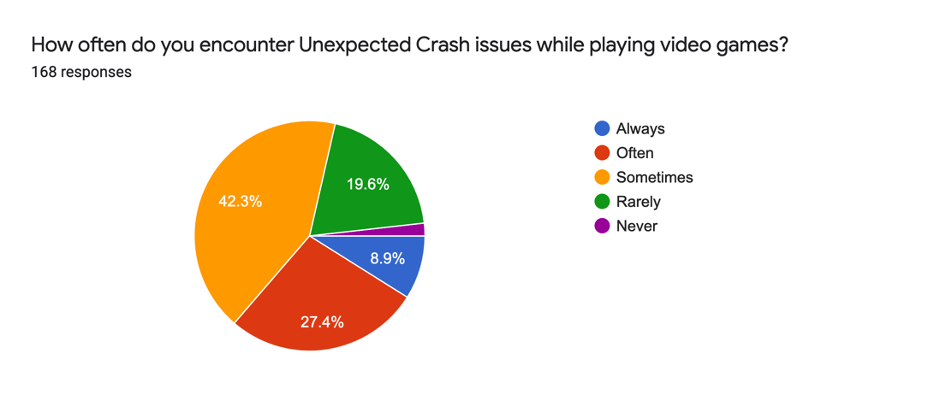}
\caption{\label{fig:fcrash}Unexpected Crash  faults.}
\end{subfigure}
\begin{subfigure}{.18\textwidth}
\centering
\includegraphics[clip, trim=3cm 0.5cm 8cm 2cm,width=1.2\textwidth]{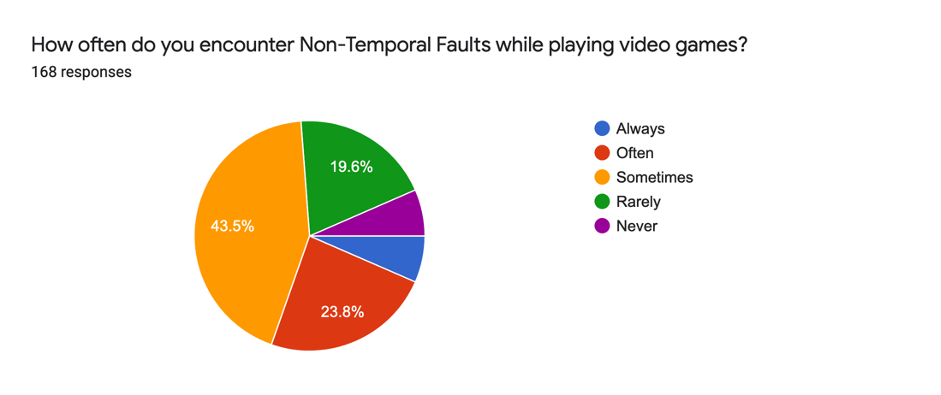}
\caption{\label{fig:fnont}Non Temporal Faults.}
\end{subfigure}
\begin{subfigure}{.18\textwidth}
\centering
\includegraphics[width=\textwidth]{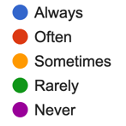}
\caption{\label{fig:freqleg}Legend.}
\end{subfigure}
\caption{\label{fig:bug-freq}Frequency of occurrence of game bugs according to survey responses.}
\end{figure}

\begin{figure}[H]
\begin{subfigure}{.18\textwidth}
\centering
\includegraphics[clip, trim=3cm 0.5cm 8cm 2cm,width=1.2\textwidth]{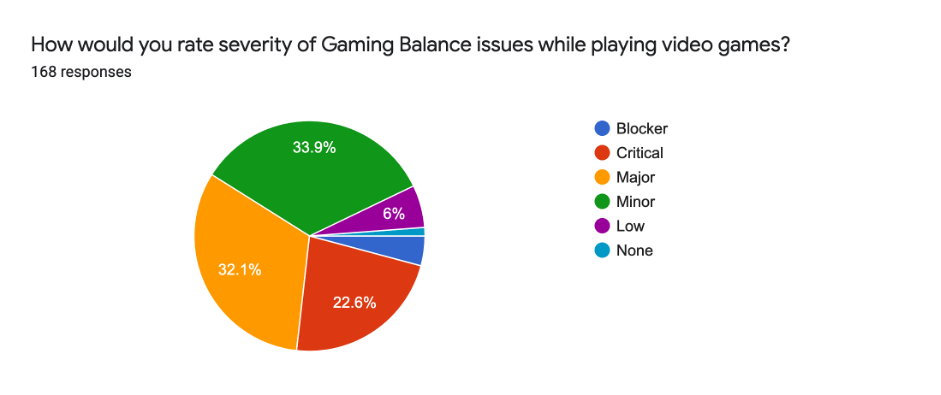}
\caption{\label{fig:sgb}Gaming Balance}
\end{subfigure}
\begin{subfigure}{.18\textwidth}
\centering
\includegraphics[clip, trim=3cm 0.5cm 8cm 2cm,width=1.2\textwidth]{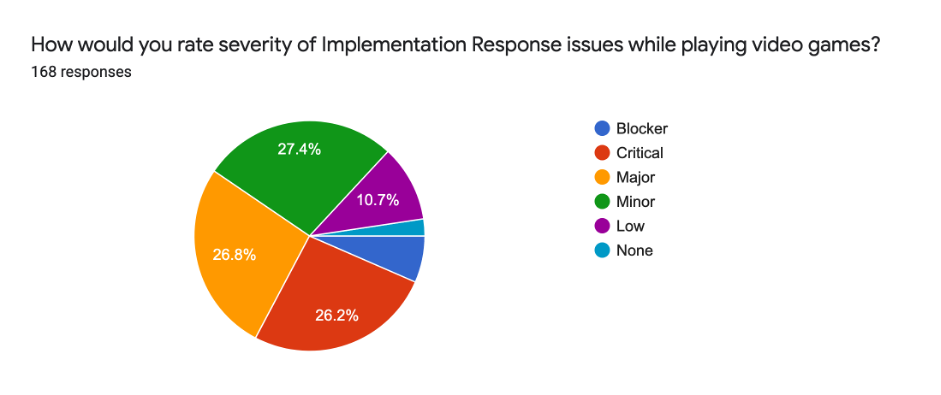}
\caption{\label{fig:simplss}Impl. Response Faults}
\end{subfigure}
\begin{subfigure}{.18\textwidth}
\centering
\includegraphics[clip, trim=3cm 0cm 8cm 2cm,width=1.2\textwidth]{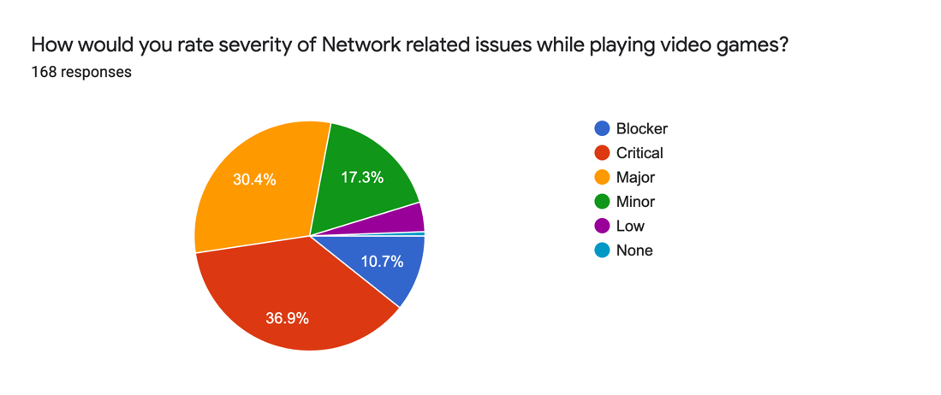}
\caption{\label{fig:snets}Network Faults}
\end{subfigure}
\begin{subfigure}{.18\textwidth}
\centering
\includegraphics[clip, trim=3cm 0cm 8cm 2cm,width=1.2\textwidth]{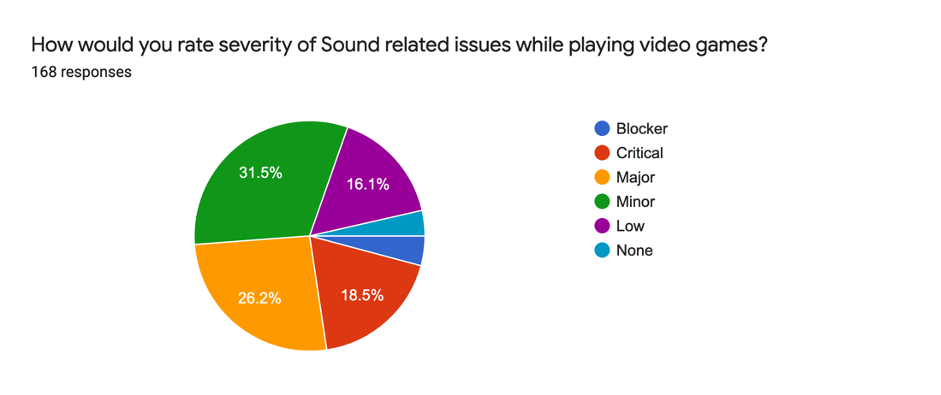}
\caption{\label{fig:ssound}Sound Faults}
\end{subfigure}
\begin{subfigure}{.18\textwidth}
\centering
\includegraphics[clip, trim=3cm 0.5cm 8cm 2cm,width=1.2\textwidth]{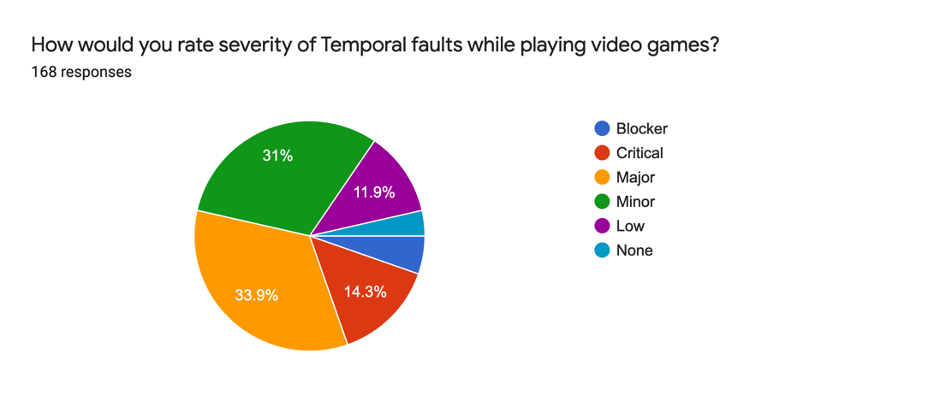}
\caption{\label{fig:sttemppp}Temporal Faults.}
\end{subfigure}

\begin{subfigure}{.18\textwidth}
\centering
\includegraphics[clip, trim=3cm 0.5cm 8cm 2cm,width=1.2\textwidth]{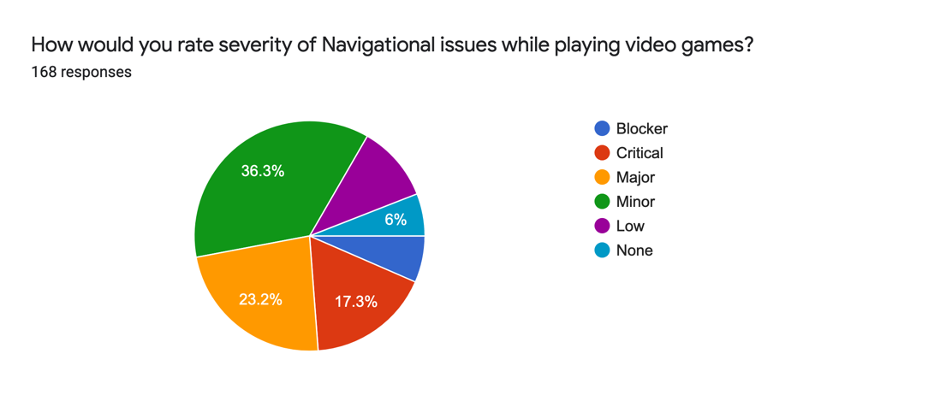}
\caption{\label{fig:snavii}Navigational Faults.}
\end{subfigure}
\begin{subfigure}{.18\textwidth}
\centering
\includegraphics[clip, trim=3cm 0.5cm 8cm 2cm,width=1.2\textwidth]{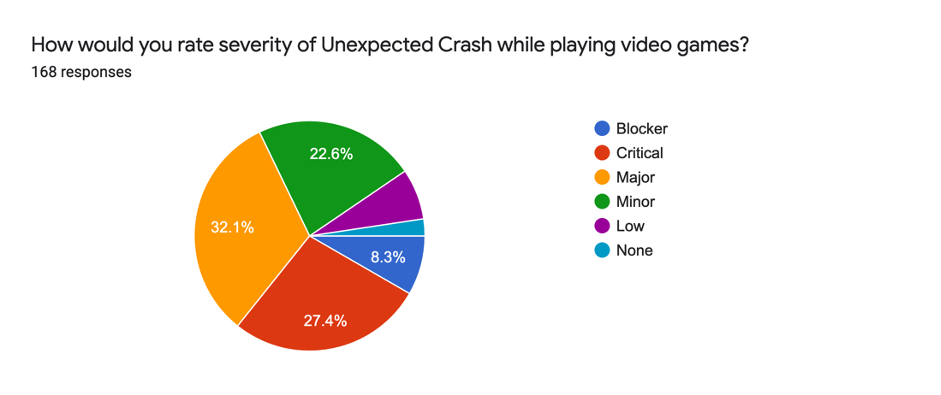}
\caption{\label{fig:scrash}Unexpected Crash  faults.}
\end{subfigure}
\begin{subfigure}{.18\textwidth}
\centering
\includegraphics[clip, trim=3cm 0.5cm 8cm 2cm,width=1.2\textwidth]{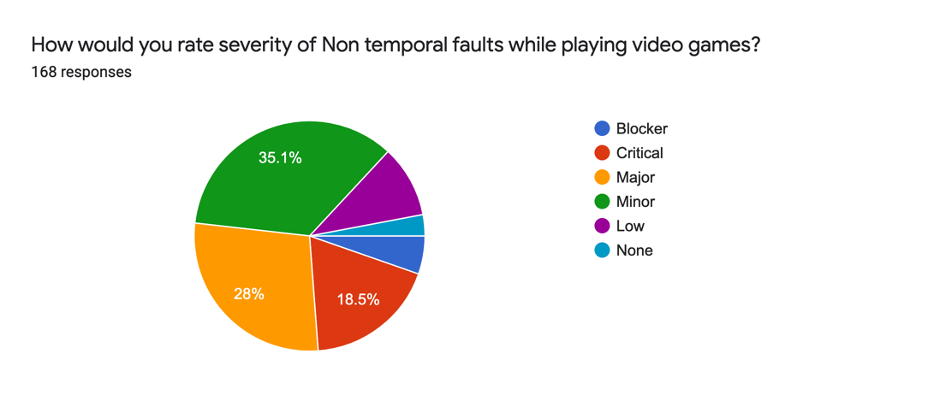}
\caption{\label{fig:snont}Non Temporal Faults.}
\end{subfigure}
\begin{subfigure}{.18\textwidth}
\centering
\includegraphics[width=0.7\textwidth]{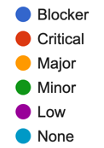}
\caption{\label{fig:sevleg}Legend.}
\end{subfigure}
\caption{\label{fig:bug-severity}Severity of game bugs according to survey responses.}
\end{figure}

\begin{figure} [h]
\begin{subfigure}{.18\textwidth}
\centering
\includegraphics[clip, trim=3cm 0.5cm 8cm 2cm,width=1.2\textwidth]{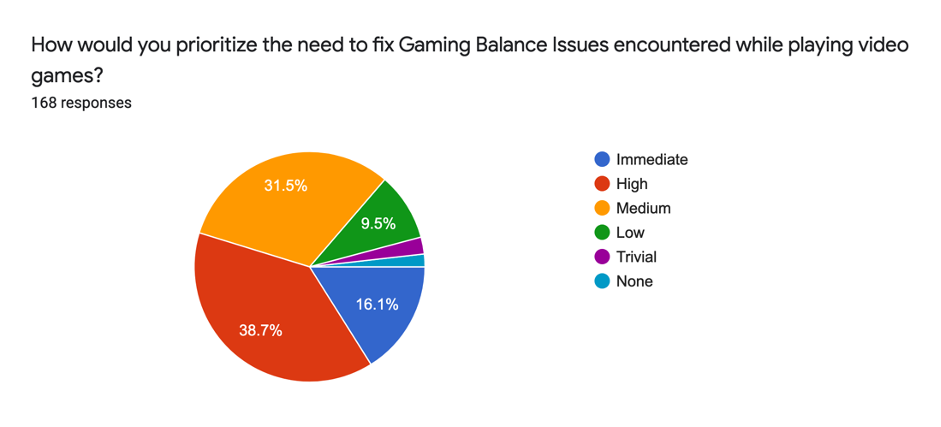}
\caption{\label{fig:pgb}Gaming Balance}
\end{subfigure}
\begin{subfigure}{.18\textwidth}
\centering
\includegraphics[clip, trim=3cm 0.5cm 8cm 2cm,width=1.2\textwidth]{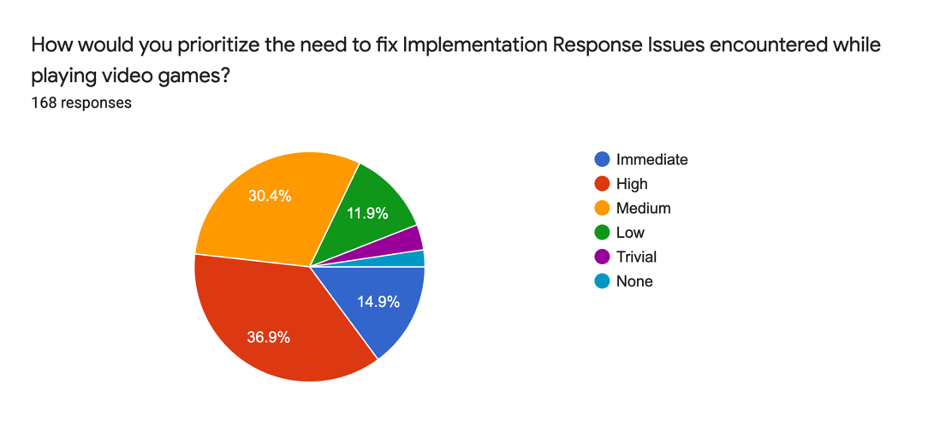}
\caption{\label{fig:pimplss}Impl. Response Faults}
\end{subfigure}
\begin{subfigure}{.18\textwidth}
\centering
\includegraphics[clip, trim=3cm 0cm 8cm 2cm,width=1.2\textwidth]{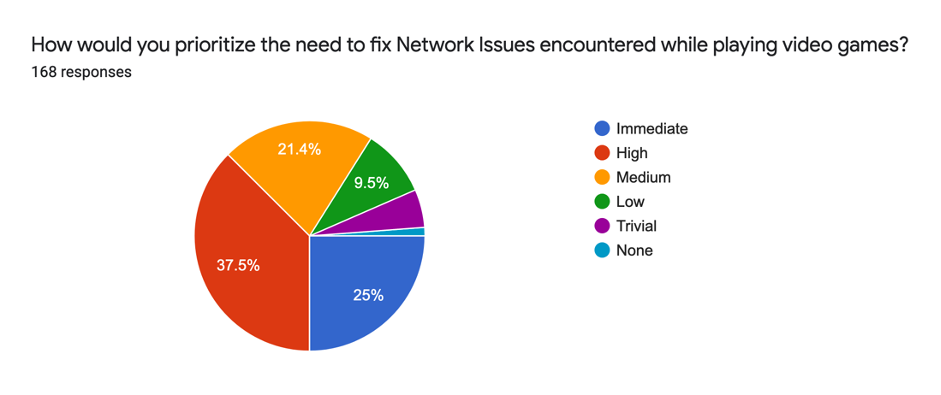}
\caption{\label{fig:pnets}Network Faults.}
\end{subfigure}
\begin{subfigure}{.18\textwidth}
\centering
\includegraphics[clip, trim=3cm 0cm 8cm 2cm,width=1.2\textwidth]{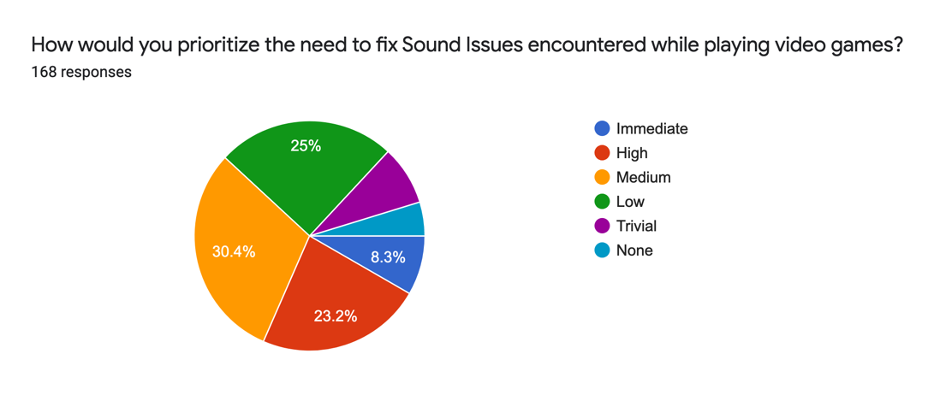}
\caption{\label{fig:psound}Sound Faults.}
\end{subfigure}
\begin{subfigure}{.18\textwidth}
\centering
\includegraphics[clip, trim=3cm 0.5cm 8cm 2cm,width=1.2\textwidth]{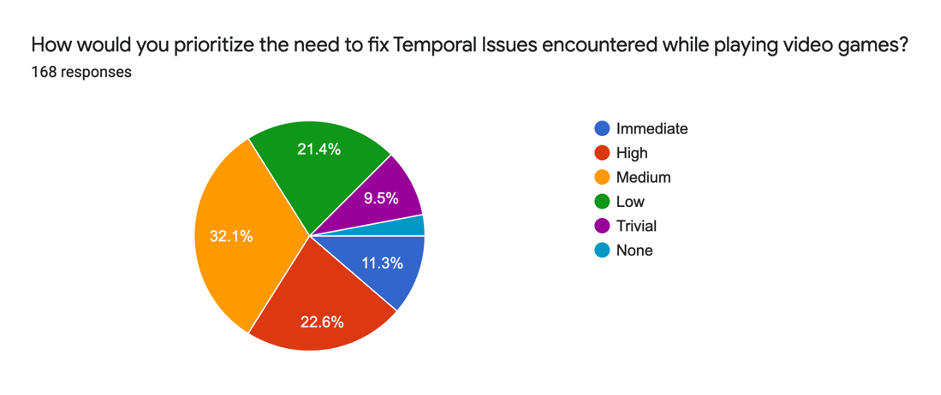}
\caption{\label{fig:pttemppp}Temporal Faults.}
\end{subfigure}
\begin{subfigure}{.18\textwidth}
\centering
\includegraphics[clip, trim=3cm 0.5cm 8cm 2cm,width=1.2\textwidth]{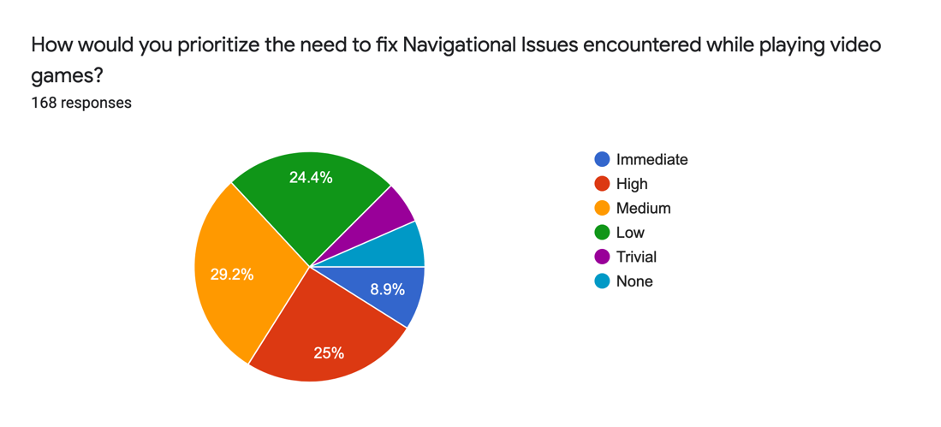}
\caption{\label{fig:pnavii}Navigational Faults.}
\end{subfigure}
\begin{subfigure}{.18\textwidth}
\centering
\includegraphics[clip, trim=3cm 0.5cm 8cm 2cm,width=1.2\textwidth]{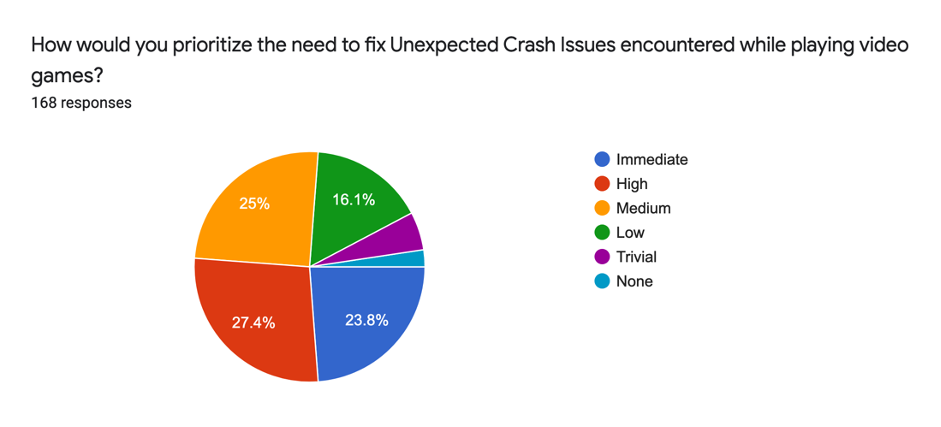}
\caption{\label{fig:pcrash}Unexpected Crash  faults.}
\end{subfigure}
\begin{subfigure}{.18\textwidth}
\centering
\includegraphics[clip, trim=3cm 0.5cm 8cm 2cm,width=1.2\textwidth]{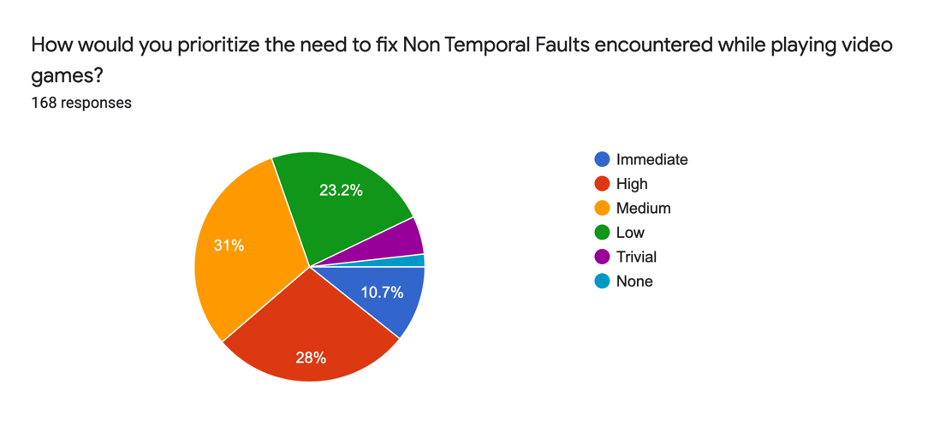}
\caption{\label{fig:pnont}Non Temporal Faults.}
\end{subfigure}
\begin{subfigure}{.18\textwidth}
\centering
\includegraphics[width=0.7\textwidth]{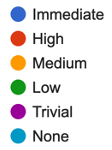}
\caption{\label{fig:prileg}Legend.}
\end{subfigure}
\caption{\label{fig:bug-priority}Bug fixing Priority of game bugs  according to survey responses.}
\end{figure}

\begin{figure}
    \centering
    \includegraphics[width=1
    \textwidth]{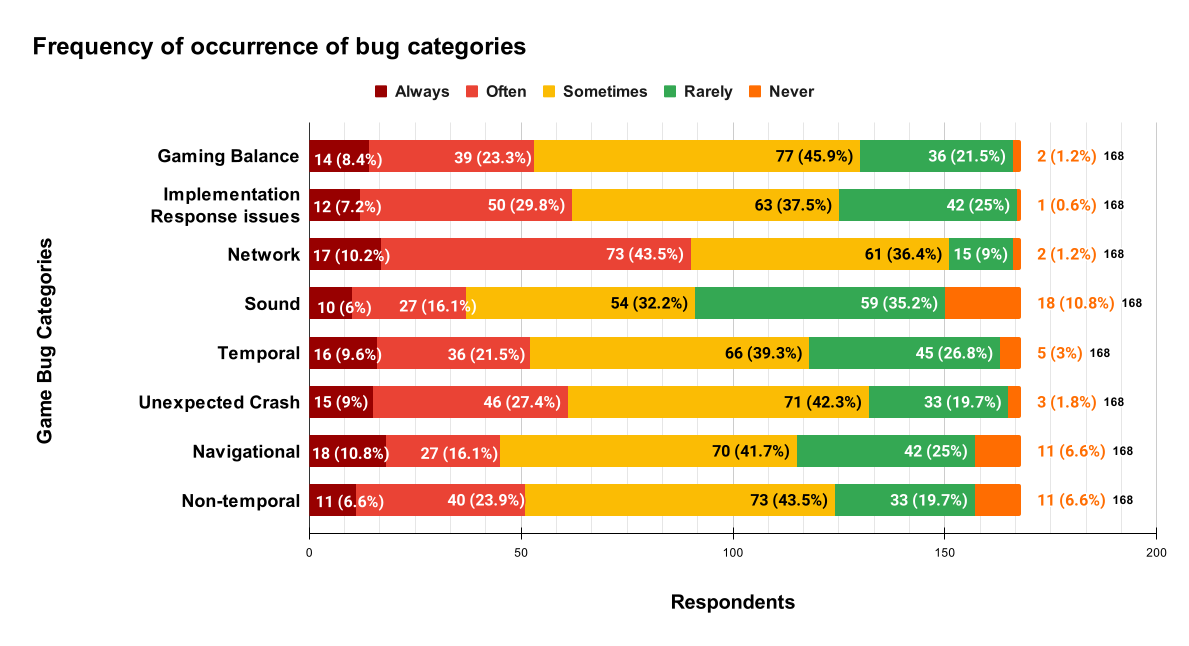}
    \caption{Frequency of occurrence of bug categories according to survey respondents.}
    \label{fig:freq}
\end{figure}

\begin{figure}[H]
    \centering
    \includegraphics[width=0.9
    \textwidth]{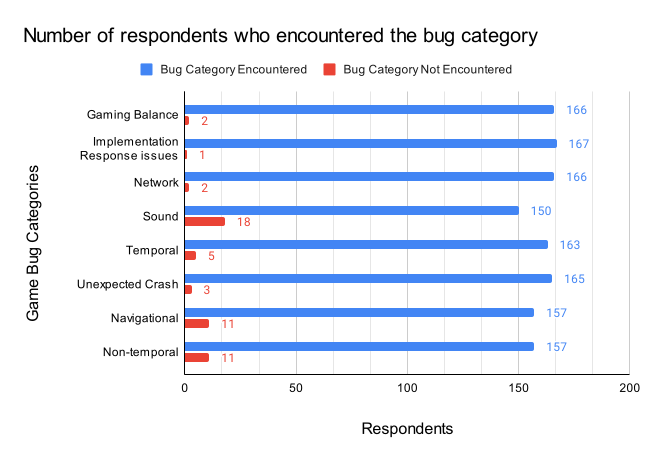}
    \caption{Respondent Count vs. Game Bug category.}
    \label{fig:ecounterfreq}
\end{figure}

\begin{figure}[H]
    \centering
    \includegraphics[width=1
    \textwidth]{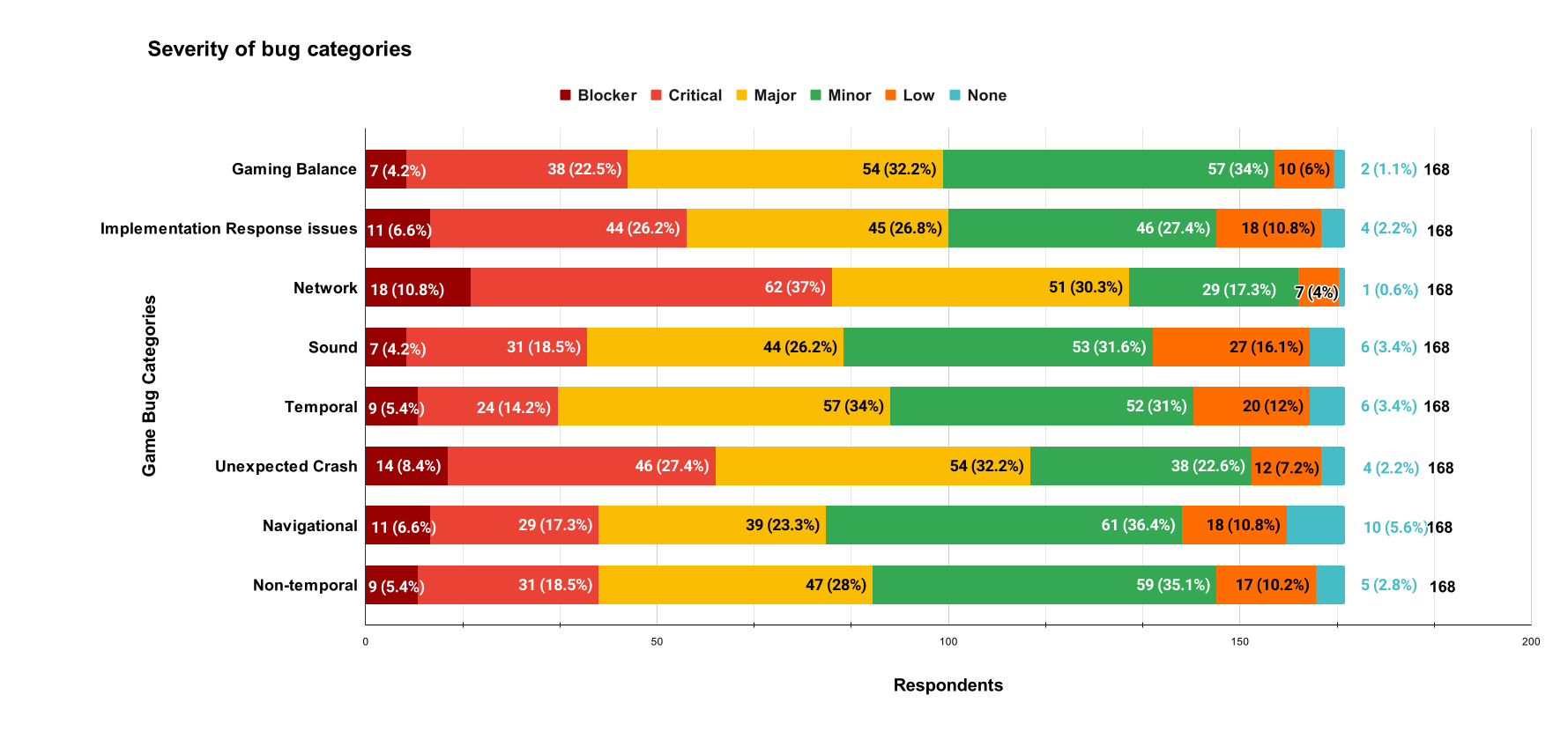}
    \caption{Severity of bug categories according to survey respondents.}
    \label{fig:sev}
\end{figure}

\begin{figure}[H]
    \centering
    \includegraphics[width=1
    \textwidth]{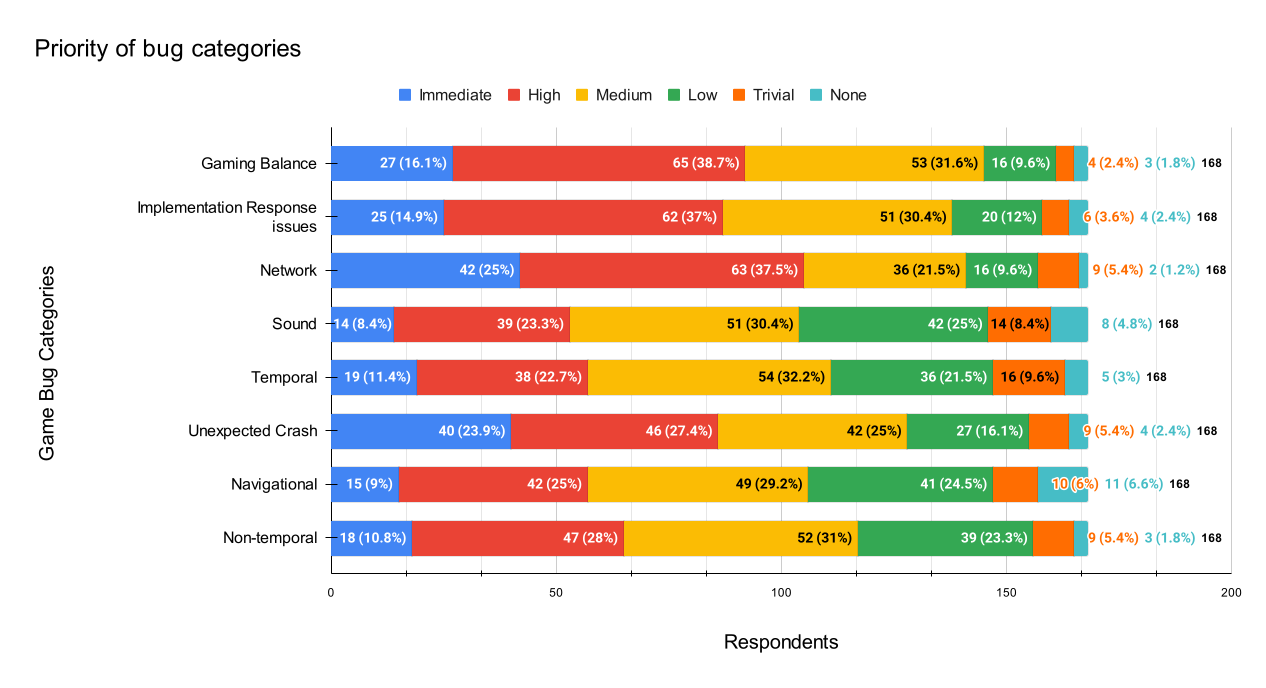}
    \caption{Priority to fix of bug categories according to survey respondents.}
    \label{fig:pri}
\end{figure}

\begin{figure}[H]
    \centering
    \includegraphics[width=0.9
    \textwidth]{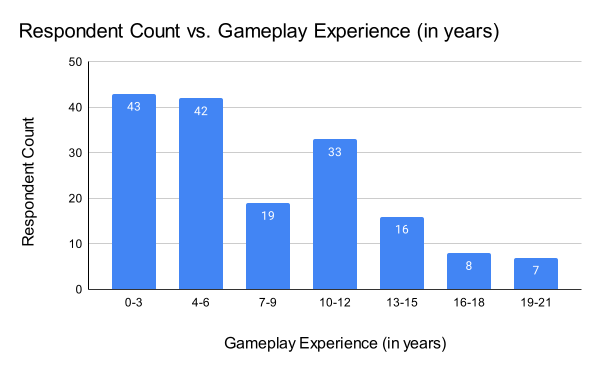}
    \caption{Respondent Count vs. Gameplay Experience (in years).}
    \label{fig:gameplay}
\end{figure}
\begin{figure}[H]
    \centering
    \includegraphics[width=0.9
    \textwidth]{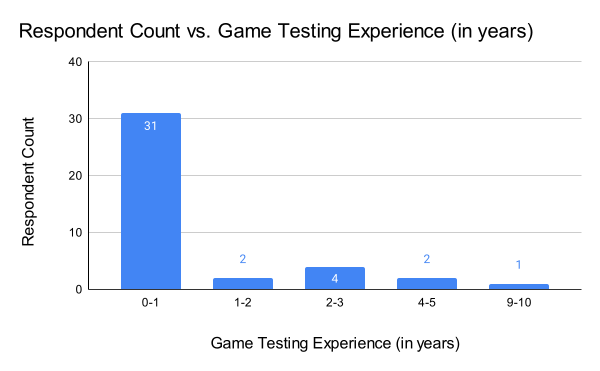}
    \caption{Respondent Count vs. Game Testing Experience (in years).}
    \label{fig:gametesting}
\end{figure}
\begin{figure}[H]
    \centering
    \includegraphics[width=0.9
    \textwidth]{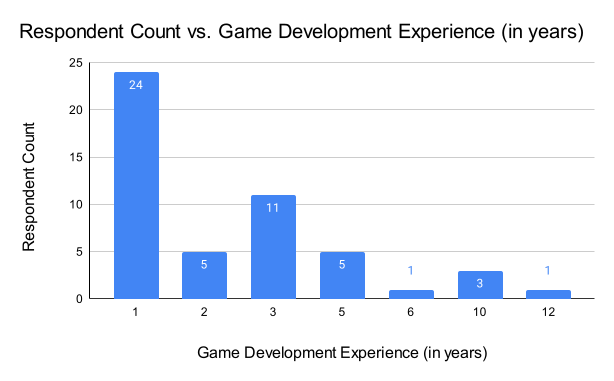}
    \caption{Respondent Count vs. Game Development Experience (in years).}
    \label{fig:gamedev}
\end{figure}

\begin{figure}
    \centering
    \includegraphics[width=0.8
    \textwidth]{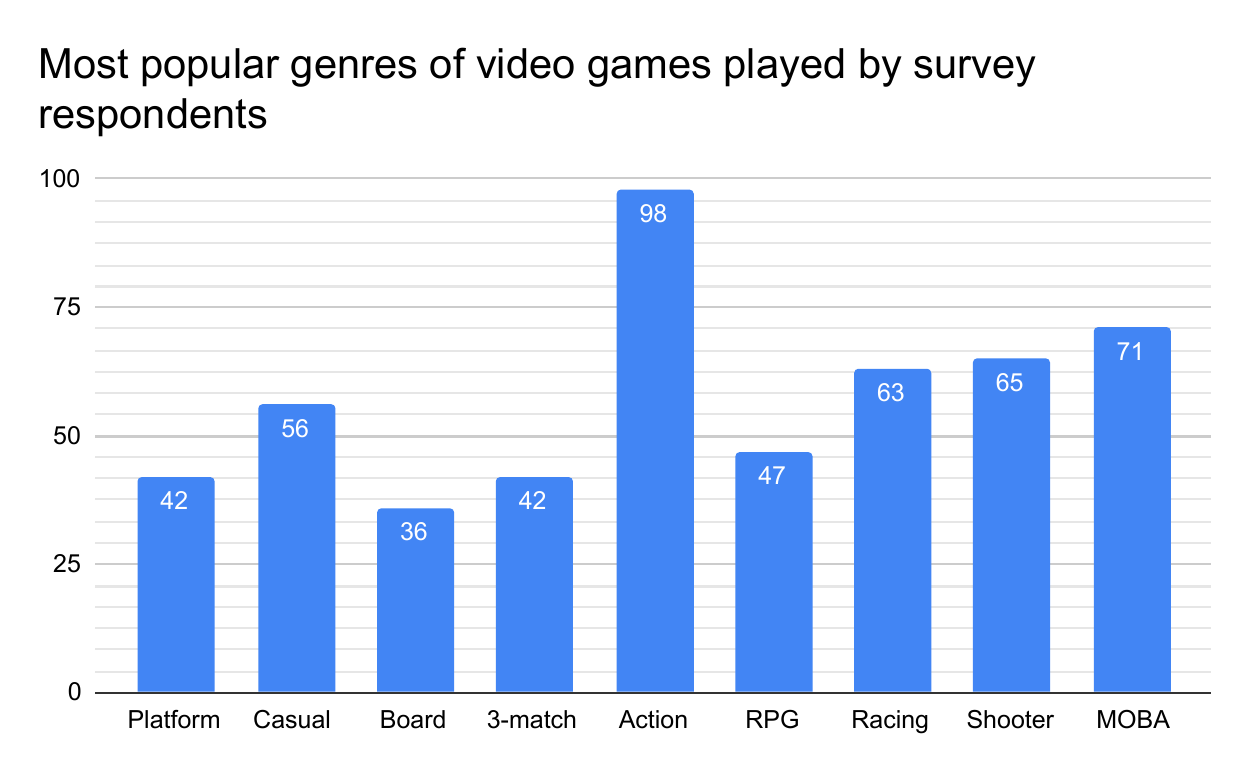}
    \caption{Most popular genres of video games played by survey respondents.}
    \label{fig:popgen}
\end{figure}

\begin{figure}
    \centering
    \includegraphics[width=0.8
    \textwidth]{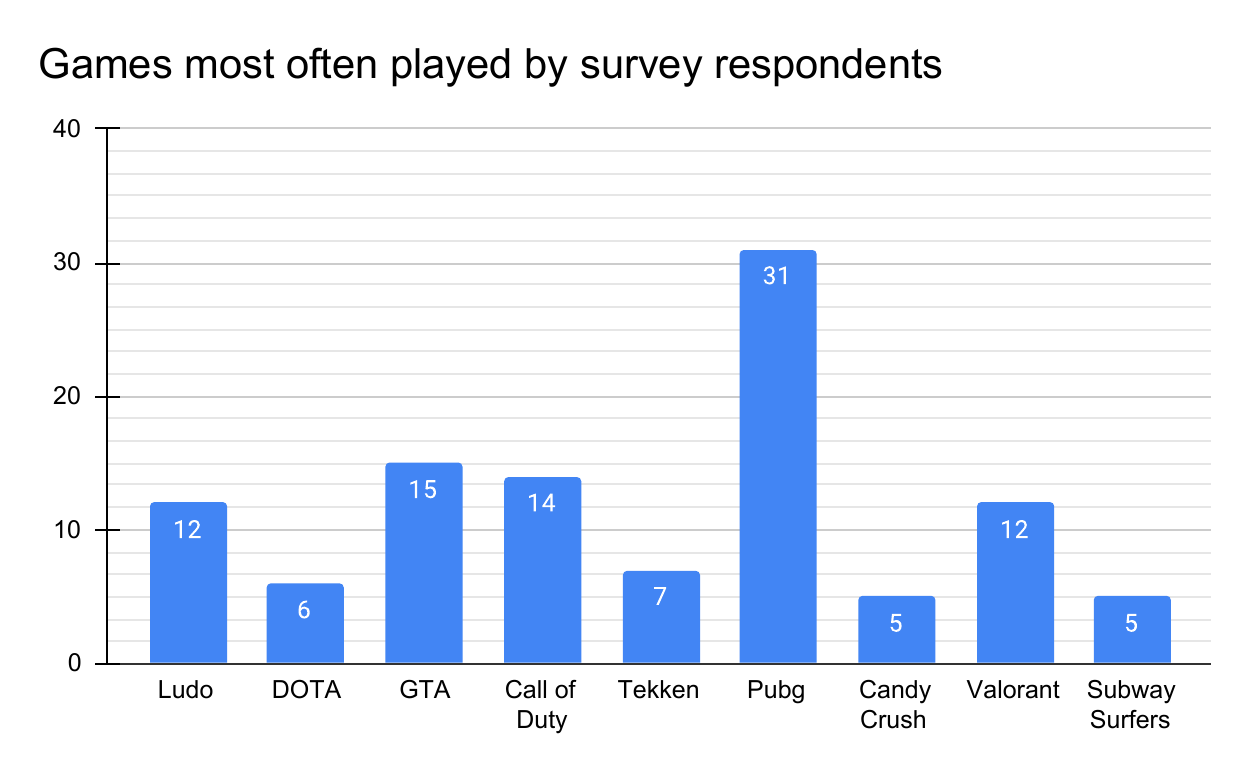}
    \caption{Games most often played by survey respondents.}
    \label{fig:popgame}
\end{figure}

\begin{figure} [H]
\begin{subfigure}{\textwidth}
\centering
\includegraphics[width=0.65\textwidth]{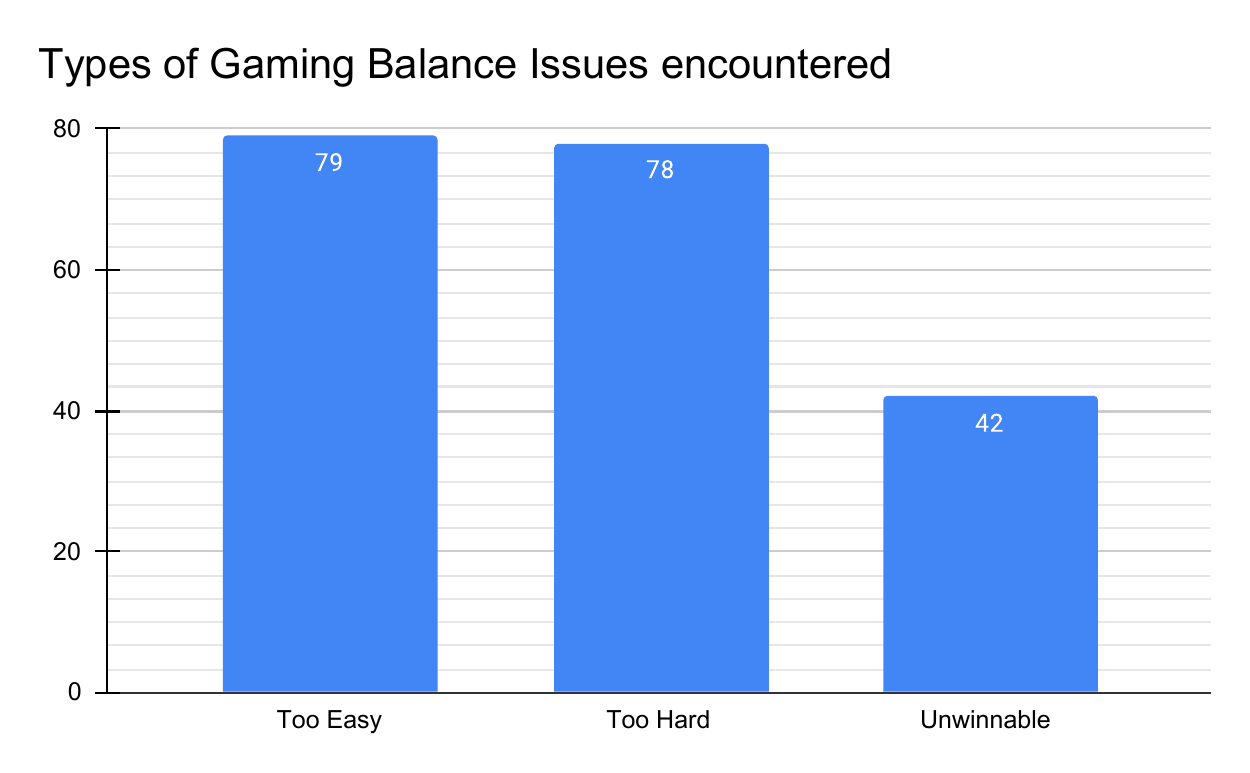}
\caption{\label{fig:fsgb}Gaming Balance}
\end{subfigure}

\begin{subfigure}{1\textwidth}
\centering
\includegraphics[clip, trim=0.4cm 0.9cm 0cm 0cm, width=0.75\textwidth]{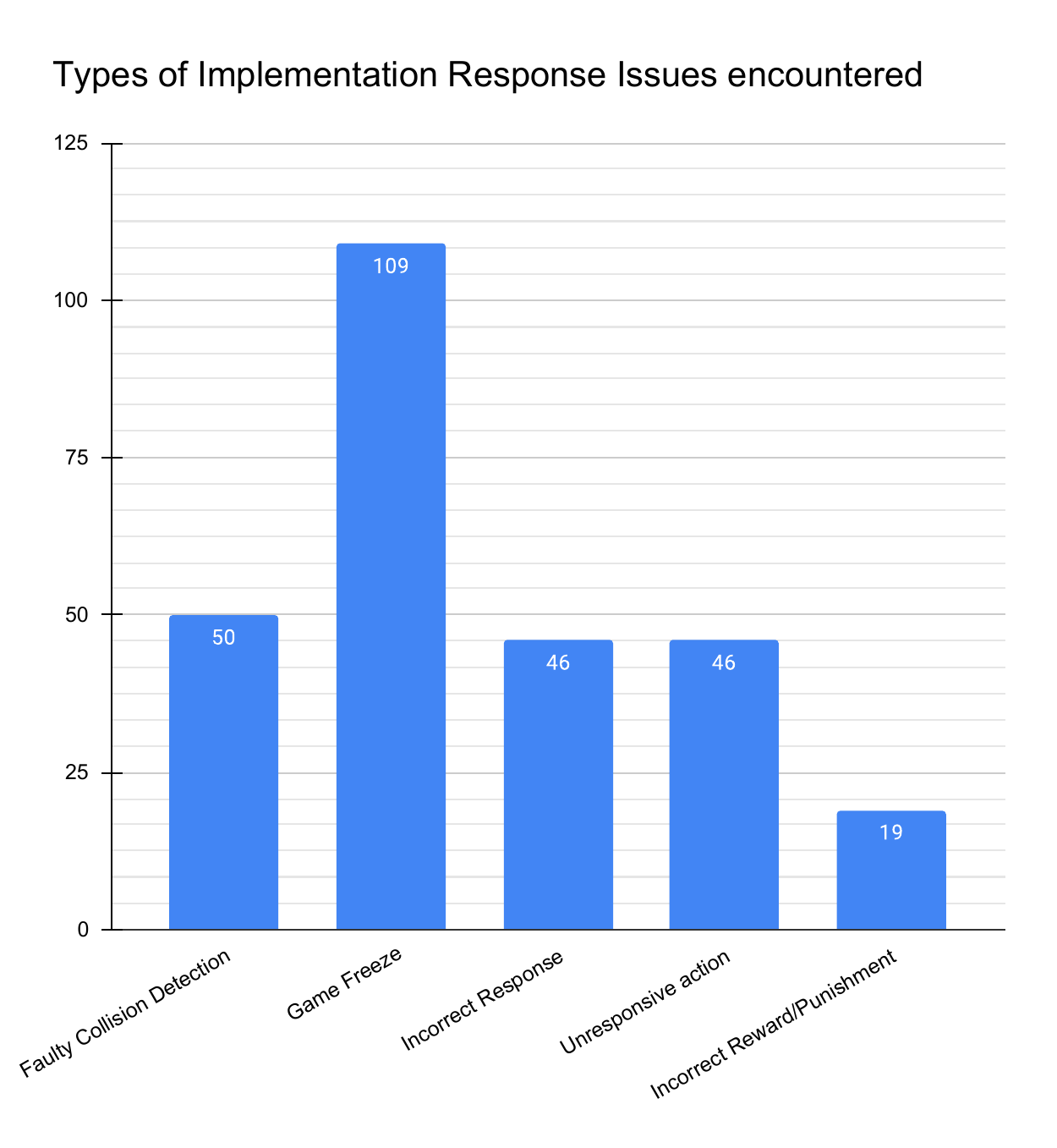}
\caption{\label{fig:fsimpl}Implementation Response Faults}
\end{subfigure}

\caption{\label{fig:bug-freq}Types of different bug taxonomy categories encountered according to survey respondents.}
\end{figure}

\begin{figure} [H]\ContinuedFloat

\begin{subfigure}{\textwidth}
\centering
\includegraphics[width=0.65\textwidth]{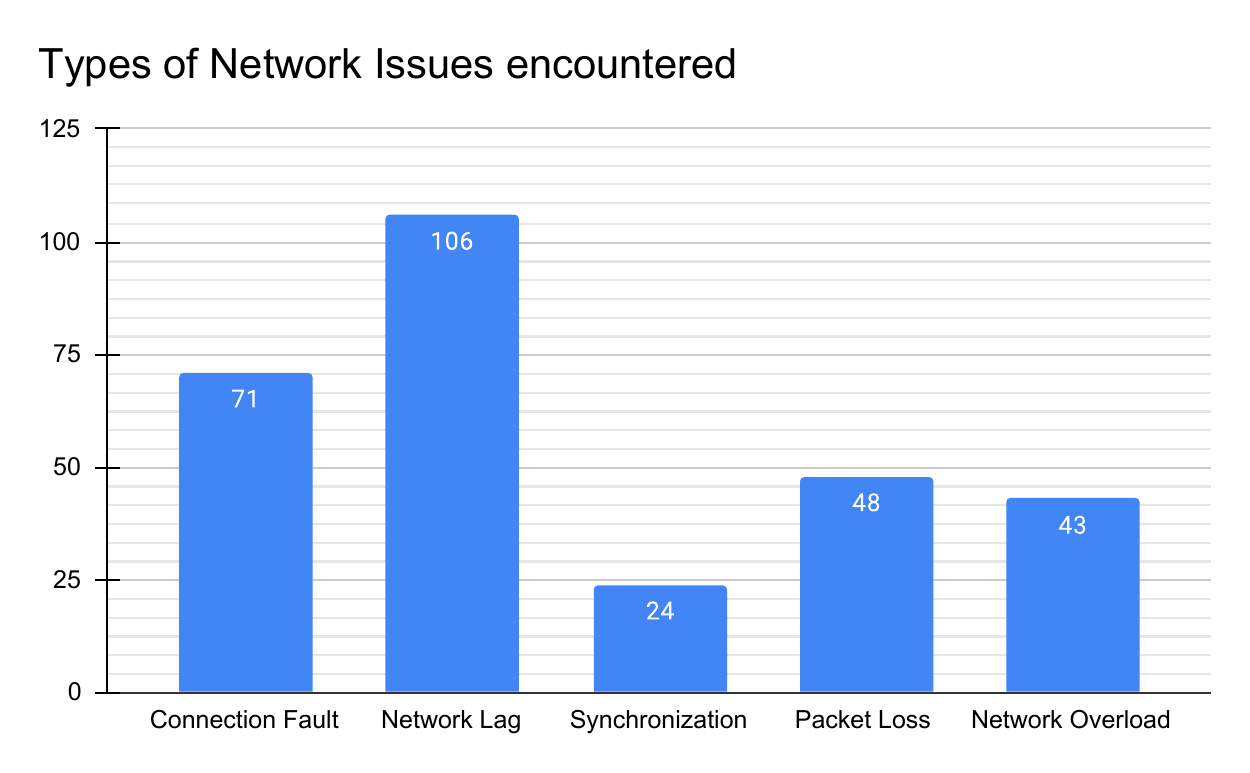}
\caption{\label{fig:fsnet}Network Faults}
\end{subfigure}

\begin{subfigure}{\textwidth}
\centering
\includegraphics[width=0.65\textwidth]{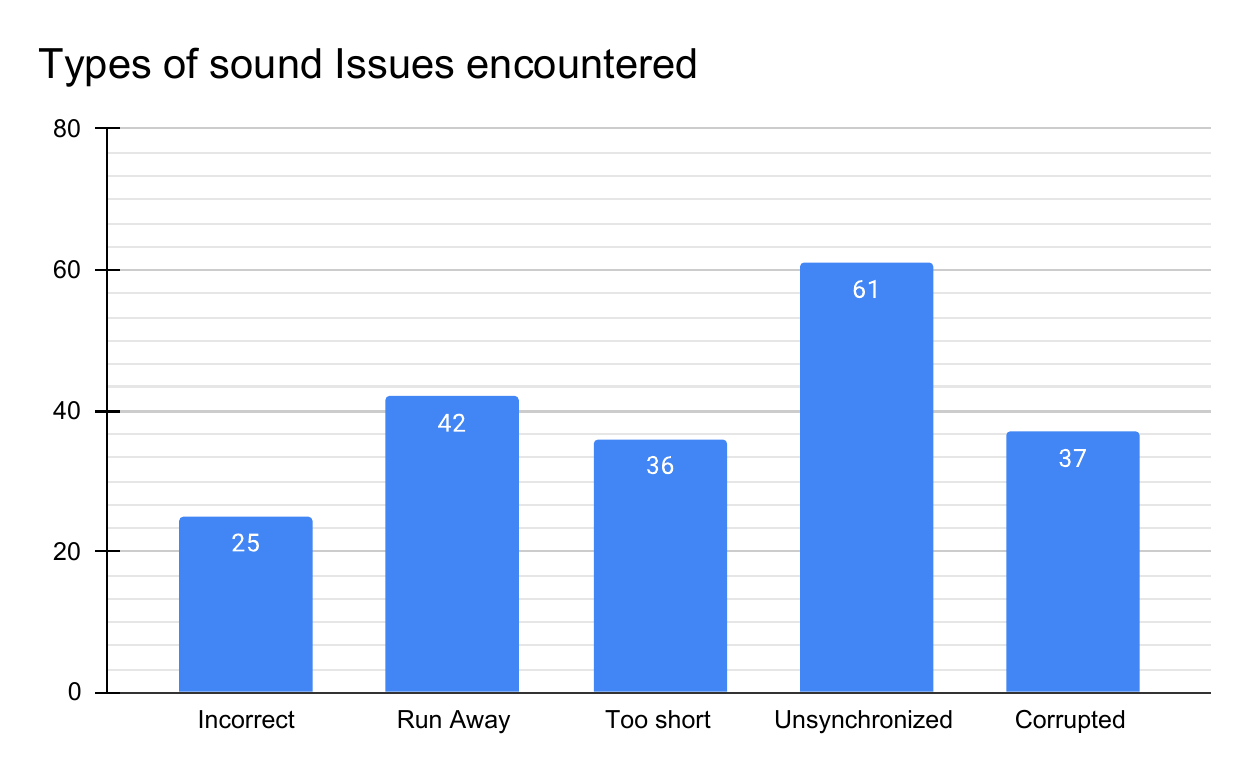}
\caption{\label{fig:fss}Sound Faults}
\end{subfigure}

\begin{subfigure}{\textwidth}
\centering
\includegraphics[width=0.65\textwidth]{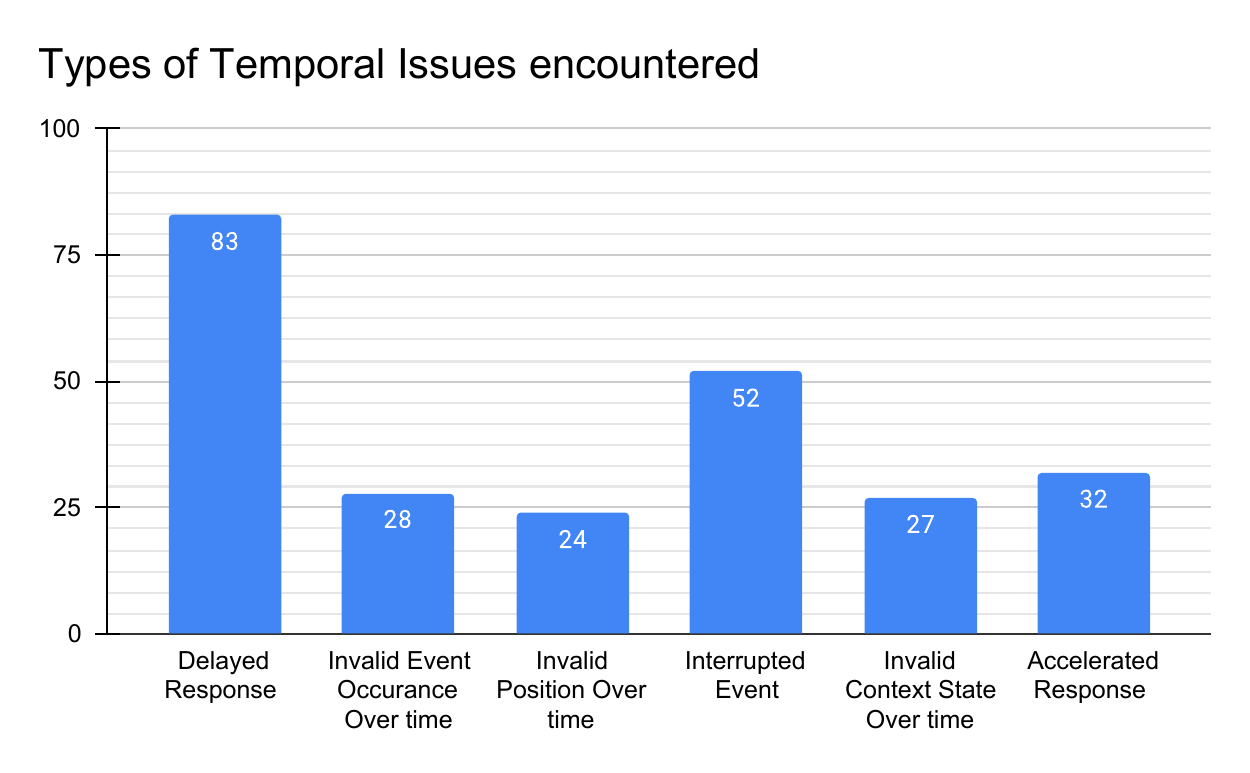}
\caption{\label{fig:fst}Temporal Faults}
\end{subfigure}

\caption{\label{fig:bug-freq}Types of different bug taxonomy categories encountered according to survey respondents (continued from previous page.).}
\end{figure}

\begin{figure} [H]\ContinuedFloat

\begin{subfigure}{\textwidth}
\centering
\includegraphics[width=0.65\textwidth]{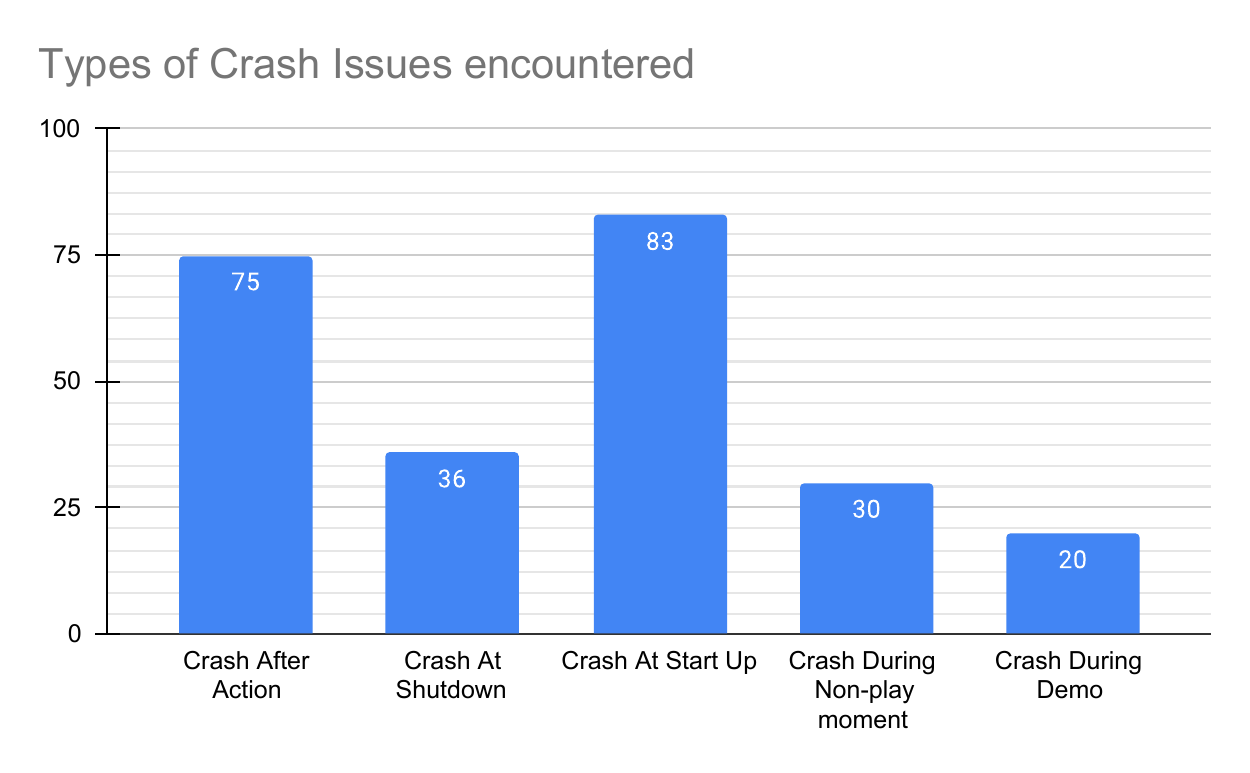}
\caption{\label{fig:fscr}Unexpected Crash}
\end{subfigure}

\begin{subfigure}{0.9\textwidth}
\centering
\includegraphics[width=0.65\textwidth]{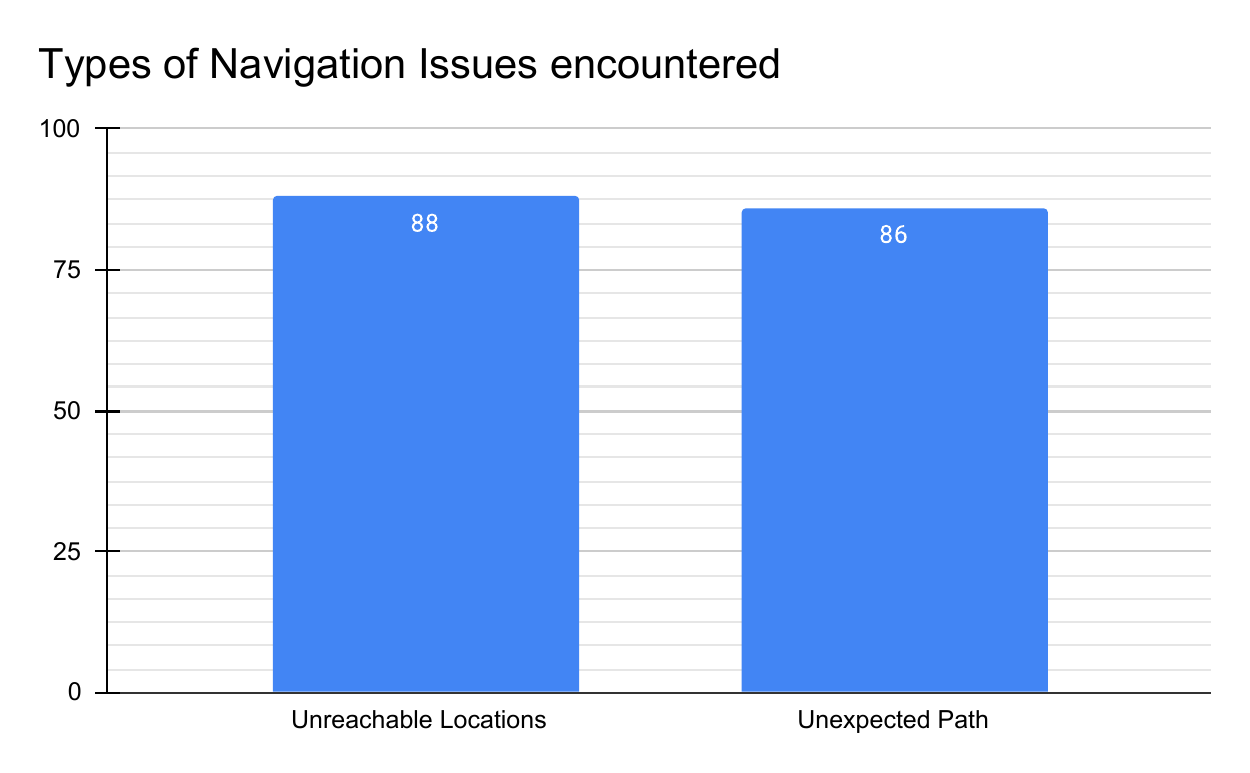}
\caption{\label{fig:fsnavi}Navigational Faults}
\end{subfigure}

\begin{subfigure}{1\textwidth}
\centering
\includegraphics[clip, trim=0.7cm 1.5cm 0cm 0cm, width=0.75\textwidth]{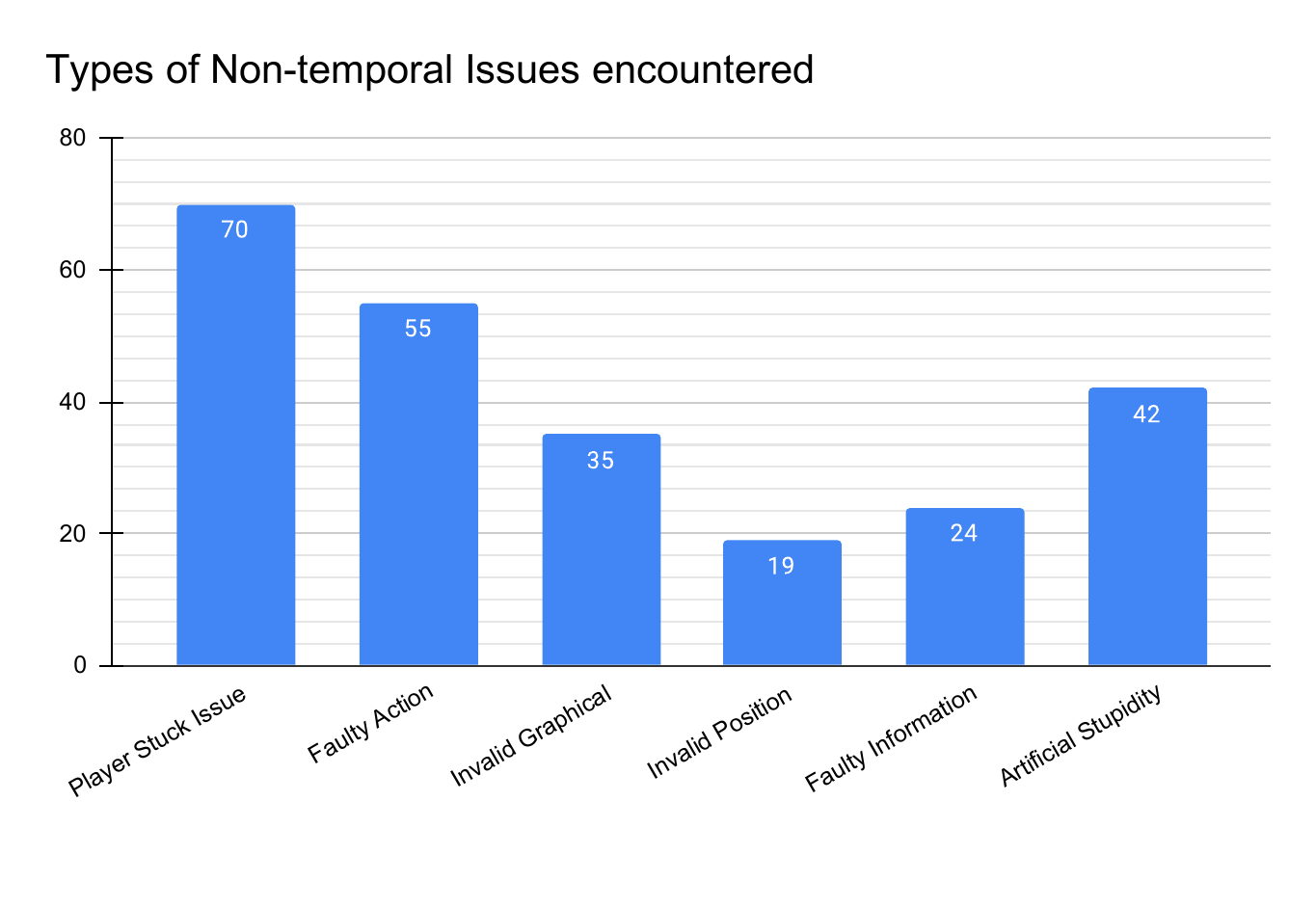}
\caption{\label{fig:fst}Non-Temporal Faults}
\end{subfigure}

\caption{\label{fig:bug-freq}Types of different bug taxonomy categories encountered according to survey respondents (continued from previous page).}
\end{figure}

\end{document}